%% file: Paper5.tex
\documentclass[Journal]{IEEEtran} %draftclsnofoot

\usepackage{cite} 
\usepackage{graphicx}
\usepackage[english]{babel}
\usepackage{amsmath}
\usepackage{amsthm}
\usepackage[table,xcdraw]{xcolor}
\usepackage{subfig}
\usepackage[linesnumbered, ruled]{algorithm2e}
\usepackage[noend]{algpseudocode}
\usepackage[font=footnotesize]{caption}
\usepackage{epsfig}
\usepackage{xfrac}
\usepackage{nicefrac}
\usepackage{psfrag}
\usepackage{times}
\usepackage{balance} 
\usepackage{amssymb}
\usepackage{amsfonts}
\usepackage{bm}
\usepackage[bottom]{footmisc}
\usepackage{mwe}
\usepackage{color}
\usepackage{booktabs,multirow}
\usepackage{tabularx}
\usepackage{tikz}
\usepackage{kotex}
\usetikzlibrary{patterns,arrows}
\usepackage{physics}

\usepackage{algorithm}

\usepackage{verbatim}

\SetKwRepeat{Do}{do}{while}
\SetKwInOut{Input}{Input}
\SetKwInOut{Output}{Output}

\definecolor{color1}{RGB}{228,26,28}
\definecolor{color2}{RGB}{55,126,184}
\definecolor{color3}{RGB}{77,175,74}
\definecolor{color4}{RGB}{152,78,163}
\definecolor{color5}{RGB}{255,127,0}
\definecolor{color6}{RGB}{200,200,200}

\begin{document}

\title{Learning Emergent Random Access Protocol for LEO Satellite Networks}
%
%Emergent Random Access Protocol for LEO Satellite Networks

\author{Ju-Hyung Lee,~\IEEEmembership{Member,~IEEE,}
		Hyowoon Seo,~\IEEEmembership{Member,~IEEE,}
		Jihong Park,~\IEEEmembership{Member,~IEEE,}\\
		Mehdi Bennis,~\IEEEmembership{Fellow,~IEEE,}
        and~Young-Chai Ko,~\IEEEmembership{Senior Member,~IEEE}
% <-this % stops a space
\thanks{J.-H. Lee and Y.-C. Ko are with the School of Electrical and Computer Engineering, Korea University, Seoul 02841, Korea (e-mail: \{leejuhyung, koyc\}@korea.ac.kr).}
\thanks{J. Park is with the School of Information Technology, Deakin University, Geelong, VIC 3220, Australia (e-mail: jihong.park@\{deakin.edu.au,\ gist.ac.kr\}).}
\thanks{H. Seo and M. Bennis are with the Centre for Wireless Communications, University of Oulu, Oulu 90014, Finland (e-mail: \{hyowoon.seo, mehdi.bennis\}@oulu.fi).
}
% <-this % stops a space
% \thanks{Manuscript received April XX, 20XX; revised January XX, 20XX.}
}
\maketitle

%%%%%%%%%%%%%%%%%%%%%%%%%%%%%%%%%%%%%%%%%%%%%%%%%%%%%%%%%%%%%%%%%
%%%%%%%%%%%%%%%%%%% ABSTRACT %%%%%%%%%%%%%%%%%%%%%%%%%%%%%%%%%%%%
%%%%%%%%%%%%%%%%%%%%%%%%%%%%%%%%%%%%%%%%%%%%%%%%%%%%%%%%%%%%%%%%%
\begin{abstract}
A mega-constellation of low-altitude earth orbit (LEO) satellites (SATs) are envisaged to provide a global coverage SAT network in beyond fifth-generation (5G) cellular systems. 
LEO SAT networks exhibit extremely long link distances of many users under time-varying SAT network topology. 
This makes existing multiple access protocols, such as random access channel (RACH) based cellular protocol designed for fixed terrestrial network topology, ill-suited. 
To overcome this issue, in this paper, we propose a novel grant-free random access solution for LEO SAT networks, dubbed \emph{emergent random access channel protocol (eRACH)}. 
In stark contrast to existing model-based and standardized protocols, eRACH is a model-free approach that emerges through interaction with the non-stationary network environment, using multi-agent deep reinforcement learning (MADRL).  
Furthermore, by exploiting known SAT orbiting patterns, eRACH does not require central coordination or additional communication across users, while training convergence is stabilized through the regular orbiting patterns.
Compared to RACH, we show from various simulations that our proposed eRACH yields $54.6$\% higher average network throughput with around two times lower average access delay while achieving $0.989$ Jain's fairness index.

% Lastly, by utilizing locally observable information, we validate that the eRACH naturally emerges from a given local environment while reducing the communication overhead between user terminals, that highlighting the applicability of the proposed LEO SAT-oriented access.
\end{abstract}


%%%%%%%%%%%%%%%%%%%%%%%%%%%%%%%%%%%%%%%%%%%%%%%%%%%%%%%%%%%%%%%%%
%%%%%%%%%%%%%%%%%%% INTRODUCTION %%%%%%%%%%%%%%%%%%%%%%%%%%%%%%%%
%%%%%%%%%%%%%%%%%%%%%%%%%%%%%%%%%%%%%%%%%%%%%%%%%%%%%%%%%%%%%%%%%
\section{Introduction} 
We are at the cusp of an unprecedented revolution led by thousands of low earth orbit (LEO) satellites (SATs) in space. SpaceX has already launched more than 1,500 Starlink LEO SATs \cite{SAT_Survey1, SAT_UCL_1, Starlink} covering 12 countries in 3 different continents \cite{SAT_Industry_Comparison}. Besides, federal communications commission (FCC) has recently authorized Amazon Project Kuiper to launch half of 3,236 LEO SATs by 2026 and the rest by 2029 \cite{FCC_Kuiper}. The ramification of this LEO SAT mega-constellation is not only limited to advancing conventional SAT applications such as high-precision earth observation \cite{SAT_Survey5} but also envisaged to open up a new type of wireless connectivity, i.e., non-terrestrial networks (NTNs) in beyond the fifth-generation (5G) or 6G cellular systems \cite{SAT_Survey6}. While each SAT plays a role as a mobile base station (BS), a mega-constellation of LEO SATs has a great potential in provisioning fast and reliable connectivity to ground users anywhere in the globe, including ocean, rural areas, and disaster sites \cite{UAV_Survey1, UAV_Survey2, UAV_Survey3}.

% LEO SAT constellation, which enables large-scale networks, has emerged as a dominant paradigm for beyond 5G communication.
% The recent deployment of LEO SAT mega-constellations by industry projects and their realistic blueprint spur on this trend.
% Currently, several companies, including SpaceX, OneWeb, and Amazon, aggressively invest in the LEO SAT network while getting approvals from the federal communications commission (FCC) to operate the LEO SAT networks .
% The industry projects toward an integrated satellite-terrestrial network include the OneWeb constellation with 648 satellites, and Amazon approved by FCC to launch 3,236 spacecraft in its Kuiper constellation; however, the most effective activities are taken by SpaceX.
% Notably, SpaceX's more than 1500 deployed Starlink SATs have already supported more than 12 countries in North America, Europe, and Oceania continents \cite{SAT_Industry_Comparison}. 

Notwithstanding, each SAT BS has an extremely wide coverage (e.g., $160 - 1000$ [km] inter-SAT BS distance \cite{SAT_Survey2}), disallowing sophisticated central coordination among SAT BSs and users in real-time under limited communication and computing resources. 
Furthermore, as opposed to fixed terrestrial BSs, SAT BSs move, requiring location-specific resource management. 
Unfortunately, existing model-based and standardized protocols, such as carrier-sense multiple access (CSMA) based WiFi protocols and random access channel (RACH) based cellular protocols, cannot flexibly optimize their operations without incurring severe protocol fragmentation, not to mention a significant effort in protocol standardization for a myriad of possible scenarios.

% \tred{[JH: On second thought, mentioning 2-step without explaining what it means is not appropriate. If we cannot explain it here (+ actually there are 4 steps on p.5 including association and data transmission, which may be confused), now I think that it'd be better to mention the 2-step not here and in the definition of eRACH, but somewhere later. Here, if we really need to address 2-step, it'd be better to say grant-free in the definition.]}
To overcome the aforementioned fundamental challenges in LEO SAT networks, in this article, we develop a novel model-free random access (RA) protocol that naturally emerges from a given local SAT random access environment, dubbed \emph{emergent random access channel protocol (eRACH)}. 
In eRACH, the emergence of RA protocol is induced by deep reinforcement learning (DRL) at the ground user agents by utilizing locally observable information, not requiring inter-agent communication or centralized training.
Consequently, eRACH jointly takes into account SAT associations and RA collision avoidance in a given local environment, thereby achieving lower RA latency and high downstream communication efficiency after eRACH.

The fundamentals of eRACH are laid by \emph{protocol learning} via multi-agent DRL (MADRL). Several recent works have underpinned the effectiveness of protocol learning \cite{MARL_Foerster1, Semantic_Hoydis1, Semantic_Hoydis2} in that the learned protocols can adapt to various environments while achieving higher communication efficiency than those of model-based protocols. However, these benefits come at the cost of additional inter-agent communication for MADRL, which are non-negligible under non-stationary network topologies, questioning the feasibility of protocol learning for LEO SAT networks.
% the feasibility of such protocol learning is significantly challenged under 
In fact, MADRL agents in \cite{MARL_Foerster1, Semantic_Hoydis1, Semantic_Hoydis2} learn to communicate by exchanging local state information of each agent using a few bits, where the information exchange between agents is often referred to as a \emph{cheap talk}. 
Such cheap talks may no longer be cheap under a non-stationary topology of an LEO SAT network, which may require a large amount of local state information for RA and SAT BS associations. 
% For instance, user location information may suffice for fixed terrestrial BSs, but RA decisions and associations in LEO SAT network may additionally require SAT BS locations that change over time.

% In this respect, our proposed eRACH is designed based on extensive testing of locally observable information candidates and their impacts on MADRL and LEO SAT network performances. 
In this respect, our proposed eRACH is designed based on locally observable information.
We extensively test which of the locally observable information candidates are essential on eRACH training. 
Surprisingly, it is shown that eRACH does not even require cheap talks. 
Instead, eRACH exploits \textit{1)} the expected SAT location that is known a priori owing to the pre-determined SAT orbiting pattern and \textit{2)} the collision events that are inherently observable without additional costs.
While training eRACH, we have validated that the expected SAT location contains sufficient information on the network topology, as long as the variance between the expected and the actual SAT locations is less than 1~[km].
Furthermore, we have realized that the collision event contains sufficient information on how crowded each SAT BS is. 
Given \textit{1)} and \textit{2)}, thanks to the periodic orbiting pattern of LEO SAT, the problem of MADRL frequently revisits the almost identical environment, ensuring to discover an optimal protocol carrying out SAT BS association and RA decisions.

% \tred{
% Thanks to the ultra-wide coverage and periodical orbital move of the LEO SAT network, it can support numerous scattered nodes on Earth, including urban as well as rural areas.
% However, the scattered innumerable nodes in the ground are infeasible to be fully coordinated, such that have led to strong demand for next-generation multiple access schemes.
% However, in the scenarios of LEO SAT networks, the ground nodes within a SAT coverage can be vast and the scattered innumerable nodes are hardly coordinated, which calls for new designs of random access (RA) protocol. 
% That is, we need a new form of multiple access for integrated terrestrial and aerial networks.
% }

%%%%%%%%%%%%%%%%
% \subsection{Contributions and Paper Organization}
% The main contributions of this work are listed as follows.
We summarize our contributions in this paper as follows.
\begin{itemize}
% \item For multiple access of LEO SAT networks, a novel random access protocol called eRACH is designed that optimizes the access actions of terrestrial nodes. 
% % so as to maximize the system throughput
% % Considering throughput, collision, and access delay, we analyze the performance of the proposed access protocol.
% Analyzing the proposed access protocol, in the MAC layer perspective, the collision rate and access delay are considered, while in the physical layer perspective, the achievable throughput taking account of a periodical move of LEO SAT constellation is considered.
% Especially, we consider a robust achievable throughput model with taking account of a periodical orbital move of the LEO SAT constellation, which is distinctive with existing works considering only a success of access. 
% To the best of our knowledge, this is the first work on LEO SAT network-oriented random access protocol.

\item We propose a novel emergent RA protocol for a ground-to-LEO SAT communication network, dubbed eRACH. 
To the best of our knowledge, this is the first of its kind to show that a new protocol that collaborates between multi-agents can emerge by utilizing just locally observable information, especially suited for LEO SAT networks.
This is done by developing an actor-critic-based neural network architecture and a fully distributed MADRL framework without any inter-agent communication (see \textbf{Algorithm~\ref{Algorithm}} in Sec.~\ref{Body}). 

\item To provide an upper-bound performance, we additionally introduce a cheap-talk inter-agent communication into eRACH, and propose a variant of eRACH, termed eRACH-Coop (see Sec.~\ref{SimulationSetting}). 

\item 
Numerical results corroborate that while eRACH-Coop achieves the highest average network throughput with the lowest collision rate at the cost of the cheap talk overhead, eRACH still achieves up to $6.02$x and $54.6 \%$ higher average network throughput with $10$x and $2$x lower average RA latency than slotted ALOHA and conventional cellular RACH, respectively (see \textbf{Table \ref{Table_Proposed}} in Sec.~\ref{Numerical Result}). 

\item
Furthermore, the distributed operations and the limited local information of eRACH inherently promote equal RA opportunities for all users, yielding 23.6\% higher Jain's fairness than eRACH-Coop (see \textbf{Table \ref{Table_Proposed}} in Sec.~\ref{Numerical Result}).

% This suggests efficient access decision considering the orbital motion of LEO satellites provides the throughput and access delay improvements.
% Furthermore, it also highlights the importance of ... in enabling high-throughput LEO SAT communication.
\end{itemize}

The remainder of this article is organized as follows. 
In Sec.~\ref{Background}, we first summarize the RA protocols for the traditional stationary SAT and for the emerging non-stationary SAT networks.
In Sec.~\ref{System Model}, network model, system scenario, and performance matrices are presented. 
Then, the emergent contention-based RA for the LEO SAT networks, called eRACH, is proposed, which is addressed by our proposed multi-agent Actor-Critic based algorithm in Sec.~\ref{Body}.
In Sec.~\ref{Numerical Result}, simulation results are provided, followed by concluding remarks in Sec.~\ref{conclusion}.

\textit{Notation:}
Throughout this paper, we use the normal-face font to denote scalars, and boldface font to denote vectors.
We use $\mathbb{R}^{D\times 1}$ to represent the $D$-dimensional space of real-valued vectors.
We also use $\|\cdot\|$ to denote the $L^2$-norm, which is a Euclidean norm, and use $(\cdot)^{\dag}$ to represents conjugate transpose.
$\nabla_{\mathbf{x}} f(\mathbf{x})$ denotes the gradient vector of function $f(\mathbf{x})$, i.e., its components are the partial derivatives of $f(\mathbf{x})$.
$\mathbf{I}_{N}$ is the identity matrix of size $N$.

\begin{figure}
  % \vspace{-1em}
  \centering
  \includegraphics[width=1\columnwidth]{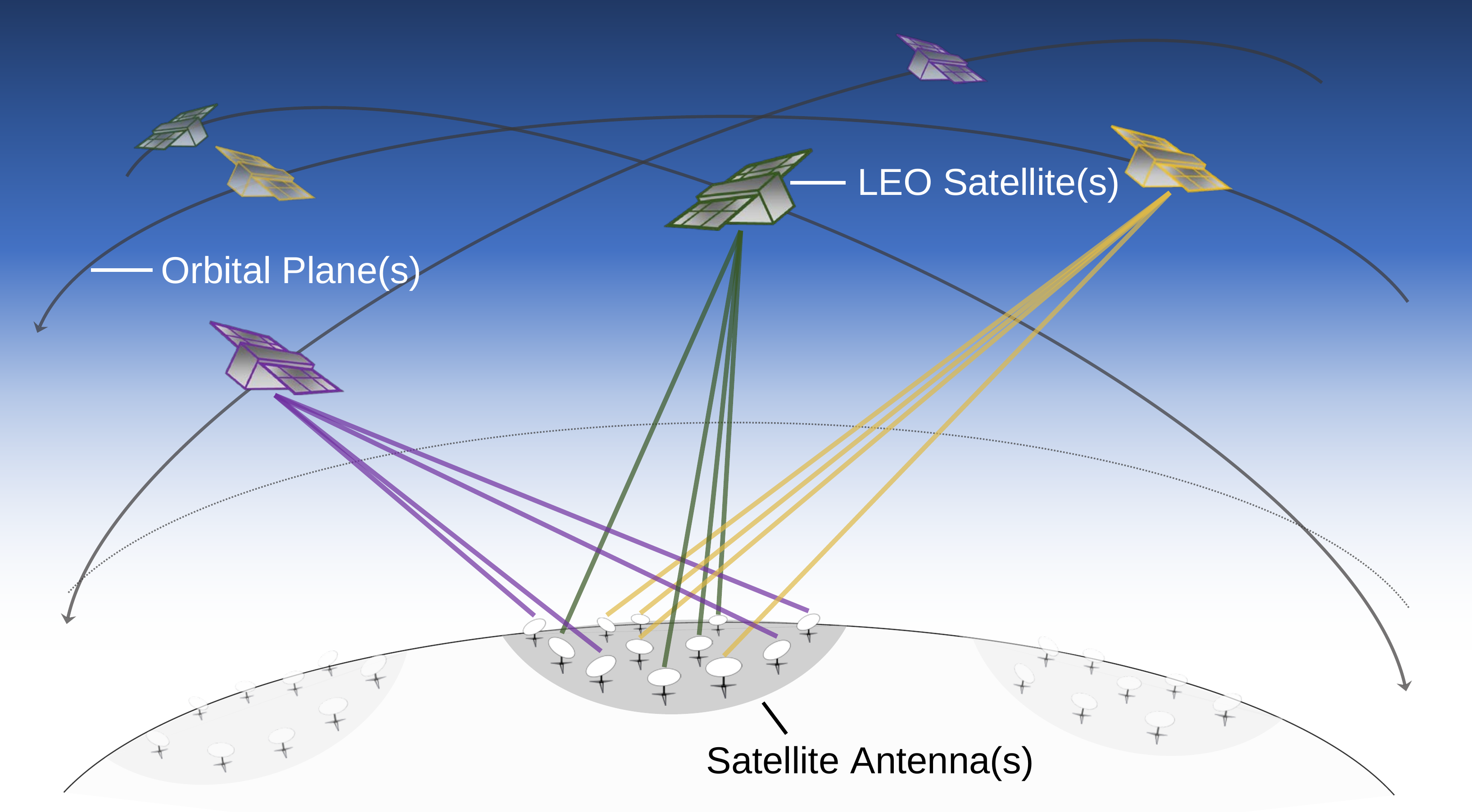}
  \caption{An illustration of random access in LEO satellite networks.}
  \label{Illustration}
  % \vspace{-.5em}
\end{figure}

%%%%%%%%%%%%%%%%

\section{Related Works and Backgrounds}\label{Background}
% \tred{[JH: Revise the merged texcts and remove redundant details; focus more on explaining the related works and backgrounds in the context of SATCOM; cite more `SATCOM RA related' works]}

% \tred{[JH: Merge too short paragraphs; If possible, make each subsection balanced. Currently A is too long and C is too short.]}

% \subsection{\tredtred{Emergent Protocol Design}}
% \tredtred{[LJH: Need an idea for where to use this paragraph.]}
% \tred{[HW:Is \emph{stationary satellites} right term? or geostationary satellites?]}

\subsection{Random Access for Satellite Networks}

% \tred{[JH: Thoroughly revise this subsection. It is currently not well structured (e.g., too short paragraphs). The related aspects are discussed only in the last paragraph. Explain more about RACH in NTN.]}

%%% 기존의 전통적인 SAT protocol은 정지궤도 위성 플랫폼에 맞추어서 디자인되어있다. 

In traditional SAT networks, most of the access protocols have been developed for geosynchronous equatorial orbit (GEO) SATs, also known as geostationary SATs \cite{Survey_RA+SAT}.
% Since GEO SATs orbiting at the altitude up to $36000$ [km] played a key role in the traditional SAT networks, most of the access protocols for SAT networks were studied focusing on GEO SATs based platform \cite{Survey_RA+SAT}.
% For the stationary satellite, several works designed a new form of multiple access.
Slotted ALOHA scheme is widely used for GEO SAT network systems owing to its simple implementation. 
Various types of slotted ALOHA have been further proposed, taking into account the characteristics of GEO SAT, including large propagation latency \cite{ALOHA_SAT1, ALOHA_SAT2}.
However, these ALOHA-based access protocols fail to support many user terminals due to a lack of collision control.
%  and its stationarity
% Moreover, long round trip time (RTT) latency of SAT networks collision event is significant issue.
% Considerable RTT latency in the SAT networks prevents the adoption of the traditional multiple access techniques.
Designing multiple access protocols for SAT networks, long round trip time (RTT) latency limits the adoption of conventional access techniques, such as CSMA/CA.
Such long RTT latency is induced by long one-way propagation delay in GEO SAT network (approximately $120$ [ms]) and in LEO SAT network (approximately $1$ - $5$ [ms]).
Particularly for GEO SAT networks, centralized reservation schemes are employed which avoid failed access attempts prior to the packet transmission to cope with such long RTT latency \cite{Survey_RA+SAT}.
% ALOHA-based access is the traditional multiple access protocol in SAT networks \cite{ALOHA_SAT1, ALOHA_SAT2}.
However, for LEO SAT networks which connect a myriad of scattered terrestrial user terminals, the coordinated method requires a significant system overhead.
% applying the coordinated method to LEO SAT networks, which connect a myriad of scattered terrestrial user terminals, requires a significant system overhead.

In contrast to stationary GEO SAT networks, emerging LEO SAT networks, wherein numerous LEO SATs orbit rapidly at a speed of around $7.6$ [km/s] (see Fig. \ref{Illustration}), are dynamic and require frequent and fast handovers \cite{3GPP_NTN, P4C_JH}, calling for another channel access considering not only long RTT latency but also the inherent dynamics of LEO SAT networks.
In this regard, the contention-based random access channel (RACH), adopted for cellular networks (e.g., 5G NR, 4G LTE/LTE-A), is also considered for LEO SAT networks \cite{3GPP_NR_RACH, RA_SHW}.
The contention-based RACH procedure is used to initialize and synchronize prior to allocating resources in downlink (DL) and uplink (UL) channels.
In the RACH, a sender randomly transmits short packet preambles to the receiver and then waits for a positive response from the receiver prior to transmitting the complete message. 
%%% contention-based RA procedure 에 대한 설명
% The RACH procedure is used by UE to initiate a data transfer and obtain UL timing information.
Particularly, the RACH protocol is initiated by transmission of a randomly chosen preamble through physical RACH (PRACH) for the preamble transmission. 
% The preambles used in the RACH protocol follow Zadoff-Chu (ZC) sequences which have a low autocorrelation property, and the number of available preambles are determined by the length of the sequences or the bandwidth of PRACH. 
The basic RACH is designed to follow the four-step message exchanging procedure, that is, \textit{1)} PRACH transmission, \textit{2)} random access response (RAR), \textit{3)} radio resource control (RRC) connection request, and \textit{4)} an RRC connection setup (and resolution).
By the four-step procedure and utilizing Zadoff-Chu (ZC) sequences, RACH is able to control collision events while attempting to access many users. 
Despite this, RACH is not ideally suited for LEO SAT networks as it does not takes into account SAT associations  \cite{3GPP_NR_NTN_SSB, 3GPP_NR_NTN_TA}.
% The maximum delay difference between nodes connected to its network is larger than the longest format of the RACH preambles, which is for accommodating a delay difference between terminals; it requires major modifications, e.g., the extension of the RAR window length, i.e., \textit{rar-WindowLength}, and the 2-step RACH \cite{3GPP_NR_NTN_TwoStepRACH}.
As such, existing access protocols have some limitations for non-stationary LEO SAT networks.
To design LEO SAT-oriented access protocol, we need to consider the following two conditions:
% the following should be considered:
\textit{1)} Collisions needs to be managed efficiently without centralized access control; and \textit{2)} SAT associations and backoff mechanisms need to be jointly optimized to achieve low access delay and high spectral efficiency.
% To this end, we newly propose an access protocol, particularly for LEO SAT networks, called eRACH.
The current standardized protocols and archetypical model-based algorithms face several challenges, mainly due to the non-stationarity of the LEO SAT network topology.
Recently, a model-free protocol has been studied in the context of emergent communication, using model-free DRL algorithms, e.g., Actor-Critic \cite{DRL_A3C}, $Q$-learning \cite{DRL_DQN}, and deep deterministic policy gradient (DDPG) \cite{DRL_DDPG}, which can be an alternative solution for the time-varying network topology.
% They rely on real samples from the environment and never use generated predictions of the next state and next reward to alter behavior (although they might sample from experience memory, which is close to being a model).
% This model-free protocol has recently been studied in the context of emergent communication.

\subsection{Emergence of Protocol through MADRL}
% \tred{[JH: Explain the difficulty in designing model-based RA protocols for non-stationary network topologies. Alternatively, one can learn a model-free protocol, which has recently been studied in the context of emergent communication. Describe the limitations of the existing emergent communication and protocol learning frameworks under SATCOM scenarios, and explain how our proposed method can overcome such limitations.
% ]}
%%% Foerster
Pioneered by \cite{MARL_Foerster1, MARL_Foerster2}, the topic of emergent communication has arisen firstly in the deep learning literature.
In \cite{MARL_Foerster1}, Foerster et al. studied how languages emerge and how to teach natural languages to artificial agents. Spurred by this trend of emergent communication, Hoydis et al. adopted the idea of emergent communication into cellular systems in \cite{Semantic_Hoydis1, Semantic_Hoydis2}.
Here, MAC signaling is interpreted as a language spoken by the user equipment (UE) and the base station (BS) to coordinate while pursuing the goal of delivering traffic across a network.
Therein, two distinct policies (i.e., a channel access policy and a signaling policy) are trained by using the off-the-shelf  learning-to-communicate (L2C) techniques (e.g., DIAL and RIAL in \cite{MARL_Foerster1}).
This approach has shown the potential of the emergent protocol design in their particular scenario.  

Despite the novelty of \cite{Semantic_Hoydis1, Semantic_Hoydis2}, the superior performance of the learned protocol largely depends on the learning of cheap-talk exchanged between network nodes, which incurs additional communication and learning overhead.
Moreover, the proposed method assumes centralized training that requires each network node to obtain full-observable information; it is an obstacle to actual system application.
% Moreover, the results show applicability only in static networks.
Besides, in these works, the protocol was optimized for only stationary networks.
Thus, it is still questionable whether a protocol for non-stationary networks can emerge without relying too much on information exchange between agents.

\input{SECTION_3.tex}

% \begin{equation}
% h_{i_{k},j}[n]=\sqrt{\frac{\beta_{0}}{s_{i_{k},j}[n]^2}}, 
% \label{Channel}    
% \end{equation}
% where $\beta_{0}$ represents the received power at the reference distance $d_0=1$ [m]. 

% Accordingly, the transmission rate in [bps] for the slot $n$ can be expressed as
% \begin{equation}
% R_{i_{k},j}[n] = B \log_2\left( 1+ \dfrac{\gamma_0}{\| s_{i_{k},j}[n] \|^2}  \right) \ \mathrm{[bps]}, 
% \label{Rate}    
% \end{equation}
% where $B$ represents the RF bandwidth, and $\gamma_0 = \frac{\beta_{0}  P}{\sigma^2}$ indicates the reference SNR with constant transmission power $P$ and noise variance $\sigma_{RF}^2$.

% \begin{align}
% D = \dfrac{\delta p_{\mathrm{c}}}{1 - p_{\mathrm{c}}}, \label{AccessDelay}
% \end{align}
% Here, the probability of collision is calculated by $p_{\mathrm{c}} = \frac{C}{N}$ with the number of access trial $N$.

%%%%%%%%%%%%%%%%%%%%%%%%%%%%%%%%%%%%%%%%%%%%%%%%%%%%%%%%%%%%%%%%%
%%%%%%%%%%%%%%%%%%%%%%% Body1 %%%%%%%%%%%%%%%%%%%%%%%%%%%%%%%%%%%
%%%%%%%%%%%%%%%%%%%%%%%%%%%%%%%%%%%%%%%%%%%%%%%%%%%%%%%%%%%%%%%%%

\input{SECTION_4.tex}

%%%%%%%%%%%%%%%%%%%%%%%%%%%%%%%%%%%%%%%%%%%%%%%%%%%%%%%%%%%%%%%%%% 
%%%%%%%%%%%%%%%%%%%%%%% Numerical Result %%%%%%%%%%%%%%%%%%%%%%%%%
%%%%%%%%%%%%%%%%%%%%%%%%%%%%%%%%%%%%%%%%%%%%%%%%%%%%%%%%%%%%%%%%%%

\input{SECTION_5.tex}

%%%%%%%%%%%%%%%%%%%%%%%%%%%%%%%%%%%%%%%%%%%%%%%%%%%%%%%%%%%%%%%%%%%%%
%%%%%%%%%%%%%%%%%%%%%%%%%%%%%%%%%%%%%%%%%%%%%%%%%%%%%%%%%%%%%%%%%%%%%
%%%%%% Conclusion %%%%%%
%%%%%%%%%%%%%%%%%%%%%%%%%%%%%%%%%%%%%%%%%%%%%%%%%%%%%%%%%%%%%%%%%%%%%
%%%%%%%%%%%%%%%%%%%%%%%%%%%%%%%%%%%%%%%%%%%%%%%%%%%%%%%%%%%%%%%%%%%%%

\section{Conclusion} \label{conclusion}
In this article, we proposed a novel RA for LEO SAT networks. 
To cope with the challenges incurred by its wide coverage and time-varying network topology, we proposed a model-free RA protocol that emerges from the given LEO SAT network environment via MADRL, dubbed eRACH. 
By simulations, we validated that eRACH better reflects the time-varying network topology than model-based ALOHA and RACH baselines, thereby achieving higher average network throughput with lower collision rates. 
Furthermore, eRACH is robust to SAT BS positioning errors, enabling its operations with known periodic patterns of SAT BS locations. 
Lastly, eRACH can flexibly adjust and optimize throughput-collision objectives in various user population scenarios.
Extending the current throughput and collision objectives, considering a fairness-aware objective could be an interesting topic for future research. 
It is also worth investigating highly scalable MADRL frameworks to address complicated scenarios with more orbital lanes and users at different altitudes, such as high altitude platform systems (HAPS) and other GEO and MEO SATs.

% proposes an emergent RA protocol, coined eRACH, 
% investigated RA for LEO SAT networks and proposed an LEO SAT-oriented emergent RA protocol, called eRACH; this emergent protocol is based on the MADRL solution maximizing the throughput while minimizing the collision.
% A main observation of the experiments is that trained eRACH, which emerges from given local environment, understands the non-stationary LEO SAT networks.
% Besides, training of the eRACH is viable only with trained even with prior knowledge. 
% Moreover, this protocol flexibly optimizes the access protocol according to its objectives and deployment scenarios.
% Those facts corroborated the proposed eRACH protocol's effectiveness, which efficiently decides RA collision avoidance and SAT-UT associations. 

%%%%%%%%%%%%%%%%%%%%%%%%%%%%%%%%%%%%%%%%%%%%%%%%%%%%%%%%%%%%%%%%%%%%%%%%%%%%%%%%%%%%%%%%%%%%%%%%%%%%%%%%%%%%%%%%%%%%%%%%%%%%%%%%%%%%%%%%%%%%%%%%%%%%%%%%%%%%%%%%%%%%%%%%%%%%%%%%%% BIBLIOGRAPHY %
%%%%%%%%%%%%%%%%%%%%%%%%%%%%%%%%%%%%%%%%%%%%%%%%%%%%%%%%%%%%%%%%%%%%%%%%%%%%%%%%%%%%%%%%%%%%%%%%%%%%%%%%%%%%%%%%%%%%%%%%%%%%%%%%%%%%%%%%%%%%%%%%%%%%%%%%%%%%%%%%%%

\bibliographystyle{IEEEtran} 
\bibliography{Reference_Paper5_Journal}
%%%%%%%%%%%%%%%%%%%%%%%%%%%%%%%%%%%%%%%%%%%%%%%%%%%%%%%%%%%%%%%%%%%%%%%%%%%%%%%%%%%%%%%%%%%%%%%%%%%%%%%%%%%%%%%%%%%%%%%%%%%%%%%%%%%%%%%%%%%%%%%%%%%%%%%%%%%%%%%%%%%%%%%%%%%%%%%%%

%%%%%%%%%%%%%%%%%%%%%%%%%%%%%%%%%%%%%%%%%%%%%%%%%%%%%%%%%%%%%%%%%%%%%%%%%%%%%%%%%%%%%%%%%%%%%%%%%%%%%%
%%%%%%%%%%%%%%%%%%%%%%%%%%%%%%%%%%%%%%%%%%%%%%%%%%%%%%%%%%%%%%%%%%%%%%%%%%%%%%%%%%%%%%%%%%%%%%%%%%%%%%

\end{document}

%% file: SECTION_3.tex
\section{System Model} \label{System Model}
% This section describes the LEO satellite network model under study and the random access scenario therein. Moreover, the figure of merits for LEO SAT random access is provided.

% \tred{\textbf{[Requires a notation table.]}}
% \begin{table}   
%   \centering
%   \resizebox{1\columnwidth}{!}{\begin{minipage}[t]{0.5\textwidth}
%   \caption{Notation of Variables.}
%   % \vspace{1mm}
%   \centering
%   \input{Table/Table_Notation}
%   \label{table_Notation}
%   \end{minipage}}
% \end{table} 

\subsection{Network and Channel Models}
Consider a set $\mathcal{K}$ of orbital planes around the earth, sets $\mathcal{I}_k$ of LEO SATs orbiting on the orbital plane $k$ for all $k\in \mathcal{K}$, and a set $\mathcal{J}$ of SAT user terminals (UTs) deployed on the ground inside an area $A$. The position of UT $j \in\mathcal{J}$ is expressed as a 3-dimensional real vector on Cartesian coordinates denoted by $\vb*{q}_{j} = (q^x_{j},q_{j}^{y},q_{j}^{z})\in \mathbb{R}^3$, and similarly, the position and velocity of SAT $i \in \bigcup_{k\in\mathcal{K}}\mathcal{I}_k$ at time $t\geq 0$ is denoted by $\vb*{q}_{i}(t) = (q_{i}^{x}(t),q_{i}^{y}(t),q_{i}^{z}(t)) \in \mathbb{R}^{3}$ and $\vb*{v}_i(t) = (v_{i}^{x}(t),v_{i}^{y}(t),v_{i}^{z}(t)) \in \mathbb{R}^3$, respectively, for all $i\in \bigcup_{k\in\mathcal{K}}\mathcal{I}_k$. 
Suppose the number of SATs on each orbital plane is equal to each other given as $|\mathcal{I}_k| = I$ for all $k\in\mathcal{K}$, and assume all SATs are moving in uniform circular motion with the same orbital period $T$, while the arc length between any two neighboring SATs on the same orbital plane is equal to each other.

Consider that time is discretized in slots of length $\tau$ and let $\vb*{q}_{i}[0]$ be the initial position of the SAT $i \in \bigcup_{k \in \mathcal{K}}\mathcal{I}_k$ at time $t = 0$. Then, by following the discrete-time state-space model \cite{UAV_SCA_YZ, P4C_JH}, the position of SAT $i$ at time $t = m\tau$  can be expressed as
\begin{align}\label{C_LEO_q}
\vb*{q}_{i}(m \tau) \approx \vb*{q}_{i}(0) + \tau \sum_{m' = 1}^{m}\vb*{v}_i(m'\tau) + \vb*{n}_i,
\end{align}
where $\vb*{n}_i$ is the additive random noise representing the perturbation on the $i$-th satellite position and attitude determination error \cite{SAT_PositionError}, whose entries are independent and identically distributed zero-mean Gaussian random variables with $\mathbb{E}[\vb*{n}_i \vb*{n}_i^{\dag}] = \sigma_{i}^{2} \mathbf{I}_{3}$.
% Note that the SAT $i$ returns to near its original position after the orbital period, e.g., $\mathbf{q}_{i}[N] \simeq \mathbf{q}_{i}[0]$.

Communication channels between SATs and UTs follow the characteristics of ground-to-space (or space-to-ground) RF link \cite{SAT_Ch1}. 
% Let $s_{i_k,j}$ be the line-of-sight distance, referred to as slant range, from SAT $i_k \in \mathcal{I}_k$ to UT $j \in \mathcal{J}$ for all $k\in\mathcal{K}$.
% \tred{The deterministic propagation models are adopted under the position of SAT and attenuation conditions. [HW:Please restate this more specifically and clearly.]}
The channel gain between satellite $i \in \bigcup_{k \in \mathcal{K}}\mathcal{I}_k$ and UT $j\in\mathcal{J}$ at time $t$ is expressed as
\begin{equation}\label{channel_RF}
h_{i,j}(t) = \tilde{h}_{i,j}(t) \sqrt{\gamma_{i,j}(t)},
\end{equation}
where $\gamma_{i,j}(t)$ and $\tilde{h}_{i,j}(t)$ are the effects of large-scale (e.g., path loss and shadowing) and small-scale fading at time $t$, respectively, and $\mathbb{E}[|\tilde{h}_{i,j}(t)|^2]= 1$ for all $t\geq 0$.
% Note that in the space-to-ground channel model, although the ionosphere causes significant impairments to the radio signal (including Doppler, polarization rotation, scintillation, and group delay), its absorption for frequencies above 70 [MHz] is negligible \cite{SAT_Ch1}.
% \tred{Doppler effect is be compensated owing to the periodicity of satellites orbiting. \tred{[HW:Should be explained]}}

% Unlike the ground-to-ground link, for ground-to-space (or space-to-ground) link, the link distance $d$ is determined by the slant range of the two stations, which represents the line-of-sight distance between them.
% We consider the channel model for the SAT~$i$ and the UT~$j$ based on the slant range $s_{i_{k},j}$.

% Since the channel characteristics of ground-to-space link, we assume LoS links as in \cite{SCA_YZ}. 

The large-scale fading is modeled based on the tractable line-of-sight (LoS) probability model \cite{SAT_Ch1, UAV_Ch_YZ} with shadowing and blockage effects. Note that in the LoS probability model, the large-scale fading follows generalized Bernoulli distribution of two different events; the channel is LoS or non-LoS (NLoS) with a certain probability. The probability of event that a channel between SAT $i$ and UT $j$ at time $t$ being LoS is modeled as 
\begin{align}\label{eq:LoSprobability}
\varphi^{\text{LoS}}_{i,j}(t) = \frac{1}{1+L_{1}\mathrm{exp}[-L_{2} (\theta_{i,j}(t) - L_{1})]},
\end{align}
% \tred{\textbf{[HW:Please find other constant instead of $C$ and $D$. Especially, $C$ is overlapping with collision.]}}
where $L_{1}$ and $L_{2}$  are the environmental parameters depending on the propagation condition \cite{UAV_Rotary_YZ} and 
\begin{align}\label{eq:elevation_angle}
\theta_{i,j}(t) = \frac{180}{\pi}\mathrm{sin}^{-1} \!\left(\frac{q_{i}^{z}(t) - q_{j}^{z}}{||\vb*{q}_i(t)-\vb*{q}_j||_2}\!\right)
\end{align}
is the angle of SAT $i$ and UT $j$ referred to as an elevation angle.

Meanwhile, depending on whether the channel is LoS or NLoS, different large-scale fading $\gamma_{i,j}(t)$ can be expressed as
\begin{equation}
  \gamma_{i,j}(t) = \begin{cases}
    \beta_{o}||\vb*{q}_{i}(t) - \vb*{q}_{j}||_2^{-\alpha}, & \text{for LoS channel},\\
    \kappa \beta_{o}||\vb*{q}_{i}(t) - \vb*{q}_{j}||_2^{-\alpha}, & \text{for NLoS channel},\\
  \end{cases}
\end{equation}
where $\beta_{o}$ is the average power gain at the reference distance $d_o=1$ [m], $\alpha$ is the path loss exponent and $\kappa$ is the attenuation scaling factor in the NLoS link \cite{UAV_Ch_YZ}. Note that $\tilde{h}_{i_{k},j}$ has its randomness from both random occurrence of LoS and NLoS as well as the small-scale fading. Accordingly, the expected channel gain over both randomness is given by
\begin{equation}
\mathbb{E}\!\left[|h_{i,j}(t)|^2 \!\right] = \tilde{\varphi}^{\text{LoS}}_{i,j}(t)\beta_{o}||\vb*{q}_{i}(t) - \vb*{q}_{j}||_2^{-\alpha},
\end{equation}
where $\tilde{\varphi}^{\text{LoS}}_{i,j}(t) = \varphi^{\text{LoS}}_{i,j}(t)+\kappa(1-\varphi^{\text{LoS}}_{i,j}(t))$.
% and $\theta_{i_{k}j}[n] = \frac{180}{\pi}\mathrm{sin}^{-1} \Big(\frac{\dot{H}_{i_{k}}}{l_{i_{k}j}[n]} \Big)$ is the elevation angle in degree

\begin{figure}
  % \vspace{-1em}
  \centering
  \includegraphics[width=1\columnwidth]{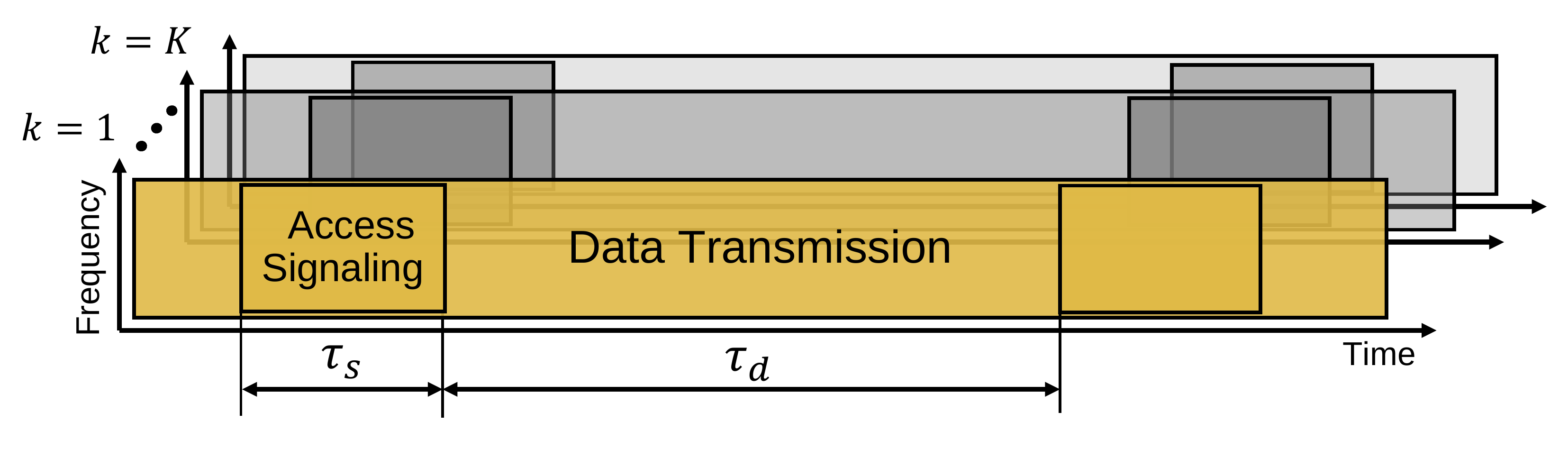}
  \caption{Time and frequency resource structure of the considering random access scenario.}
  \label{Illustration_Frame}
  \vspace{.5em}
\end{figure}

\subsection{LEO SAT Random Access Scenario} \label{Sec2_RA}
% \tred{[JH: the time definition is unclear. Clarify the relationship between $t$ and $n$. Define $n$ clearly.]}
% \tred{[JH: Do not abuse notations (without clarification). Still many notation definitions are missing, e.g., $K$ (should be $K:=|\mathcal{K}|$), $N$, $T$, ..., as well as their relationships]}

Consider an LEO satellite random access (RA) scenario, where UTs attempt to access to LEO satellite networks for radio resource grant. Suppose that UTs always have data to transmit in their infinite length queue, and thus, they always have intentions to RA at every opportunity they have. Each UT has information of the periodic position of SATs on each orbital plane and attempts to access only to the closest SAT on each orbital plane. For the sake of convenience, we assume that the closest SAT on each orbital plane is the same for every UT at the same time point. According to the network model, note that the time duration that one SAT becomes the closest among all SAT on each orbital plane is $\frac{T}{I}$, where $T$ is the orbiting period and $I$ is the number of SATs on each orbital plane. Suppose that there are $N$ RA opportunities during the interval $\frac{T}{I}$ and they are synchronized over all orbital planes as shown in Fig.~\ref{Illustration_Frame}. The time duration of each RA opportunity is $\frac{T}{IN}$, which incorporates RA signaling duration $\tau_s$ and data transmission duration $\tau_d$, i.e., $\tau_s+\tau_d = \frac{T}{IN}$, for some $\tau_s$ and $\tau_d$ such that ${\tau_s}/{\tau},{\tau_d}/{\tau} \in \mathbb{Z}^+$. For simple description, we focus only on $N$ RA opportunities in the rest of this section and suppose the first opportunity starts at $t = 0$.
% Here, we consider time is discretized.
Here, the time duration of each $n$-th RA opportunity is discretized with $\tau$, i.e., $t = n \tau, \forall n\in \{1,2,\dots,N\}$.

At each RA opportunity, each UT chooses whether to access or backoff, and then further decide the SAT to which it will access. 
Such set of actions is simply denoted by $\lbrace \texttt{BACKOFF},1, \ldots, K \rbrace$ in which $K:=|\mathcal{K}|$ is the number of orbital planes. 
The RA action of UT $j\in\mathcal{J}$ at RA opportunity $n$ is denoted by 
\begin{equation}
    a_{j}[n] \in \lbrace \texttt{BACKOFF},1, \ldots, K \rbrace. \label{AccessAction}
\end{equation}
% $a_{j}[n] \in \lbrace \texttt{BACKOFF},1, \ldots, K \rbrace$ for all $n\in \{1,2,\dots,N\}$. 
Note that $a_{j}[n] = \texttt{BACKOFF}$ means that the UT $j$ does not access at the $n$-th opportunity and waits for the next one. 
Moreover, those who attempt to access additionally choose a preamble uniform-randomly, that is
\begin{equation}
    p_{j}[n] \in \lbrace 1,2,\dots,P \rbrace.
\end{equation}
Here, each preamble is associated with $P$ resources that the SATs can grant during the data transmission duration.

The RA signaling is done in two steps. First, those UTs that determine to access send preambles to the corresponding SATs. Secondly, SATs send feedback to confirm whether there were collisions or not for each chosen preamble. UTs that have chosen collided preambles fail to access, while those of UTs that have chosen preambles without collision succeed to access. 

% Once the UT decides the access action, $a_{j}[n]$, the UT randomly selects the preamble signatures and then transmits the PRACH preamble sequence, as in 5G NR \cite{RACH_5G} and 4G LTE/LTE-A \cite{RA_SHW}.
% Recall that when a UE transmit a PRACH Preamble, it transmits with a specific pattern and this specific pattern is called a \textit{signature}. 
% In each LTE cell, total 64 preamble signatures are available and UE select randomly one of these signatures.

\subsection{Collision Rate, Access Delay, and Network Throughput}
The performance of RA in LEO SAT network is evaluated with collision rate, access delay and average network throughput, as explained in the following.

\subsubsection{Collision Rate}
Denote the collision indicator of UT $j\in\mathcal{J}$'s RA at opportunity $n \in \{1,\dots,N\}$ by $c_{j}[n]$, and define it as
\begin{align}
c_{j}[n] \! = \!\begin{cases}
    0, & (a_{j}[n],p_{j}[n])\! \neq\! (a_{j^{'}}[n],p_{j'}[n])\ \forall j' \! \in\! \mathcal{J}\backslash j\\
    1, & \text{otherwise},
  \end{cases}\label{Collision}
\end{align}
if $a_j[n] \neq \texttt{BACKOFF}$, and otherwise we have $c_j[n] = 0$.
Then, the \emph{collision rate} is defined as:
%is defined with the collision ensemble proportion $C[n] = $ as
\begin{align}
C = \frac{1}{|\mathcal{J}|} \sum_{n=1}^{N}\sum_{j \in \mathcal{J}}{c_{j}[n]}.
\end{align}
% Note that the collision information is $c_{j}[n]=1$, if a collision occurs in access action of UT $j$ for time slot $n$; otherwise, $c_{j}[n]=0$.

% \begin{align}
% c[n] = \sum_{j \in \mathcal{J}}{c_{j}[n]}. \label{collision}
% \end{align}
% The collision information is $c_{j}[n]=1$ if a collision occurs in access action of UT $j$ for time slot $n$, otherwise $c_{j}[n]=0$.
% With the collision information, the collision ensemble proportion is epx $C[n] = \sum_{j \in \mathcal{J}}{c_{j}[n]}/J$
% The collision rate is defined with the collision ensemble proportion $C[n] = \sum_{j \in \mathcal{J}}{c_{j}[n]}/J$ as
% \begin{align}
% C = \sum_{n=1}^{N}{C[n]}.
% \end{align}

\subsubsection{Access Delay}
% Define the access indicator of UT $j$ at time slot $n$ as
% \begin{align}
% \eta_{j}[n] = \begin{cases}
%     (1 - c_{j}[n]), & a_{j}[n] \neq \texttt{BACKOFF} \\
%     0, & a_{j}[n] = \texttt{BACKOFF}
%   \end{cases}. \label{Access}
% \end{align}
% \tred{\textbf{[HW:Where do we use this? Let us erase it if you don't need this except for the one defining $N_{access}$ below.]}}
Define the access indicator of UT $j\in\mathcal{J}$'s RA at opportunity $n \in \{1,\dots,N\}$ as
\begin{align}
\eta_{j}[n] = \begin{cases}
    (1 - c_{j}[n]), & a_{j}[n] \neq \texttt{BACKOFF} \\
    0, & a_{j}[n] = \texttt{BACKOFF}
  \end{cases}. \label{Access}
\end{align}
% \tred{\textbf{[HW:Where do we use this? Let us erase it if you don't need this except for the one defining $N_{access}$ below.]}}
Let $N_{s}$ be the number of successful accesses of all UTs out of $N$ access attempts, which is given as
\begin{align}
N_{a} = \frac{1}{|\mathcal{J}|}\sum_{n=1}^{N}\sum_{j\in\mathcal{J}}\eta_j[n].
\end{align}
Then, the \emph{average access delay} is given as:
\begin{align}
D = \frac{(N_{a}-N) (m_s+m_d)\tau}{N_{a}} + m_s\tau. \label{AccessDelay}
\end{align}

\subsubsection{Network Throughput}
% In order to analyze the spectral efficiency of the access protocols more rigorously, rather than assuming the number of successful access request as normalized throughput that is the conventional approach \cite{RA_AccessDelay}, we here consider the channel and throughput model between the SAT and the accessed UT. 
According to the defined channel model, the uplink (UL) throughput from UT $j\in\mathcal{J}$ to SAT $i\in \bigcup_{k\in\mathcal{K}} \mathcal{I}_k$ can be expressed as
\begin{align}
R_{i,j}(t) = B \log_2\left( 1+ \frac{ \Gamma \left|h_{i,j}(t)\right|^{2}}{\sigma_{n}^2} \right) \mathrm{[bps]},
\end{align}
where $B$ represent the bandwidth, $\sigma_{n}^2$ is the noise variance, and $\Gamma$ is the UL transmission power.

Note that UT can transmit data only when its access is successful. Thus, when $a_j[n] \neq \texttt{BACKOFF}$, the attainable throughput at UT $j \in \mathcal{J}$ in the $n$-th RA opportunity can be expressed as
\begin{align}
R_{j}[n] = \eta_j[n] \tau_d \sum_{t = (n-1)(\tau_s+\tau_d) + \tau_d}^{n(\tau_s+\tau_d)} R_{i_{a_j[n]},j}(t), \label{AchievableRate} 
\end{align}
where $i_{k} \in \mathcal{I}_{k}$ denotes the most closest SAT from UT $j$ among all SATs on the orbital plane $k\in\mathcal{K}$. Otherwise if $a_j[n] = \texttt{BACKOFF}$, $R_{j}[n] = 0$. 
% \tred{$i_{k} = i_{k}^{j}[n], k = a_{j}[n]$ [HW:What is this? If we don't need it please erase it.]}
Consequently, the time-average network throughput $R$ within $N$ RA opportunities can be given as
\begin{align}
R = \frac{1}{N}\sum_{n = 1}^{N}\sum_{j\in\mathcal{J}}R_j[n].    
\end{align}
% Hereafter, $R$ is referred as average network throughput for convenience.

In what follows, under the LEO SAT RA scenario and with the performance metrics including collision rate, access delay, and network throughput, we propose an emergent RA protocol for LEO SAT networks.

% \tred{[JH: `attainable' sounds weird; sum of rates over time sounds also weird; instead, use the time-average sum throughput, which does not change the formulation as $1/N$ becomes negligible]}
% \tred{[JH: Try not to conclude a section, subsection, paragraph with an equation. Add more text afterwards.]}

%% file: SECTION_4.tex
\section{Emergent Random Access Protocol for \\LEO Satellite Networks} \label{Body}
This section introduces the emergent contention-based RA protocol for LEO SAT networks.
Moreover, we explain step-by-step how the protocol is specifically designed.

%%%%%%%%%%%%%%%%%%%%%%%%%%%%%%%%%%%%%%%%%%%%%%
\subsection{Emergent RA Protocol for LEO SAT}
% \tred{[HW:First we need to explain briefly what is our aim and how eRACH should work, instead of first explaining the optimization problem.]}
% \tred{[HW:Please do not use semni-colon if possible, all of its current uses can be replaced by commas and periods in my opinion. Current usage looks bit weird.]}

We propose a novel RA solution for LEO SAT networks, coined \emph{emergent random access channel protocol (eRACH)}.
As illustrated in Fig. \ref{Illustration}, the UT on the ground attempts to access an SAT by the contention-based RA and then transmits a data frame only when it successfully accesses to intended LEO SAT. To specifically discuss and evaluate our proposed eRACH protocol, the following optimization problem is mathematically formulated.
This problem aims to maximize the throughput while minimizing the collision rate under the constraints related to the practical conditions of LEO SAT networks:
\begin{eqnarray}  
&\displaystyle \max_{ \scriptsize \begin{array}{c} \scriptsize a_{j}[n] \end{array} } &
\sum_{n=1}^{N}  R_{j}[n] - \rho c_{j}[n] , \ \forall j\in\mathcal{J},  \label{P1}   \label{Eq:OptObj} \\ 
&\textrm{s.t.} & \eqref{C_LEO_q}, \ \eqref{Collision}, \nonumber 
\end{eqnarray}
% where $\mathcal{A}=\{ a_{j}[n] | j \in \mathcal{J}, \forall n \in \{1,\dots,N\} \}$ is the set of access actions.
% [HW: Definition of this set is incorrect. This set should be defined as the combination of all possible action choices for $N$ time slots.]
% \tred{[JH: In the problem formulation, we have only addressed collision rate and network throughput, ignoring access delays. How can we make them linked closely and explicitly?]}
where $\rho$ denotes a normalization coefficient, which strikes a balance between throughput and collision rate, and $N$ represents one orbital cycle of SAT.
% Note that by adjusting $\rho$, it strikes a balance between throughput and collision rate.
% \tred{[JH: if possible remove $\rho_R$ that is redundant]}
Here, the constraint in \eqref{C_LEO_q} represents the orbital motion of the LEO SAT constellation.
Besides, the constraint in \eqref{Collision} represents the collision in LEO SAT.
Recall that the collision occurs for UT agents that attempt to access the same LEO SAT $i_{k}$ with the same preamble signature, at the same time slot.

In a nutshell, eRACH considers the following steps:
% \textit{1)} association decision for SAT-UT,
% \textit{2)} backoff decision,
% \textit{3)} RACH access (PRACH preamble $p$ is chosen uniform-randomly),
% \textit{4)} UL data transmission (only when accessed successfully),
\begin{enumerate}
  \item Association decision for SAT-UT,
  \item Backoff decision,
  \item RACH access (PRACH preamble $p$ is chosen uniform-randomly), and
  \item UL data transmission (only when accessed successfully),
\end{enumerate}
during which, eRACH determines 1) and 2), while the rest are automatically determined according to the established protocol as described in Sec. \ref{Sec2_RA}. 
As 3) and 4) follow the standard operations, the problem of interest focuses on the joint decision on 1) and 2).
However, to optimize 1) and 2) for \eqref{P1}, traditional convex optimization methods (e.g., SCA \cite{P2J_JH}, BCD \cite{P3C_JH}) face several challenges mainly due to time-varying LEO SAT networks. 
% To this end, we utilize the MADRL method in the proposed eRACH.

To this end, we utilize MADRL algorithm in the eRACH protocol.
While training the MADRL, we only use locally observable information of UT in the LEO SAT networks. 
Further, considering the specificity of the LEO SAT networks, we carefully pick out a piece of essential by extensively testing candidates among locally observable information to minimize the complexity while retaining near-optimality.
For applying the MADRL method, first and foremost, it is necessary to design Markov decision process (MDP) model, which includes a state, action, and reward function in an environment that reflects our network scenario, which is discussed next.

% To jointly model 1) and 2) with low complexity in RL, we define an one-hot encoded action $\mathbf{a}_j[n]=\{a_0,a_1,a_2,\cdots,a_K\}$ such that $\sum_{\ell=0}^K a_\ell=1$ with $a_\ell\in\{0,1\}$, where $a_\ell$ for $\ell\neq 0$ denotes the association with the $\ell$-th SAT, and $a_0$ implies backoff.
% \tred{[JH: the action notation and definition as well as the control parameter to be optimized are confusing; First clarify (in the system model, here, or both) that we consider the following steps:
% 1) (SAT BS) association decision
% 2) backoff decision
% 3) RACH access (uniformly at random)
% 4) (downlink or UL) data transmission,
% during which the problem of interest focuses on the joint decision on 1) and 2), while 3) and 4) follow the standard operations as described in XX. To jointly model 1) and 2) with low complexity in RL, we define an one-hot encoded action $\mathbf{a}_j[n]=\{a_0,a_1,a_2,\cdots,a_K\}$ such that $\sum_{\ell=0}^K a_\ell=1$ with $a_\ell\in\{0,1\}$, where $a_\ell$ for $\ell\neq 0$ denotes the association with the $\ell$-th SAT, and $a_0$ implies backoff.
% ]}
% \tred{[JH: use $a$ for the action, and remove the redundant $w$]}
%  which works for time-varying LEO SAT networks

\subsection{MADRL Formulation for Emergent RA Protocol}\label{Sec4B}

In what follows, we first reformulate the problem \eqref{P1} of eRACH using an MDP model, and then describe how to solve this problem via MADRL.

\subsubsection{MDP Modeling} 
To recast \eqref{P1} as an MADRL problem, we model the SAT network scenario using an MDP, a discrete-time stochastic control process that mathematically characterizes decision-making in a random environment. In the MDP model, for a given state, a policy $\pi$ refers to the probability of choosing an action, and $\pi^{*}$ is the optimal policy maximizing the long-term average reward, which is the goal of MADRL. 

% By the dynamic states of UAV $j$ in \eqref{C_UAV_v&a}-\eqref{C_UAV_q&v&a} (e.g., $\mathbf{q}^{j}_{\mathrm{U}}[n]$ and $\mathbf{v}^{j}_{\mathrm{U}}[n]$) and the dynamic states of UAV $j$ in \eqref{C_LEO_q&v_i} (e.g., $\mathbf{q}^{i_{k}_{k}}_{\mathrm{L}}[n]$), it holds a Markov characteristics. 
% As such, we can map the optimization problem (P1$^{\star}$) as MDP.
% Briefly, mapping of (P1$^{\star}$) into MDP model, each system utility in the objective function (such as E2E system throughput, consumption energy of UAVs) corresponds to reward function, while the optimization variables (e.g., $\mathcal{A}, \mathcal{A}$) corresponds to action space. 
% A mapping of (P1$^{\star}$) into the environment, state, action, and reward functions in MDP model is presented in detail in the following subsections.

% \tred{[JH: For set notations, use mathcal; For vectors, use mathbf; Kepp this consistently throughout the paper]}

%%% Environment 와 observation space 관련
% \subsubsection{Environment}

\begin{figure}[t]
  % \vspace{-1em}
  \centering
  \includegraphics[width=1\columnwidth]{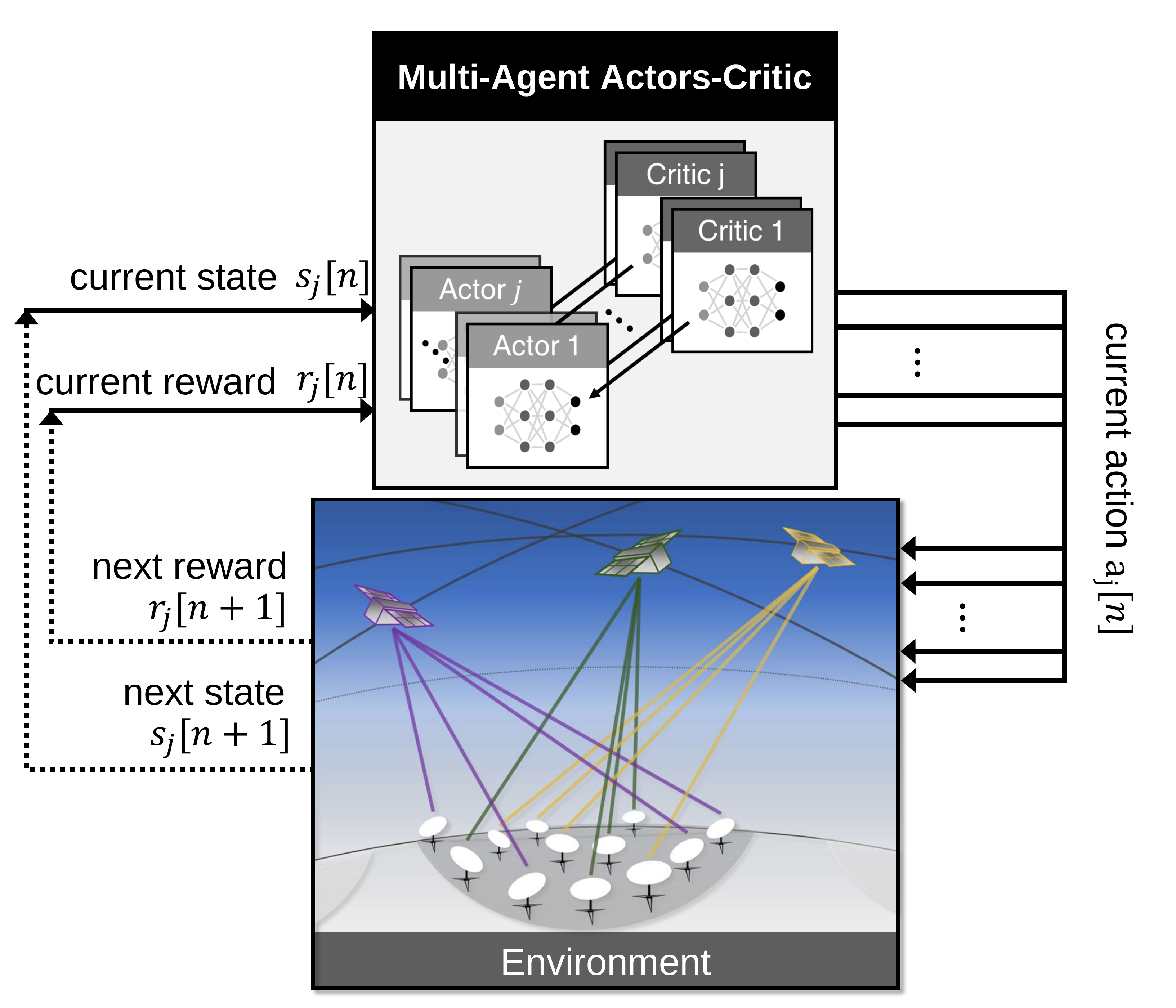}
  \caption{An illustration of the proposed eRACH training process based on multi-agent Actor-Critic method.}
  \label{Figure_Illustration_Network}
%   \vspace{.7em}
\end{figure}

\textbf{Environment}.\quad
As illustrated in Fig. \ref{Figure_Illustration_Network}, the MADRL framework under study consists of multiple UTs interacting with an environment that follows an MDP model. At each time $t$, each UT $j$ is an agent observing a state $\mathbf{s}_{j}[n]\in\mathcal{S}$, and takes an action $\mathbf{a}_{j}[n]\in\mathcal{A}$ based on a state-action policy $\pi$. Given this action, the state of the agent $j$ transitions from $\mathbf{s}_{j}[n]$ to $\mathbf{s}_{j}[n+1]$, and in return receives a reward $r_{j}[n]$ that reinforces the agent follows an optimal policy $\pi^*$. How to define these states, actions, and rewards significantly affects the performance, communication, and computing costs of eRACH as we shall elaborate next.

\begin{figure}[t]
  \centering
  \includegraphics[width=1\linewidth]{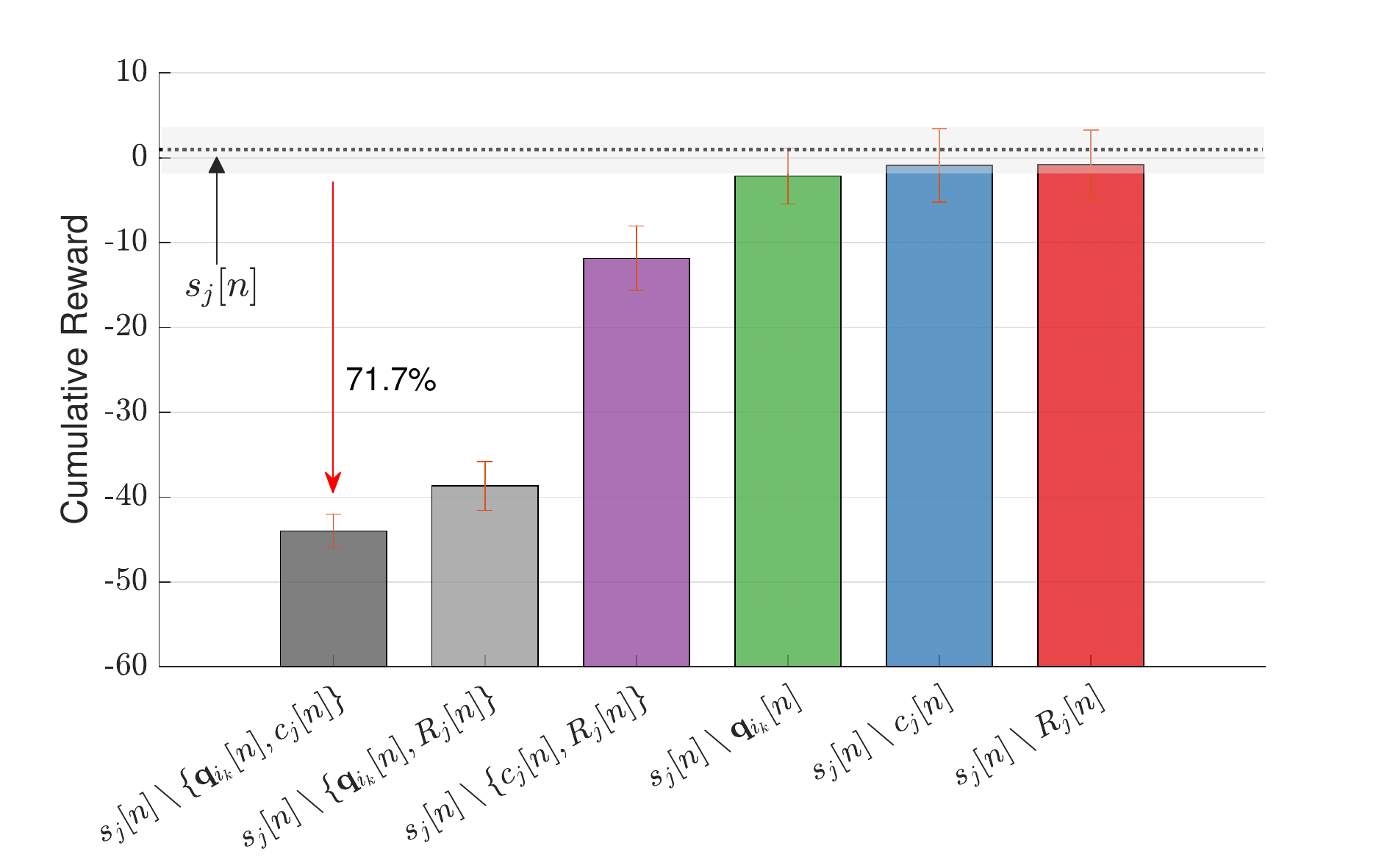}
  \caption{Comparison of cumulative rewards with different information included in the state $s_j[n]$.}
%   \tred{[HW:Text in the figure is too small and hard to acknowledge the difference between the cases. How about listing everything or make a table below and then use `o' and `x' to show what is in and what is not?]}
  \label{fig_StateReasoning}
  \vspace{.2em}
\end{figure}

%%% State space 관련
\textbf{State}.\quad
In the aforementioned MDP model, we consider the following state of UT $j$:
\begin{align}
\vb*{s}_{j}[n] &= \lbrace n, \vb*{q}_{i_{1}}[n],\dots,\vb*{q}_{i_{K}}[n], R_{j}[n], c_{j}[n], a_{j}[n\!-\!1] \rbrace, \label{State_D}
\end{align} 
%   \tred{[JH: clarify which SAT, at which lane. Use the notation clearly and consistently throughout the paper]}
where $\vb*{q}_{i_{k}}[n] \in\mathbb{R}^{I \times 3}, i\in \mathcal{I}, k\in \mathcal{K}$ denotes the position of SAT $i$ in the orbital plane $k$, $R_{j}[n]$ is the throughput, $c_{j}[n]$ is a binary indicator function of an RA collision event, and $a_j[0]$ and $\vb*{s}_{j}[1]$ are initial values chosen randomly at the beginning of Algorithm \ref{Algorithm}. 
Here, the previous action $a_{j}[n-1]$ and current time slot $n$ are used as a \emph{fingerprint}, adopting the fingerprint-based method proposed in \cite{MARL_Stabilising} to stabilize experience replay in the MADRL.  
Note that the non-stationarity of the multi-agent environment is dealt with by the fingerprinting that disambiguates the age of training samples and stabilizes the replay memory.

The state of an agent should contain sufficient information for carrying out decision-making. 
Meanwhile such information should be minimal to reduce any additional data collection overhead from the environment or other agents and to promote MADRL training convergence.
% To clarify why such information should be taken into account in the state of our MDP model, we extensively test possible local information candidates and their impacts on eRACH in our LEO SAT network environment.
To screen for important information to be included in the state among locally observable information, we extensively test possible local information candidates and their impacts on eRACH performance.
In this regard, we provide an ablation study of Fig. \ref{fig_StateReasoning} which presents the contribution of each information to the overall system. 
Each information in $s_j[n]$ has a different importance. 
As shown in this figure, eRACH mainly exploits the following information: \textit{1)} the expected SAT locations that are known a priori owing to the pre-determined LEO SAT orbiting pattern; and \textit{2)} the collision that are inherently observable. 
% In particular, the overall performance degrades $71.7$ \% without both information.
Such two locally observable information significantly contributes to the training of eRACH, which highlights eRACH does not even require cheap talks between agents. 

\begin{figure}[t]
  \centering
  \includegraphics[width=1\columnwidth]{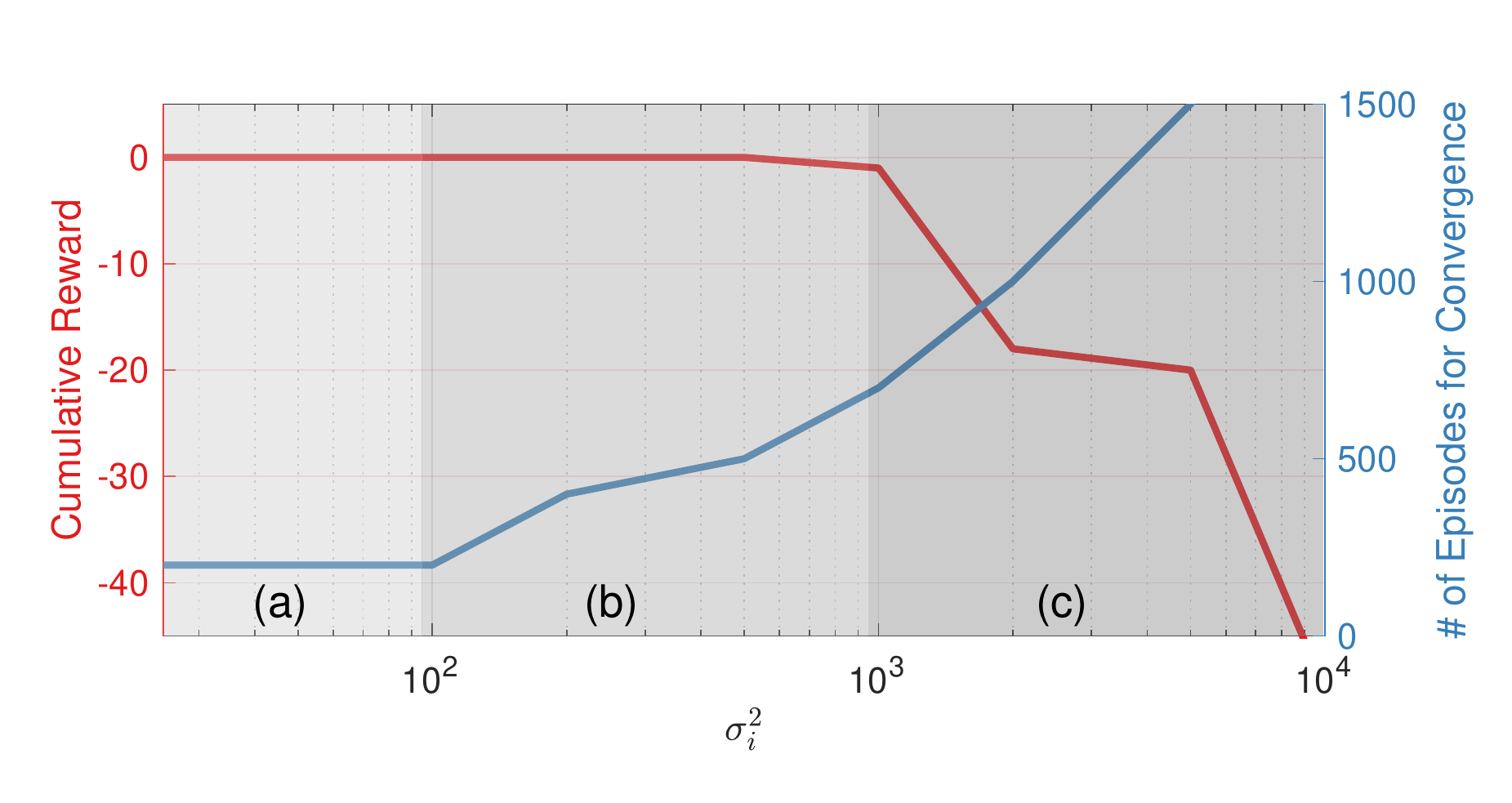}
  \caption{Robustness to SAT BS positioning errors, where shaded regions identifies the three different regimes: (a) same reward and convergence speed, (b) same reward, slower convergence, and (c) lower reward, slower convergence. Specific reward values and convergence times are reported in Table~\ref{Table_PositionError}.}
  \label{fig_PositionError}
\vspace{.2em}
\end{figure}

While training eRACH, we validate that the expected SAT location contains sufficient information on the network topology, as long as the variance between the expected and actual SAT position is less than $1$~[km] as shown in Fig. \ref{fig_PositionError}. 
Moreover, it is confirmed that the collision event incorporates sufficient information on the local environment, e.g., how crowded the network is, as will be elaborated in Sec \ref{Numerical Result}. 
% Given such two locally observable information, thanks to the LEO SAT's periodic orbiting pattern, the problem of MADRL frequently revisits the same environment, promoting to discover an emergent protocol carrying out SAT-BS association and RA decisions even without communication overhead between agents. 
As the LEO SAT periodically orbits a pre-designed trajectory, MADRL frequently revisits the same environment of SAT-UT RA, which facilitates the discovery of emergency protocols that perform SAT-UT connections and RA decisions without communication overhead between agents.

% While training of eRACH in which UT agent only requires observable information, each information in state $s_j[n]$ has different importance.
% Recall that the state of agent UT $j$ in \eqref{State_D}, $s_j[n]$, mainly includes the information of rate, collision, and position of SAT.
% As identified in the figure, the SAT position information has the most significant influence on training compared to rate or collision information.

%%% Action space 관련
\textbf{Action}.\quad
The action space $\mathcal{A}$ in our environment is related to RA.
Among SATs $i_{k}$ in orbital plane~$k$, the agent UT $j$ chooses one SAT to access by using the access action $a_{j}[n]$ in \eqref{AccessAction}. 
% Here, the value of $a_{j}[n]$ indicates the orbital plane to which the UT is trying to access, while $0$ represents the back-off action by which the agent do not attempt to access for the time slot~$n$.
For the set of access action, $\mathcal{A}$, we define an one-hot encoded action as follows
\begin{equation}
\mathbf{a}_j[n]=\{a_0,a_1,a_2,\cdots,a_K\}, \ \mathrm{s.t.} \sum_{\ell=0}^K a_\ell=1,    
\end{equation} 
where $a_\ell$ for $\ell\neq 0$ denotes the association with the $\ell$-th orbital plane, and $a_0$ implies backoff.
Note that it is assumed that the UT attempts to access the nearest SAT among SATs $\mathcal{I}_k$ in orbital plane~$k$.

% \tred{\textbf{[HW:What is the difference between this and $\mathcal{A}$ written above"]}}
% Accordingly, the actions of agent UT $j$, $\mathbf{a}_{j}[n]$ is given by
% \begin{equation}
%     \mathbf{a}_{j}[n] \in w_{j}[n], \ \forall j, n. \label{action}   
% \end{equation}

% \tred{\textbf{[HW:Please verify above mathematical expression. Seems weird.]}}
% \tred{\textbf{[HW:Why do we need $a_j[n]$?]}}

%%% Reward 관련
\textbf{Reward}.\quad
Our reward function is supposed to reinforce UTs to carry out optimal access actions that maximize the network throughput while penalizing access collision. The objective function \eqref{Eq:OptObj} captures this goal, although it is not suitable for the reward function. 
Indeed, if there is no collision, \eqref{Eq:OptObj} gives positive throughput without penalty, and otherwise imposes only collision penalty with zero throughput. This incurs too significant and frequent reward variation, hindering the training convergence. To mitigate this issue, following \cite{DRL_DQN, DRL_Normalization}, we normalize \eqref{Eq:OptObj}, and consider the reward function of UT $j$ as follows
\begin{align}
r_{j}[n] =  g(R_{j}[n] - \rho c_{j}[n]). \label{Reward_D}
\end{align}
% \tred{[JH: Unclear, revise it]}
Note that the normalization function is given by $g(Y) = \frac{Y - \mu}{\Sigma}$, where $\Sigma$ and $\mu$ are parameters for scale which shrinks the output value $Y$ between $-1$ and $1$ and for mean value, respectively. 
Here, the mean value corresponds with the network throughput of the RACH baseline in Sec.~\ref{Numerical Result}. 

It is worth noting that the state and reward of each UT agent are all locally observable information and thus the MDP model for RA protocol can be trained and executed in a fully \emph{distributed} manner.

% where the reward function is positive if our network model outperforms a baseline, or otherwise it returns a negative reward, i.e., penalty.
% we normalize the objective in the reward function by clipping the rewards and temporal difference errors to a certain range, as in \cite{DRL_DQN, DRL_Normalization}.

% \tred{[HW:Please notify that $Y$, $\Sigma$ and $\mu$ are vetors and say that they are vectors of some parameters.]}
% The mean values for the network throughput $\mu_{D}$ are the network throughput of the baseline (i.e., RACH in Sec.~\ref{Numerical Result}).
% The scaling parameter are denoted as $\sigma$, which shrinks the output value between $-1$ and $1$.
% Note that the subscripts $c$ and $d$ in the mean values and the scaling parameters stand for our proposed 'Cooperative' and 'Distributed', respectively.
% As such, the system reward for the UT~$j$  in the $n$-th time slot induced by the current state $\mathbf{s}_{j}[n]$ and action $\mathbf{a}_{j}[n]$ is defined as

% [HW:This should be more clarified. What does require for the network model to be \emph{better}?]}

%%%%%%%%%%%%%%%%%%%%%%%%%%%%%%%%%%%%%%%%%%%%%%
\subsubsection{Actor-Critic MADRL} \label{Body_3}

% \begin{figure}
%     % \vspace{-1em}
%     \centering
%     \includegraphics[width=\linewidth]{Figure2_J.eps}
%     \caption{Proposed multi-agent Actor-Critic structure for vertical multi-hop communication in LEO SAT constellation network.}
%     \label{Structure}
%     % \vspace{-1.em}
% \end{figure}

% \tred{[JH: Carefully double check the $j$ index throughout the entire section, e.g., $\pi_j$, wihch is used in Fig. 3 and the centralized-Critic MARL descriptions; Why do we omit $j$ only in $\theta$ and $\phi$ but in $a$ and $s$? I suggest not to omit this, and for the single we can simply add `Hereafter, the index $j$ identifies different actors for multi-agent scenarios, which can be omitted for a single agent case' at the first place we use the index.]}

Following the standard RL settings, the aforementioned SAT-UT RA environment is considered in which many agents interact for a given number of discrete time steps.
% [HW:Please rephrase this, looks grammatically wrong. ex. environment that interacts with RL agents for given..?] 
At each time step $n$, the agent $j$ receives a state $\mathbf{s}_{j}[n]$ and selects an action $\mathbf{a}_{j}[n]$ from some set of possible actions $\mathcal{A}$ according to its policy $\pi_{\theta_{j}}$, where $\pi_{\theta_{j}}$ is a mapping from states $\mathbf{s}_{j}[n]$ to actions $\mathbf{s}_{j}[n]$. 
% For notational simplicity, we omit the dependence of $\theta$ on $j$.
% \footnote{For notational simplicity, we omit the dependence of $\theta$ on $j$. \tred{[JH: Let's minimize the use of footnotes. You can simply put this in line]}}
In return, the agent receives the next state $\mathbf{s}_{j}[n+1]$ and receives a scalar reward $r_{j}[n]$. 
The process continues until the agent reaches a terminal state after which the process restarts. The return $\tilde{r}_{j}[n] = \sum\nolimits_{k=0}^{\infty} \gamma^{k} r_{j}[n+k]$ is the total accumulated return from time step $n$ with discount factor $\gamma \in (0, 1]$. 

% Multiple agents act in environments with the goal of maximizing their shared utility in a cooperative manner.
% The goal of the agent $j$ is to maximize the expected return from each state $\mathbf{s}_{j}[n]$.

% In our prior work \cite{DRL_UAV}, we considered a single UAV agent operating a deep-Q network (DQN) framework wherein each action of the agent is evaluated as the Q-value that is approximated using the output of a neural network (NN). While effective, DQN may not guarantee the convergence due to its off-policy learning nature, particularly under complicated tasks with large state and action dimensions \cite{ActionState}. 
% By contrast, there exist policy gradient methods in which an NN's output approximates each policy \cite{Sutton}, yet the training convergence is often too slow due to its ignoring the effort to find better actions associated with higher values. 

For our multi-agent scenario, we employ an Actor-Critic RL framework \cite{DRL_AC}, which combines the benefits of both policy gradient and value-based methods. 
The Actor-Critic framework comprises a pair of two NNs: an Actor NN seeks to take actions to obtain higher rewards based on the policy gradient method; and its paired Critic NN aims to approximate value functions more accurately via the value-based method.
% [HW:seeking better actions? rephrase the sentense.]
In particular, each UT agent $j$ has an Actor~NN and a Critic~NN and follows the synchronous advantage Actor-Critic operation \cite{DRL_AC, DRL_A3C}, in which the Critic NN updates its model parameters $\phi_{j}$ according to the policy $\pi_{\theta_{j}}$ given by the Actor NN. Meanwhile, the Actor NN updates its model parameters $\theta_{j}$ according to the value functions $V^{\pi_{\theta_{j}}}(\mathbf{s}_{j}[n]; \phi_{j})$ approximated by the Critic NN.
Specifically, the Critic NN aims to minimize the loss function
\begin{align}
    L^{j}_{\mathrm{Critic}}(\phi_{j}) = \kappa_{j}[n]^{2},
\end{align}
where $\kappa \mathbf{s}_{j}[n] = r_{j}[n+1] + \gamma V^{\pi_{\theta_{j}}}(\mathbf{s}_{j}[n+1]; \phi_{j}) - V^{\pi_{\theta_{j}}}(\mathbf{s}_{j}[n]; \phi_{j})$ is referred to as the advantage of an action \cite{DRL_TDerror}. The Critic NN model parameters are then updated as
% $\phi \leftarrow \phi + \beta_{c} \kappa[n] \nabla_{\phi} V^{\psi_{\theta}}(s[n]; \phi)$, 
\begin{equation}
    % \phi_{j} \leftarrow \phi_{j} + \beta_{c} \kappa_{j}[n] \nabla_{\phi_{j}} V^{\pi_{\theta_{j}}}(\mathbf{s}_{j}[n]; \phi_{j}),
    d \phi_{j} \leftarrow d \phi_{j} + \beta_{c} (R - V^{\pi_{\theta_{j}}}(\mathbf{s}_{j}[n]; \theta_{j}))\nabla_{\phi_{j}} V^{\pi_{\theta_{j}}}(\mathbf{s}_{j}[n]; \phi_{j}),
\end{equation}
where $\beta_{c}$ is the hyperparameter of Critic NN, which is related to the value estimate loss. Meanwhile, the Actor NN aims to minimize the loss function
\begin{equation}
    L^{j}_{\mathrm{Actor}}(\theta_{j}) = -\kappa_{j}[n] \log\pi(\mathbf{s}_{j}[n] | \mathbf{s}_{j}[n];\theta_{j}).
\end{equation}
Hereafter, the index $j$ identifies different actors for multi-agent scenarios, which can be omitted for a single agent case. Consequently, the Actor NN model parameters are updated~as 
% $\theta \leftarrow \theta + \beta_{e}\kappa[n]\nabla_{\theta}\log\pi(a[n] | s[n];\theta)$, 
\begin{equation}
    % \theta_{j} \leftarrow \theta_{j} + \kappa_{j}[n]\nabla_{\theta_{j}}\log\pi(\mathbf{s}_{j}[n] | \mathbf{s}_{j}[n];\theta_{j}), 
    d \theta_{j} \leftarrow d \theta_{j} + \nabla_{\theta_{j}} \log{\pi(\mathbf{s}_{j}[n] | \mathbf{s}_{j}[n]; \theta_{j})(R - V^{\pi_{\theta_{j}}}(\mathbf{s}_{j}[n]; \theta_{j}))}.
\end{equation}
% \tred{[HW:Please check whether use of $i$ is right or not in the equation above and in the algorithm table.]}
where $H(\pi)$ is the entropy of the policy and $\beta_{e}$ is the hyperparameter of controlling the relative contributions of the entropy regularization term.
% the Actor NN related to the entropy regularization term. 
The NN parameters are updated via gradient descent and backpropagation through time, using Advantage Actor-Critic as detailed by \cite{DRL_A3C}.
% Details of training, including the use of entropy regularization and a combined policy and value estimate loss, are described by
We summarize the eRACH training process for each UT agent $j$ in Algorithm \ref{Algorithm}.

\begin{algorithm}[] 
  \small
  \caption{eRACH Training Algorithm} \label{Algorithm}
    Initialize step counter $n \leftarrow 1$ \\
    Initialize episode counter $E \leftarrow 1$ \\
    \Repeat{$E > E_{\mathrm{max}}$}{
        Reset gradients: $d\theta_{j} \leftarrow 0$ and $d \phi_{j} \leftarrow 0$ \\
        $n_{\mathrm{start}} = n$ \\
        Get state $\mathbf{s}_{j}[n]$ \\
        \Repeat{$\mathrm{terminal} \ \mathbf{s}_{j}[n]$}{
            Perform access action $\mathbf{s}_{j}[n]$ according to policy $\pi(\mathbf{s}_{j}[n] | \mathbf{s}_{j}[n]; \theta_{j})$ \\
            Receive reward $r_{j}[n]$ and new state $\mathbf{s}_{j}[n+1]$ \\
            $n \leftarrow n + 1$
        } 
        $R = \begin{cases}
        0, & \text{for terminal} \ \mathbf{s}_{j}[n]\\
         V^{\pi_{\theta_{j}}}(\mathbf{s}_{j}[n]; \theta_{j}), & \text{for non-terminal} \ \mathbf{s}_{j}[n]        
        \end{cases}$ \\
        \For{$i \in \{ n-1, \dots, n_{\mathrm{start}} \}$}{
            $R \leftarrow r_{j}[n] + \gamma R$ \\
            Accumulate gradients w.r.t. $\theta_{j}$: \\
            $ d \theta_{j} \leftarrow d \theta_{j} + \nabla_{\theta_{j}} \log{\pi(\mathbf{a}_{j}[i] | \mathbf{s}_{j}[i]; \theta_{j})(R - V^{\pi_{\theta_{j}}}(\mathbf{s}_{j}[i]; \theta_{j}))} $ \\
            $ + \beta_{e} \partial H(\pi(\mathbf{a}_{j}[i] | \mathbf{s}_{j}[i]; \theta_{j}) / \partial \theta $ \\
            Accumulate gradients w.r.t. $\phi_{j}$: \\
            $d \phi_{j} \leftarrow d \phi_{j} + \beta_{c} (R - V^{\pi_{\theta_{j}}}(\mathbf{s}_{j}[i]; \theta_{j}))\nabla_{\phi_{j}} V^{\pi_{\theta_{j}}}(\mathbf{s}_{j}[i]; \phi_{j})$ 
        } 
        Perform update of $\theta_{j}$ using $d \theta_{j}$ and of $d \phi_{j}$ using $\phi_{j}$ \\
        $E \leftarrow E + 1$
        }
\end{algorithm}

%% file: SECTION_5.tex
\section{Numerical Evaluations} \label{Numerical Result}

%%%%%%%%%%%%%%%%%%%%%%%% Parameter Table

This section validates the proposed emergent RA methods of eRACH for LEO SAT networks with respect to average throughout, collision rate, and delay. 
% \tred{[JH: If something has to be proposed, it should be presented before the simulation section.]
% Furthermore, considering the specificity of the LEO SAT networks, we introduce a novel state abstraction method that minimizes the complexity of decision making while retaining near-optimality.
% }

\subsection{Simulation Settings} \label{SimulationSetting}

\begin{table}[]   
  \centering
  \resizebox{1\columnwidth}{!}{\begin{minipage}[t]{0.5\textwidth}
  \caption{Simulation parameters.}
  % \vspace{1mm}
  \centering
  \input{Table/Table_Parameter}
  \label{table_Paramter}
  \end{minipage}}
\end{table}

Unless otherwise stated, UTs under study are located in the ground uniformly at random within an area of $1000 \times 1000$ [m$^2$], while two orbital planes $K=2$ circulate over UTs at the altitude of $550$ [km]. 
% \tred{[JH: what does the last 10 m mean? Is each ground UT located within a box, not a rectangle on the ground? ]}
% \tred{$\mathbf{q}_{i_{k}}[0] = \mathbf{q}^{I}_{i_{k}}$ is the initial position of SAT~$i$ in orbit plane~$k$.}
Each orbital lane consists of $22$ SATs with the orbital lane circumference of $43486$ [km], resulting in an inter-SAT distance of $1977$ [km]. 
Since the orbital lane circumference is much larger than the area of interest, each orbital lane is approximated as a line segment at the altitude of $550$ [km] \cite{Starlink}, in which $22$ SATs orbit with the orbital speed $7.59$ [km/s], are separated with the inter-SAT distance $1977$ [km], and perturbed with $\sigma_{i}$.
% The length of each episode is set as $N=2604$ which corresponds with the RA opportunity, during which an SAT with the orbital speed $7.59$ [km/s] completes one orbit.
Here, the orbital period and speed are calculated using the relations $4\pi^{2}(r_{\mathrm{E}})^{3} = T^{2}GM$ and $V^{2}r_{\mathrm{E}} = GM$, respectively, where $r_{\mathrm{E}}$ is the radius of orbit in metres; $T$, the orbital period in seconds; $V$, the orbital speed in [m/s]; $G$, the gravitational constant, approximately $6.673 \times 10^{-11}$ [m$^{3}$/kg$^{1}$/s$^{2}$]; $M$, the mass of Earth, approximately $5.98 \times 1024$ [kg].
% \tred{Here, $\sigma_{\mathcal{I}}$ represents the randomness of SAT's position which is related to the perturbation of LEO SAT and the attitude determination error \cite{SATPosionError}.}

Unless otherwise stated, the simulation environments and parameters are as follows: each RA opportunity is opened in every $\tau = \tau_{s} + \tau_{d} = 100$ [ms];
the RA signaling duration and data transmission duration are set as $\tau_{s}=10$ [ms] and $\tau_{d}=90$ [ms], respectively;
data transmission is conducted for $\tau_{d}$ time only if the attempt of the access succeeds;
the objective function in our MDP model corresponds to the un-discounted accumulated rewards with $\rho = 1$ over an episode up to $N = \frac{T}{I \tau} = 2604$, which corresponds to the RA opportunity; and $J=5$ five UT agents  are considered wherein two orbital planes $K=2$ circulate over UTs with resources $P=2$, given the MADRL environment mentioned above.

% We assume that $2 - 54$ preamble signatures are utilized to see the LEO SAT random access scenario. 

\begin{table*}[t]
%   \hspace{5pt}
\centering
\resizebox{2.\columnwidth}{!}{\begin{minipage}[h]{1.72\columnwidth}
\centering
\caption{Throughput, collision rate, and access latency of eRACH, compared with eRACH-Coop, RACH, and Slotted ALOHA.
% Comparison over different RA schemes in LEO SAT networks. 
% \tred{[JH: What is $\rho$ for eRACH? Clarify this at least in the main body. Present the values under $\rho>0$; For all caption titles, be more specific.]}
}
\label{Table_Proposed}
\input{Table/Table_Proposed}
  \vspace{.3em}
\end{minipage}}
\end{table*}

% \begin{table}[t]
% %   \hspace{5pt}
% \centering
% \resizebox{.95\columnwidth}{!}{\begin{minipage}[h]{0.8\columnwidth}
% \centering
% \caption{Fairness and cheap talk communication cost comparison between eRACH and eRACH-Coop
% % Comparison over eRACH with or  Here, Cheap Talk represents whether each UT exchanges local state information with a few bits with other UTs. 
% }
% % \tred{[JH: Clarify how much bits are needed; If possible, combine this table with Table II after filling in some missing values.]}
% \label{Table_Proposed_Fair_Cheap}
% \input{Table/Table_Proposed_Fairness+CheapTalk}
% \end{minipage}}
% \end{table}

% \tred{[JH: Throughout the paper, make it clear that what we propose is either (distributed) eRACH or both distributed and cooprative eRACH. Accordingly, revise the intro and all other sections.]}
Throughout this section, we consider two benchmark RA schemes and our proposed eRACH with or without cooperation among agents as listed below.
\begin{enumerate}

\item \textbf{Slotted ALOHA} is a traditional contention-based channel access protocol.
UTs agree upon the discrete slot boundaries. 
At the beginning of the access slot, i.e., $\tau_{s}$, each UT uniform-randomly chooses the SAT to access.
For each SAT, if more than one UT attempts to access at the beginning of a slot, collisions occur.

\item \textbf{RACH} is another conventional contention-based channel access protocol used in NR and LTE.
RACH uniform-randomly selects a SAT to access and additionally chooses the preamble $p_j[n]$ at the beginning of the access slot, i.e., $\tau_{s}$.
When a collision occurs, UT waits for a uniformly distributed backoff time and again repeats the process from the beginning. 
Here, the backoff range follows a discrete uniform distribution as $\tau_{b} \sim DU(1, W\tau)$, where $W$ is the backoff window size assumed fixed at $10$. 
As in Release 16 of NR \cite{3GPP_NR_RACH_Rel16}, we consider that RA signaling is done in two steps, i.e., 2-step RACH.

% First, those of UTs that decided to access send preambles to the corresponding SATs. Second, SATs send feedback to confirm whether if there were collision or not for each chosen preamble. The UTs that have chosen collided preambles fail to access, while those of UTs that have chosen preambles without collision succeed to access. 

% and $\delta$ represents the RACH duration.

\item \textbf{eRACH-Coop} is another variant of our proposed RA scheme, in which each UT can select an optimal action while cooperatively communicating with other UT agents.
Unlike \emph{eRACH} where each distributed UT agent uses partially observable information, for \emph{eRACH-Coop}, cooperative UT agents use the full observability through cheap-talk with other agents.
% also not fully observable of the environment, but is
% By utilizing the additional information, \textit{Cooperative} can achieve a better performance. 
% Here, the performance of it can be regarded as upper-bound result.

In particular, the cooperative UT agents use the network throughput as a corresponding reward by the previous access action of each agent, and the collision rate which involves all the collision information of each UT as state. 
The reward and state for the cooperative agent $j$ is given by
\begin{align}
r^{\mathrm{C}}_{j}[n] &= g({\textstyle\sum}_{j \in \mathcal{J}} R[n]), \label{Reward_C} \\   
% \mathbf{s}_{j}[n] &= \lbrace \mathbf{q}^{i_{k}}_{\mathrm{L}}[n], R_{j}[n], {\textstyle\sum}_{j \in \mathcal{J}} R[n], c_{j}[n], C[n], n \rbrace. \label{State_C}
% \mathbf{s}^{\mathrm{C}}_{j}[n] &= \lbrace n, \vb*{q}_{i_{1}}[n],\dots,\vb*{q}_{i_{K}}[n], R_{j}[n], c_{j}[n], a_{j}[n\!-\!1], {\textstyle\sum}_{j \in \mathcal{J}}R_{j}[n], {\textstyle\sum}_{j \in \mathcal{J}}c_{j}[n] \rbrace. \label{State_C}
\mathbf{s}^{\mathrm{C}}_{j}[n] &= \lbrace \mathbf{s}_{j}[n], {\textstyle\sum}_{j \in \mathcal{J}}R_{j}[n], {\textstyle\sum}_{j \in \mathcal{J}}c_{j}[n] \rbrace. \label{State_C}
% && r_{j}[n] = g(R_{j}[n]), \forall j. \label{Reward_D} 
\end{align}
Here, this reward structure is used in the centralized training and decentralized execution (CTDE). 
During the centralized training, the centralized reward can be used to observe the throughput and the collision event of other UT agents.
Note that each agent has its own Actor-Critic network and decide the optimal access action for itself via trained policy.
Here, cheap talk is necessary during both the training and exploitation phases.

\item \textbf{eRACH} is our proposed RA scheme with using Actor-Critic framework, wherein each UT agent optimizes its access action in the environment in a fully \emph{distributed} manner. 
Each UT has partial state observability.
% agent only observes a part of the state that is local to its operation. 
The agents interact through the environment but do not communicate with other agents. 
The learning can be cast as a partially observable MDP (POMDP).
% , and applying its learning algorithm independently

% Instead we can assume that each agent only observes a part of the state that is local to its operation. If the agents interact through the environment but do not otherwise communicate, then each agent is learning to behave in a partially observable MDP (POMDP), ignoring the distributed nature of the interaction, and applying its learning algorithm independently.
\end{enumerate}
% i) all frames have the same time length, 
It is worth noting that in eRACH-Coop, each UT utilizes information from other UTs (see \eqref{Reward_C} and \eqref{State_C}) unlike the other baselines.
Accordingly, eRACH-Coop can achieve better performance by using additional information. 
Here, one can regard its performance as an upper-bound result.

The following conditions are considered comparing the baselines: \textit{1)} all UTs have enough data to be transmitted; \textit{2)} if collisions occur at a time slot for an SAT, all attempted UTs fail to access at the time slot to the SAT; and \textit{3)} the RA signaling is done in a two step procedure (see Sec.~\ref{Sec2_RA}).

\begin{figure}[t]
  \centering
  \includegraphics[width=\linewidth]{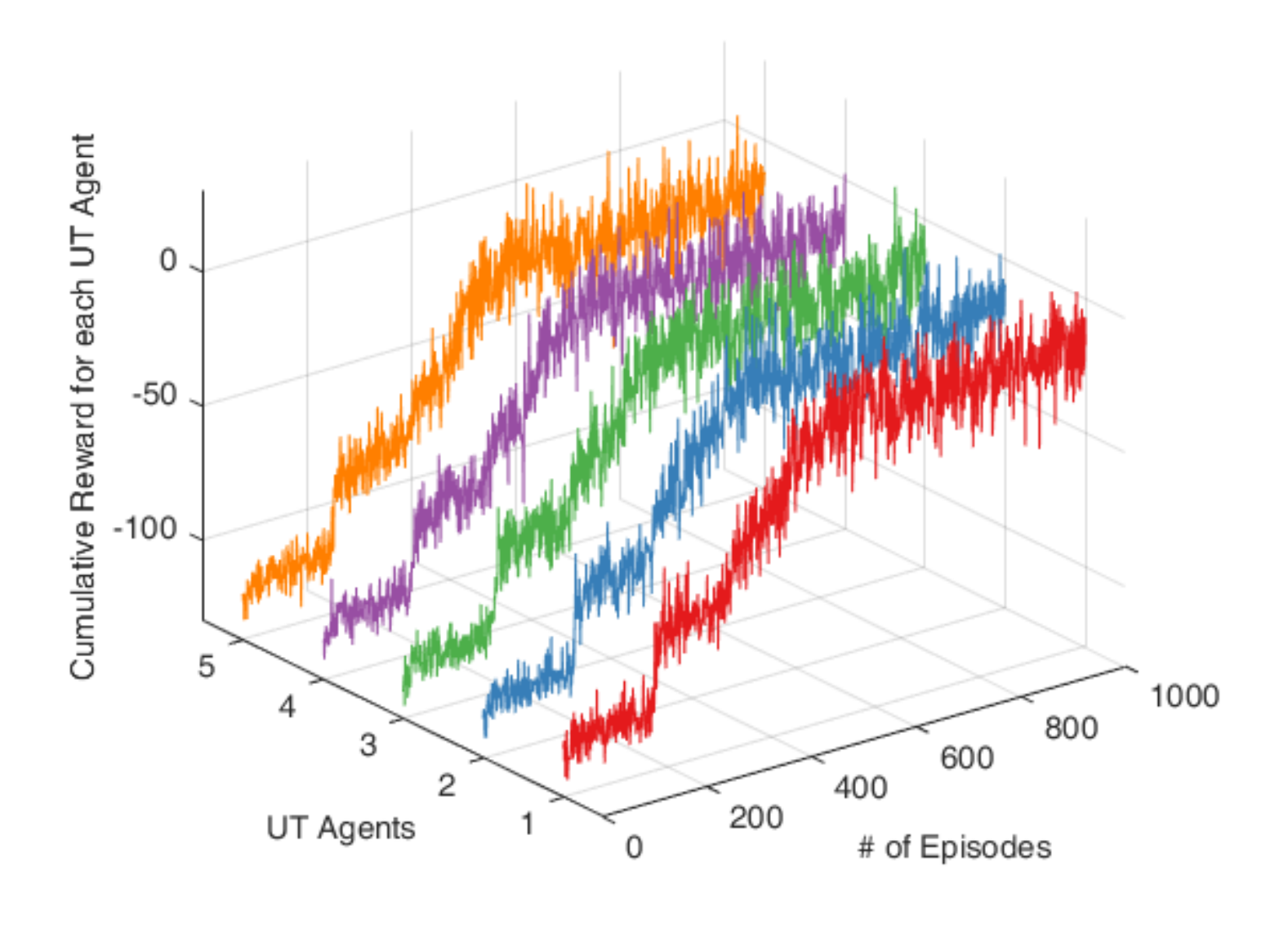}
  \caption{The convergence of eRACH for each UT agent ($J=5$).}
  \label{fig_Convergence_Agents}
% \vskip -5pts
\end{figure}

\begin{figure}[t]
  \centering
  \includegraphics[width=\linewidth]{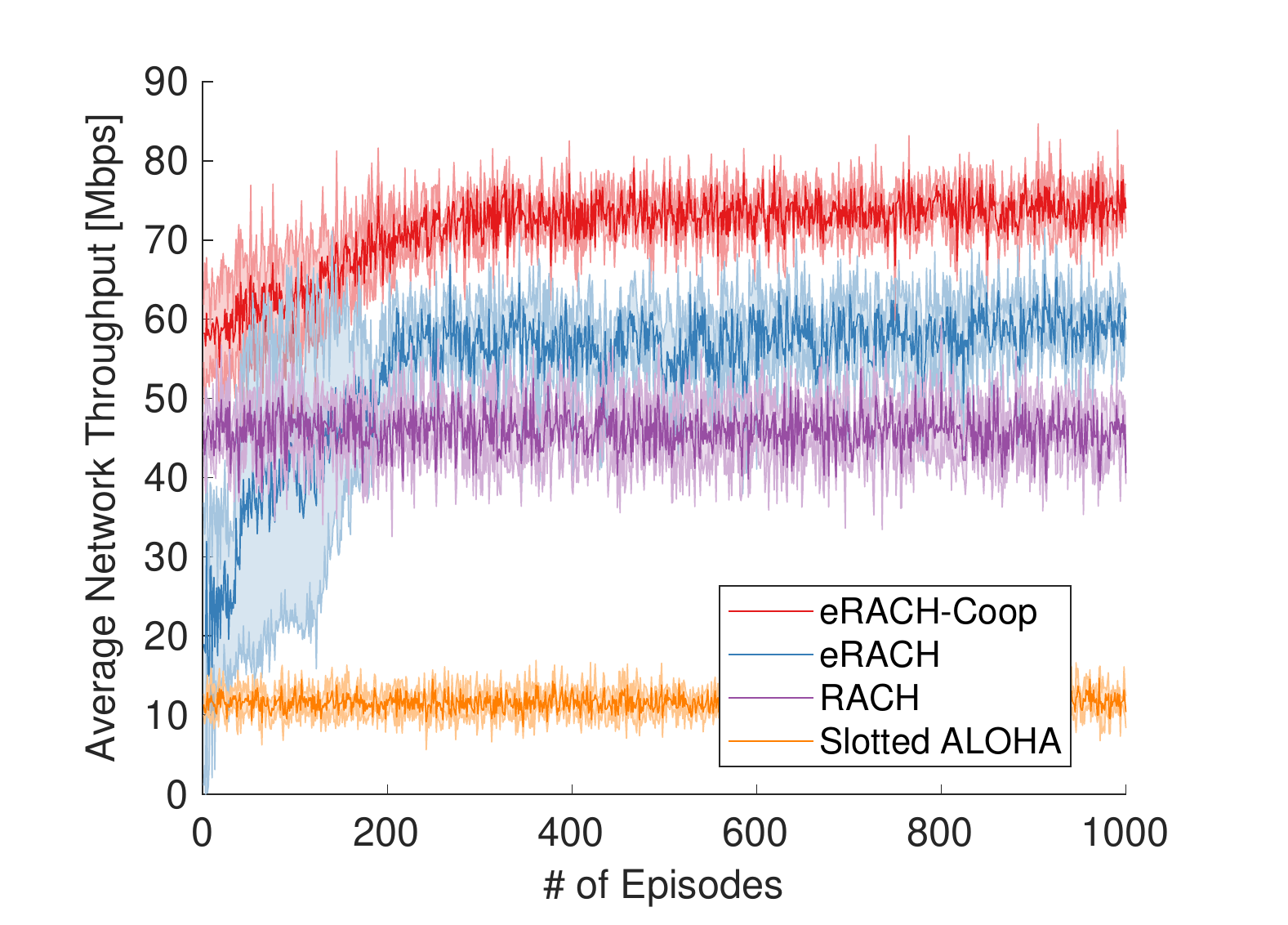}
  \caption{Network throughput for different RA schemes.}
  \label{fig_Thr_Itr}
% \vskip -5pts
\end{figure}

% access action policy of eRACH. The color of each box in the figures indicates the probability of collision, and the darker the color, the higher the collision rate. The results marked by yellow lines in (a) and (b) corresponds with Fig. \ref{Illustration_Snapshot}.}
% \tred{[JH: Throughout the paper, for every subfigure caption, put a period at the end.]}

%%%%%%%%%%%%%%%%%%%%%%%%%%%%%%%%%%%%%%%
\subsection{Model Architecture and Training}

% Following the Actor-Critic MADRL framework \cite{MA_Foerster_1, Lowe, MA_Foerster_2}, each agent has an Actor~NN and a Critic~NN. 
For all Actor and Critic NNs, we identically consider a $4$-layer fully-connected multi-layer perceptron (MLP) NN architecture. 
Each MLP has $2$ hidden layers, each of which has the $128\times 128$ dimension with the rectified linear unit (ReLU) activation functions. Both Actor and Critic NNs are trained using RMSprop with the learning rate $0.0001$, batch size $2604$, $2604$ iterations per episode, training episodes $1000$, and $2604$ iterations per update. 
The simulations are implemented using TensorFlow Version 2.3. 
Main parameters on simulation settings are summarized in Table \ref{table_Paramter}.

Figs. \ref{fig_Convergence_Agents} and \ref{fig_Thr_Itr} plot the convergence of eRACH.
Firstly, in Fig. \ref{fig_Convergence_Agents}, it is provided the training convergence behavior of Actor and Critic NNs for the five UT agents, in which the solid curves denote cumulative reward for each UT agent.
% For the proposed eRACH Actor-Critic MADRL framework in Algorithm \ref{Algorithm}, we provide the training convergence behavior of the Actor and Critic NNs for five UT agents in Fig. \ref{fig_Convergence_Agents}, in which the solid curves denote cumulative reward for each UT agent.
% and the shared areas correspond to the maximum deviations during $5$ simulation runs. 
% We note that our cumulative reward directly represent the network throughput, such that we can validate the convergence of the proposed Actor-Critic MADRL through the result of network throughput over iterations.
The results show that each Actor and Critic NN for $J=5$ converge within about $500$ episodes.
% whereas for $3$ agents, the convergence requires around $50000$ episodes.
% For large-scale systems, it is thus necessary to accelerate the training convergence. In this regard, periodically averaging the Actor NN parameters (i.e., federated learning \cite{Google:FL19}) or outputs (i.e., federated distillation \cite{MLPCD}) across agents could be an interesting topic for future study. 
% Note that the cumulative reward does not directly represent the system utilities, i.e., network throughput, collision, and delay, which are evaluated in the following subsections.
Besides, Fig. \ref{fig_Thr_Itr} compares eRACH and eRACH-Coop with conventional RACH and slotted ALOHA over the number of training episodes, in which shaded areas correspond to the maximum deviations during $5$ simulation runs.
% , to identify the effectiveness of access protocol in throughput.
Note that our cumulative reward corresponds to the average network throughput, such that we can validate the convergence of the proposed Actor-Critic MADRL-based method in Algorithm \ref{Algorithm} with this figure.

As identified in this figure, eRACH and eRACH-Coop converge and outperform the conventional RACH scheme only within around $200$ training episodes.
It suggests that the observed data during only one day is enough to train the eRACH UT agent.

\begin{table*}[t]
  \centering
 \resizebox{1\linewidth}{!}{\begin{minipage}[h]{\linewidth}
  \centering
    \caption{Top-view snapshots of SAT BS associations and preamble resource utilization (Resource Util.) under eRACH and RACH for $4$ consecutive time slots, where associated and backed-off UTs are drawn with or without solid lines, respectively ($K=2$, $I=22$, $J=5$, $P= 2$). For the same period, the RA snapshots are illustrated in Fig. \ref{Snapshot}.
    % solid-colored lines and no-line represent the access actions of each UT agent and backoff, respectively, which correspond (a) RACH and (d) eRACH in Fig. \ref{Snapshot}. Resource~Util. represents the utilization of the available resources $P$, that is (Achievable throughput by the access action)/(Achievable throughput by an optimal action).
    }
	\label{Illustration_Snapshot}
	\input{Table/Table_Snapshot}
  \end{minipage}}
\end{table*}
% \tred{[JH: Good idea to have the bar plot below each snapshot. Can we also report other performance measures in Tables II and III here?]}

\begin{figure*}
\centering
\subfloat[RACH. \label{a-Snapshot}]{\includegraphics[width=0.33\linewidth]{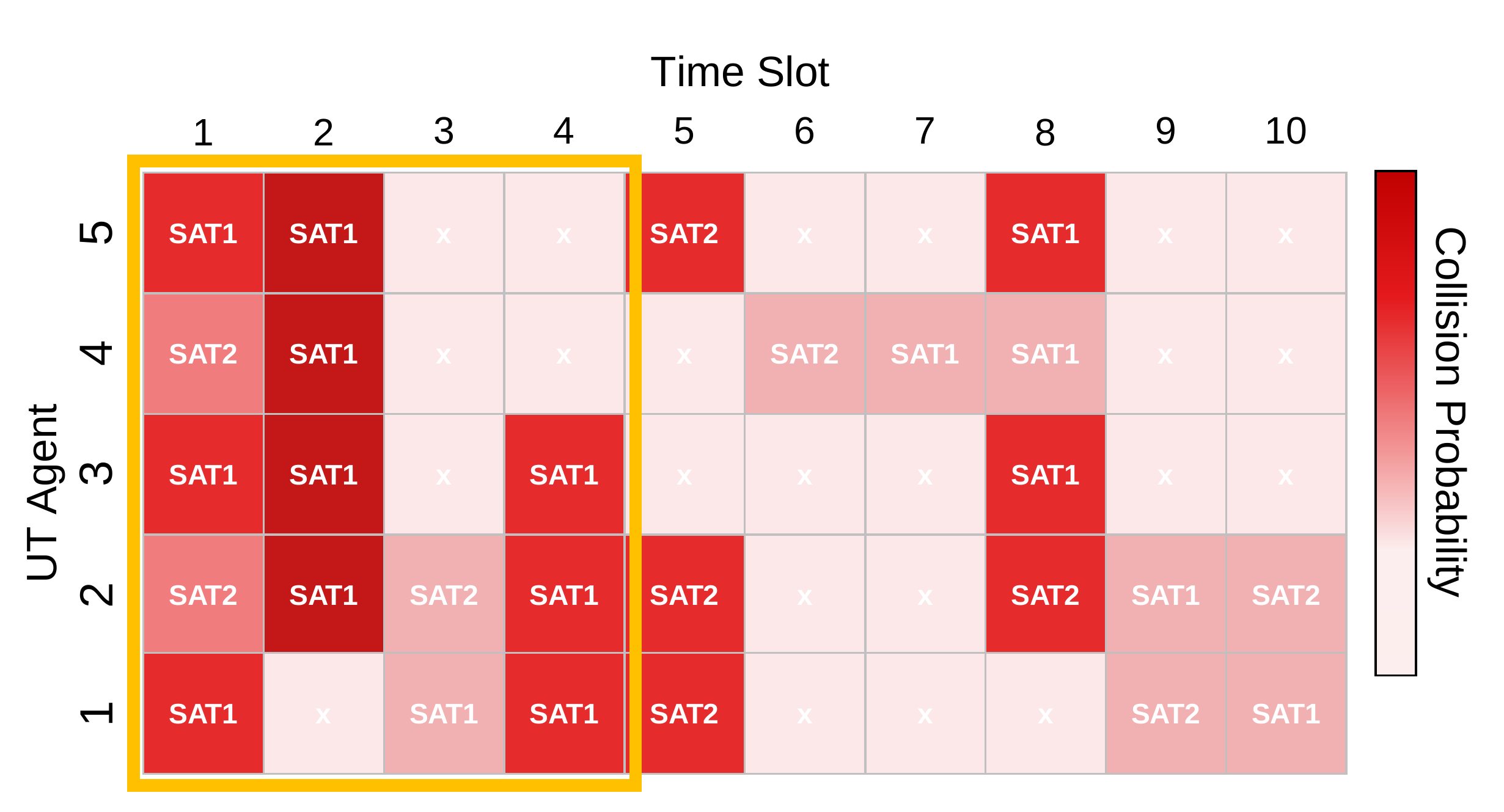}}
% \hfill \\
\subfloat[eRACH. \label{c-Snapshot}]{\includegraphics[width=0.33\linewidth]{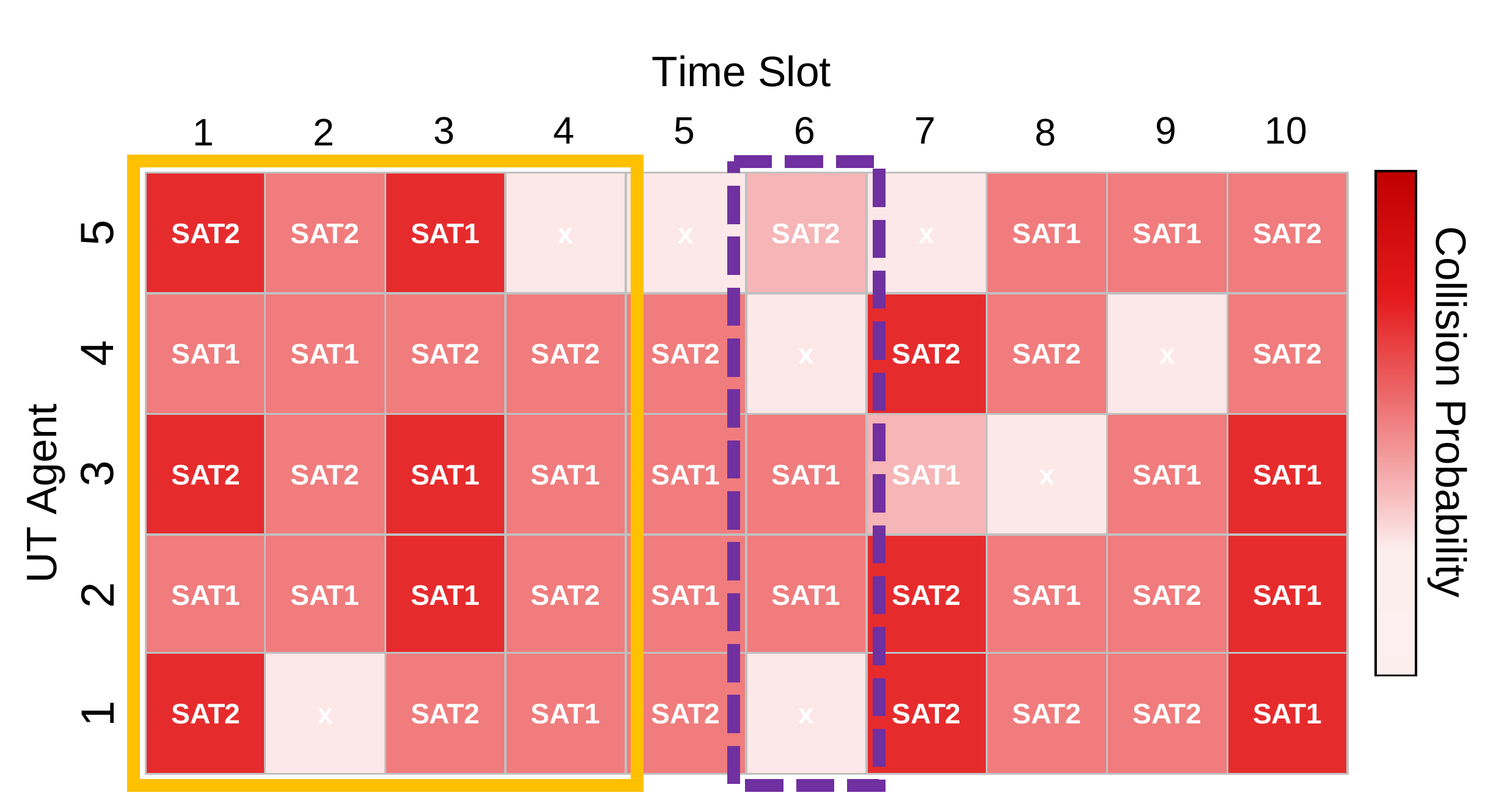}}
% \hfill \\
\subfloat[eRACH-Coop. \label{d-Snapshot}]{\includegraphics[width=0.33\linewidth]{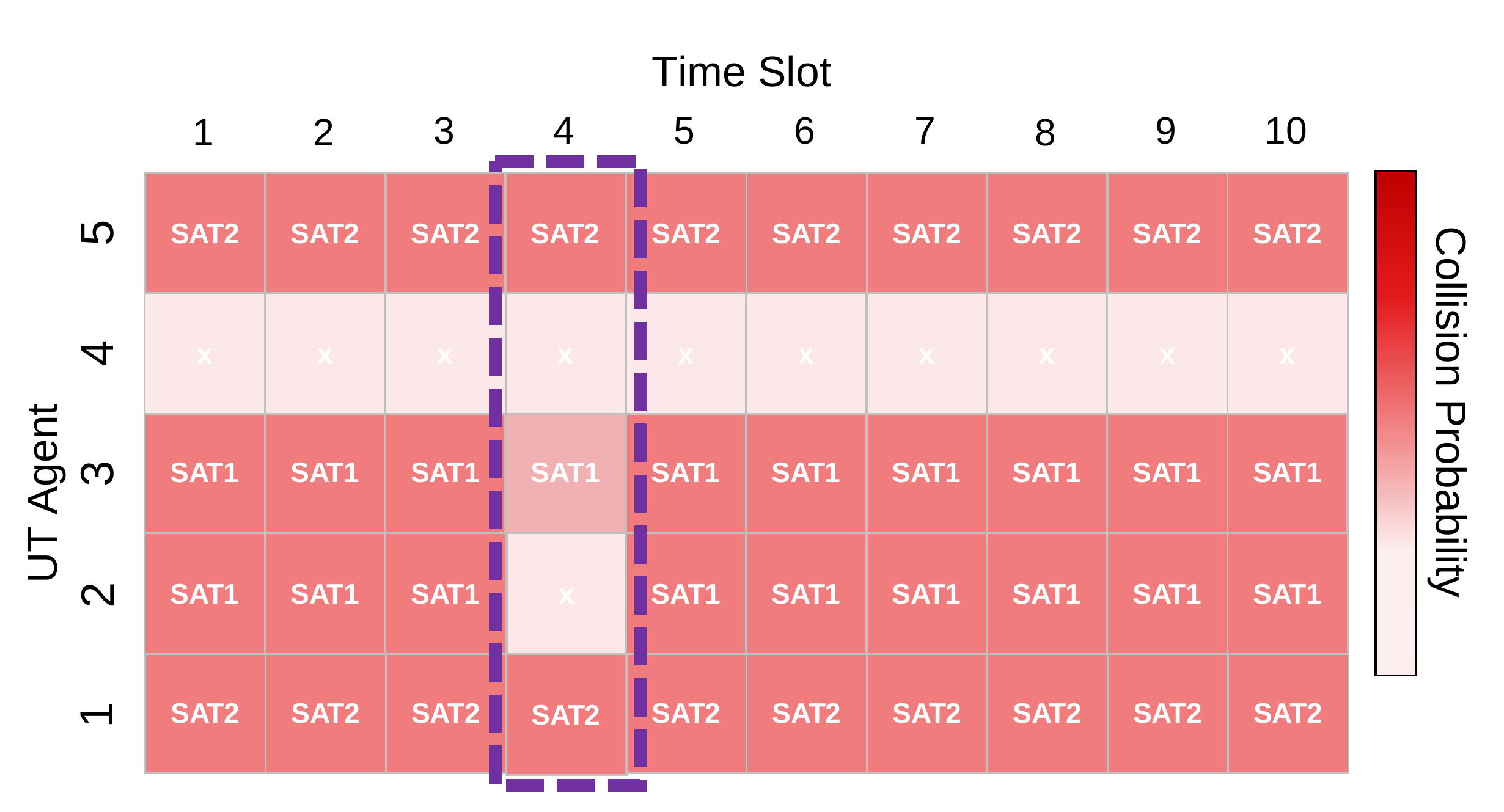}}
% \hfill 
% \subfloat[eRACH (\textit{Distributed - Rate Max}) \label{b-Snapshot}]{\includegraphics[width=0.8\linewidth]{Figure_Result/SnapShot_Distribute_Rate.pdf}}  \hfill \\

\caption{RA snapshots under eRACH, eRACH-Coop, and RACH for 
$10$ consecutive time slots, where $\{\textsf{SAT1},\textsf{SAT2}\}$ identifies SAT BS associations, and the red color of each box indicates the collision rate (the darker, the higher). For the first $4$ consecutive time slots in (a) and (b), the SAT BS association snapshots are illustrated in Fig. \ref{Illustration_Snapshot} ($K=2$, $I=22$, $J=5$, $P=2$).}
% \tred{[JH: Add periods for all subfigure captions; collision rate $\to$ collision rate]}
\label{Snapshot}
\vspace{.2em}
\end{figure*}

%%%%%%%%%%%%%%%%%%%%%%%%%%%%%%%%%%%%%%%%%%%%%%%%%%%%%%%%%%%%%%%%%%%%%%%%%%%%%%
\subsection{RA Performance Analysis}
% \tred{[JH: Refer to this revised section as an example, and revise other subsections in a similar way]}

% To validate the effectiveness of eRACH, we compare eRACH  in terms of network throughput, collision avoidance, and RA latency, we present comparison study in Fig. \ref{fig_Thr_Itr} and provide ablation studies in Tables \ref{Table_Proposed} and \ref{Table_Proposed_Fair_Cheap}.
% Especially, Fig. \ref{fig_Thr_Itr} and Table \ref{Table_Proposed} show the validity of eRACH as a spectral-efficient or delay-efficient RA method for LEO SAT networks.
% emphasizes the applicability of eRACH, while 
% \tred{[JH: Throughout the simulation sections, make sure that the first sentence is clear and compact. Remove any redundant introduction and unclear expressions (e.g., comparison study (wrong), ablation studies (in terms of what, for what purpose?), not all the figures/tables covered in the subsequent paragraphs are mentioned; proposed X$\to$ the proposed X, our proposed X; the proposed eRACH$\to$ (simply) eRACH; a cooperative version of eRACH (name it, eRACH-Cooperative); Consistently follow the definitions without too frequently abusing notations (e.g., throughput, average throughput$\to$ average network throughput) ]}

Fig. \ref{fig_Thr_Itr} and Table \ref{Table_Proposed} compare eRACH with other baselines, in terms of network throughput, collision rate, and access delay.
The results validate that eRACH achieves higher average network throughput with lower average access delay than the baselines.
% than the baselines in all cases, including dense and sparse network scenarios.
In particular, compared to RACH, eRACH and eRACH-Coop achieve 31.18 \% and 54.61 \% higher average network throughput, which is equivalent to 5.08x and 6.02x higher average network throughput of slotted ALOHA, respectively. 
Moreover, the results show that eRACH achieves 1.49x and 7.31x lower average access latency compared to RACH and slotted ALOHA, respectively.
% eRACH succeeds in access attempts most rapidly, i.e., lowest access delay.

% \tred{[JH: Consistently use either average collision rate or average collision rate in figures and the main body; when we say `average,' clarify the averaging is taken with respect to which source of randomness (e.g., simulation runs, channel realizations, and so forth)]}
Next, in terms of RA collision, the average collision rate  of eRACH is 1.41x lower than slotted ALOHA, yet is 4.94x higher than RACH. 
As opposed to other model-based protocols, eRACH is willing to risk collisions for yielding higher throughput and lower access latency in a given network and channel environment.
This fact suggests that eRACH optimizes access flexibly according to the importance of throughput-collision in the LEO SAT network.
Such flexibility is advantageous to best-effort services such as enhanced mobile broadband (eMBB) applications, but becomes a downside for mission-critical applications such as ultra-reliable and low-latency communication (URLLC). For the latter case, investigating the emergent protocols that strictly abide by a collision constraint are interesting and deferred to future work.

% Interestingly, the collision rate of eRACH is higher than RACH, though eRACH achieves the best throughput and access delay competing with RACH.
% This fact suggests that it can be optimal to attempt access in some scenarios for LEO SAT networks even if there is a risk of collision.
% In this regard, one may question how eRACH specifically determines access action to achieve this low access delay and high throughput, which is discussed next.

Lastly, comparing the performance between eRACH and eRACH-Coop in Table \ref{Table_Proposed}, we conclude that eRACH is a fairer protocol that achieves 1.24x higher Jain's fairness than that of eRACH-Coop. The rationale is because the fully distributed operations of eRACH inherently hide the information that may promote selfish actions. Meanwhile, there is also a room to improve the fairness of eRACH-Coop by applying a fairness-aware reward function during training and/or learning fairness-aware representations for cheap talk communication, which could be an interesting direction for future research.

\subsection{SAT Association and RA Operations}

As shown in Table \ref{Illustration_Snapshot}, eRACH efficiently utilizes resources by optimizing the association of SAT-UT while considering the non-stationary LEO SAT network.
Besides, as shown in Fig. 8, eRACH backs off flexibly thereby avoiding the collision under consideration of the given local environment, whereas when collision occurs, RACH backs off more than necessary due to a randomly selected backoff window. 
% We observe from such a comparison of eRACH and RACH that; eRACH designs the optimum association in addition to proper backoff depending on the network scenario.
% for efficient RA, it is necessary to decide on the optimum association in addition to proper backoff depending on the network scenario.

%%% explain the rationale behind such results
% \textit{Distributed} (\textit{Distributed - Rate Max}), which correspond to \eqref{Reward_D_Objective} with $\sigma=0$, selects the access actions which evenly split UT agents for each time slot, while in \textit{Distributed}, which correspond to \eqref{Reward_D_Objective} with $\sigma=2$, some distributed UT agent often backs off action to reduce collisions even at the expense of throughput.
In particular, eRACH-Coop in Fig. \ref{d-Snapshot} shows that a certain UT agent continuously backs off. 
Here, since the cooperative UT agents consider the network throughput rather than the throughput of itself in the reward function, the sacrifice by this certain agent is reasonable. 
% a certain agent can make decisions that sacrifice itself for the overall network performance.
In contrast, for eRACH in Fig. \ref{c-Snapshot}, each agent takes turns and decides to backoff.
% , even though there is no communication with the other agent
% some agents often take backoff action to reduce collisions even at the expense of throughput. 
The distributed UT agents also learn how to sacrifice for the entire network even without exchanging information between agents.
It suggests that the distributed eRACH protocol is able to emerge from a given local environment during the training process.

% the optimal policy appropriately determines the backoff action of each agent in the network depending on the network scenario.
% it is suggested that the optimal policy some agents should purposely back off (waits for access attempt) for the whole network performance in the massive access scenario.
% In that respect, this cooperative method can be seen as making the optimal decision for the overall network performance.

%%% briefly mention some limitations of the proposed method, opening up future research topics.
We, however, are aware of a few limitations of eRACH and eRACH-Coop, which are marked by a dotted purple line in Figs. \ref{c-Snapshot} and \ref{d-Snapshot}.
% In particular, the first is that most agents attempt to access the same SAT on some slot, which is collision-wise inefficient, as observed in \textit{Distributed - Rate Max} of Fig. \ref{b-Snapshot}.
In eRACH and eRACH-Coop, there are unnecessary backoff decisions that seem to be throughput-wise inefficient.
It underlines that \textit{1)} DRL-based method occasionally decides a poor action, especially when dealing with time-series data, like our environment, and \textit{2)} the fairness issue can arise in Cooperative, during the certain agent sacrifice for the sake of the entire network. 

Despite the few limitations, we observe from such comparison results that eRACH can emerge from a local environment with the following remarkable features: \textit{1)} eRACH flexibly avoids unnecessary backoff with an understanding of the given network conditions, and \textit{2)} eRACH access the optimal LEO SAT considering the periodicity of the LEO SAT constellation.
% \tred{[JH: Explaining `why' and `how' is the key, which is currently missing.]}

%%%%%%%%%%%%%%%%%%%%%%%%%%%%%%%%%%%%%%%%%%%%%%%%%%%%%%%%%%%%%%%%%%%%%%%%%%%%%%
\subsection{Ablation Study of Key Hyper Parameters}

\begin{figure}[t]
  \centering
  \includegraphics[width=\linewidth]{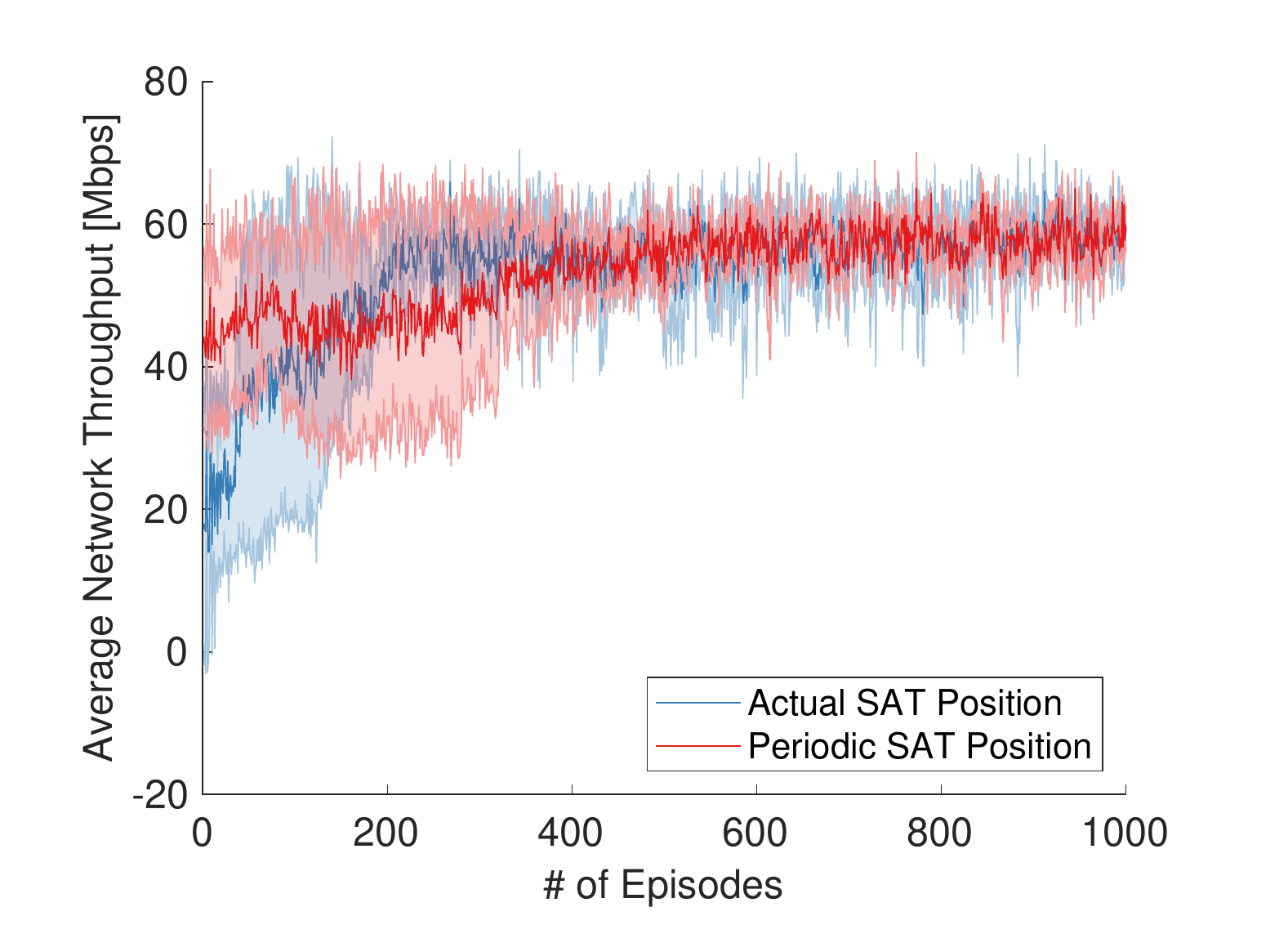}
  \caption{Impact of SAT location precision on average network throughput ($\sigma^2=100$). 
  }
  \label{fig_Thr_Itr_Actual}
\vspace{.4em}
\end{figure}
% \tred{[JH: Add one more subfigure in-between Rate-Max and Collision-Aware; Then, throughout the paper we will call them using only $\rho$. In this part, we compare the sensitivity to $\rho$. When $\rho=0$ (current Rate-Max), we cannot see any back-off behaviours that emerge as $\rho$ increases.]}

% \textbf{Observed Information vs Prior Knowledge:}
\textbf{Impact of SAT Location Precision}.\quad
While training eRACH, the information of SAT location affects dominantly the reward and convergence, as explained in Fig. \ref{fig_StateReasoning} of Sec.~\ref{Sec4B}.
However, it is challenging to find the SAT location precisely.
In this regard, to examine whether easy-to-acquire prior information can be used instead of actual observed information, we present Fig. \ref{fig_Thr_Itr_Actual} and Table \ref{Table_PositionError}.
% we here compare the eRACH, which trained with observed information and with prior knowledge.

\begin{table}[]
  %   \hspace{5pt}
    \centering          
    \caption{Impact of SAT location precision on normalized reward and the number episodes until convergence. This table is also visualized in Fig. \ref{fig_PositionError}.}
    % Normalized reward and \# of episodes for convergence over $\sigma^2$ which represents the positioning error of LEO SAT. This table corresponds with the result in Fig. \ref{fig_PositionError}. \tred{[JH: Report the exact values, without inequalities and/or approximations.}}
    \resizebox{1\columnwidth}{!}{\begin{minipage}[]{.96\columnwidth}
    \centering
    \label{Table_PositionError}
    \input{Table/Table_PositionError}
    \end{minipage}}
\end{table}

Specifically, various physical forces perturb SAT, e.g., Earth oblateness, solar and lunar gravitational effects, and gravitational resonance effects \cite{SAT_PositionError}.
In actual SAT communication systems, various methods are used to track the location of perturbed SAT, such as pointing, acquisition, and tracking (PAT) mechanism \cite{SAT_PAT} and global navigation satellite system (GNSS).
Even with the tracking and navigation systems, some orbital positioning error of LEO SAT is inevitable.
In this regard, we demonstrate the validity of the LEO SAT periodic location information, which is prior knowledge, in eRACH in Fig. \ref{fig_Thr_Itr_Actual} and Table \ref{Table_PositionError}.

% \textbf{Usefulness of LEO SAT Position Information:}
As shown in Fig. \ref{fig_Thr_Itr_Actual}, eRACH trained with the periodic LEO SAT position information converges within around $600$ episodes.
It closely approaches the one trained with actual LEO SAT position information.
% Note that the periodic LEO SAT position can be predicted utilizing SGP4, which is an open-source orbit propagator \cite{SAT_PositionError}.
However, the periodic SAT position is not always effective for training eRACH, as shown in Table \ref{Table_PositionError}.
The number of episodes for convergence, which corresponds to the training time, increases as the positional error goes.
Besides, the overall reward decreases as the positional error goes.
Nevertheless, information of periodic SAT position can still be used effectively for eRACH until the divergence point at around $\sigma^2 = 10^4$.
% These results suggest that eRACH can be sufficiently trained with only prior information, highlighting eRACH can be trained in offline as well as online manners 
These results suggest that eRACH can be sufficiently trained with only prior information, and this emphasize that eRACH can be trained by not only online RL but also offline RL \cite{DRL_OnlineOffline}.

% In particular, for pointing, acquisition, and tracking (PAT), two-line elements (TLE) \cite{TLE} and simplified general perturbations 4 (SGP4), which is open-source orbit propagator, are utilized.
% TLE+SGP4 considers secular and periodic variations due to Earth oblateness, solar and lunar gravitational effects, gravitational resonance effects, and orbital decay using a simple drag model. 
% However, the prediction by TLE+SGP4 has some orbital error, and the orbit error is increasing over time.
% The subsidiary use of GPS systems is also being considered to overcome the issue. However, there are still limitations in determining the exact position of the SAT.

%%%%%%%%%%%%%%%%%%%%%%%%%%%%%%%%%%%
% \textbf{Throughput-Collision Tradeoff:}
\begin{figure}[t]
\centering
\subfloat[Sparse Case, $d_{j, j'} = 1000$.   \label{Snapshot_Density_Sparse}]{\includegraphics[width=0.75\linewidth]{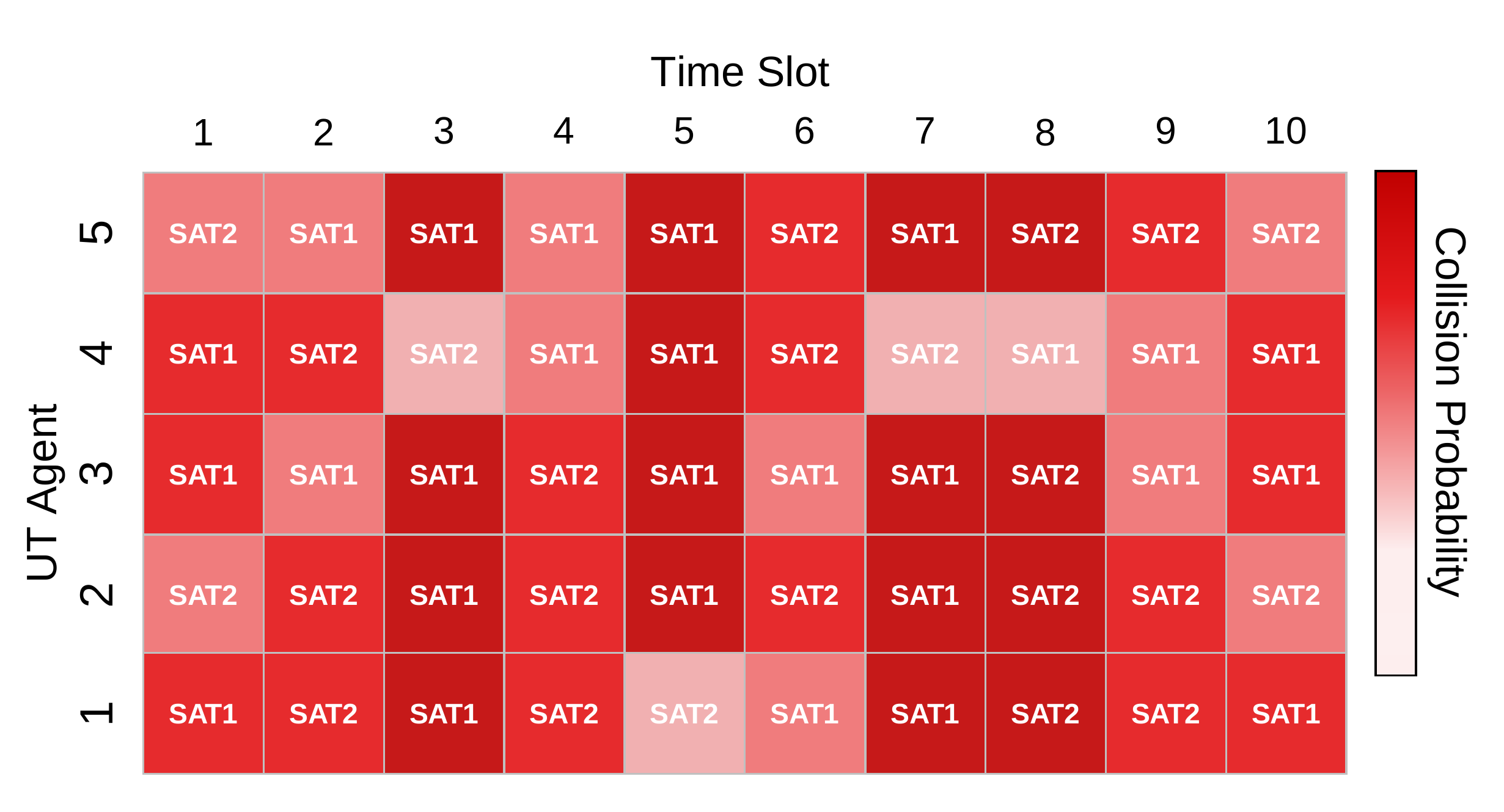}}\hfill \\
\subfloat[Dense Case, $d_{j, j'} = 10$. \label{Snapshot_Density_Dense}]{\includegraphics[width=0.75\linewidth]{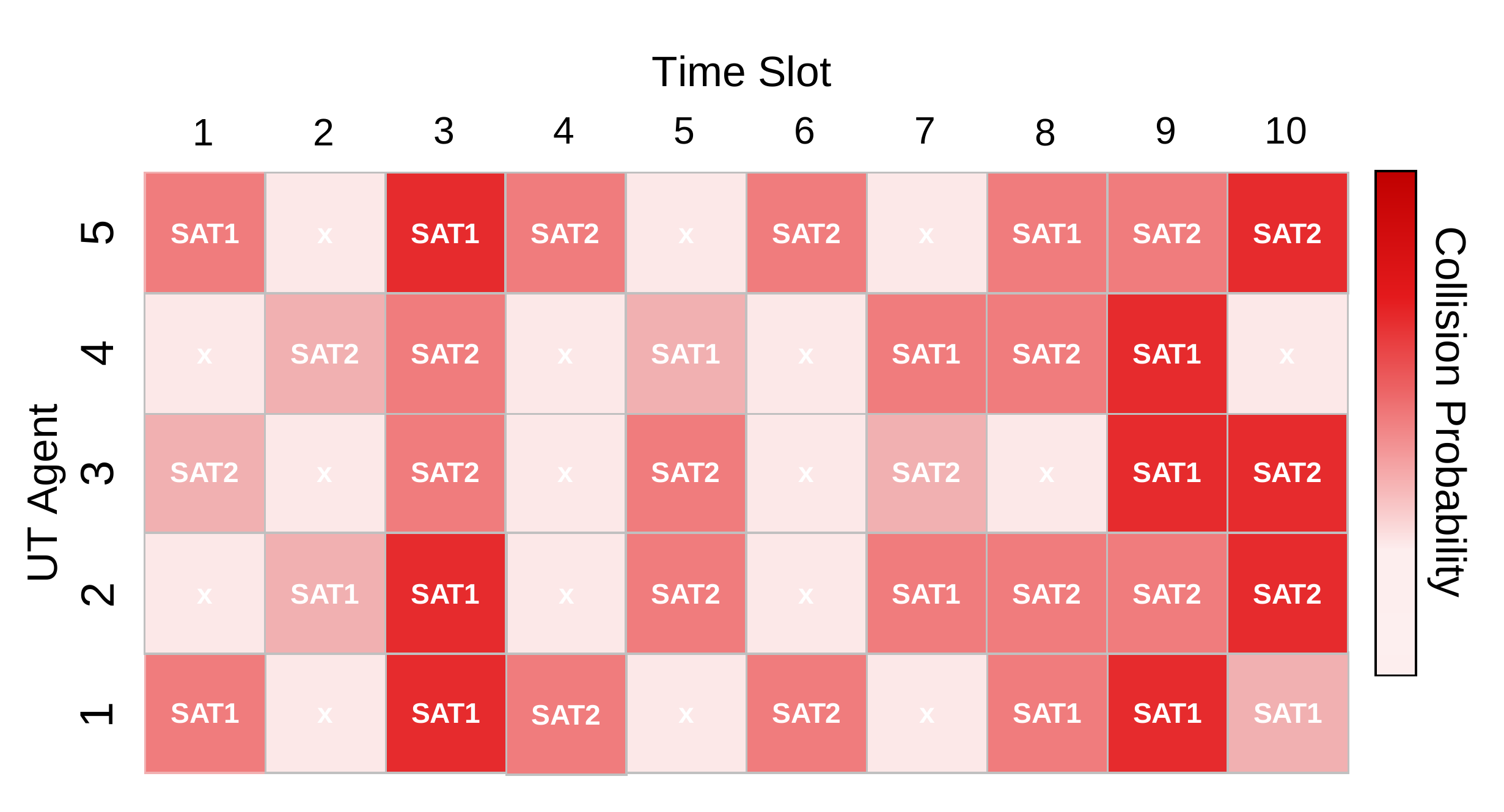}}
\caption{Impact of UT distribution on access action policy of eRACH. Here, $d_{j, j'}$ represents the average distance between UTs in [m].}
\label{Snapshot_Density}
% \tred{Snapshot of access action policy of eRACH in sparse and dense cases. The color of each box in the figures indicates the probability of collision. The darker the color, the higher the collision rate. Here, $d_{j, j'}$ represents the average distance between UTs in [m].}
\vspace{.2em}
\end{figure}

% \subsection{\tred{Adaptability in Different Objectives}}
%%%% Rate-oriented vs Collision-oriented
% \textbf{Application-Oriented Protocol:}
% \textbf{Adaptive Protocol:}
% \textbf{Adaptability in Different Objectives:}

\textbf{Impact of UT Distribution}.\quad
The performance of the access protocol depends not only on the available resources and the number of UTs which attempt to access but also on the deployment scenario, e.g., sparse and dense networks.
In this regard, we present Fig. \ref{Snapshot_Density} to demonstrate the impact of UT distribution on eRACH.
% which shows the access action of eRACH in different deployment cases, e.g., sparse and dense cases.

In the sparse networks, each UT experiences different channel conditions for SAT; thus, each UT has a particular advantageous SAT to connect to.
Here, eRACH mainly focuses on designing the optimal association with SAT rather than backoff action.
Whilst, in the dense networks, the channel difference between each UT is not noticeable since the distance between UTs is relatively close compared to the distance between SAT and UTs.

For the sparse case of Fig. \ref{Snapshot_Density_Sparse}, eRACH does not backoff during the entire time slot.
In contrast, in the dense case of Fig. \ref{Snapshot_Density_Dense}, eRACH backoff up to about $30$ \% of the total time slot and focuses on backoff action rather than optimal association.
Notably, eRACH in both cases achieves similar performance even though they chose different aspects of access action.
This fact further corroborates that eRACH, which flexibly decides the access action, can emerge from a given local network scenario without communication between agents.

\begin{figure}[t]
\centering
% \subfloat[RACH \label{a-Snapshot}]{\includegraphics[width=0.8\linewidth]{Figure_Result/SnapShot_RACH.pdf}}\hfill \\
% \subfloat[eRACH-Coop \label{d-Snapshot}]{\includegraphics[width=0.8\linewidth]{Figure_Result/SnapShot_Cooperative.pdf}}\hfill \\
\subfloat[\textit{Rate-Max} ($\rho=0$). \label{RateMax}]{\includegraphics[width=0.75\linewidth]{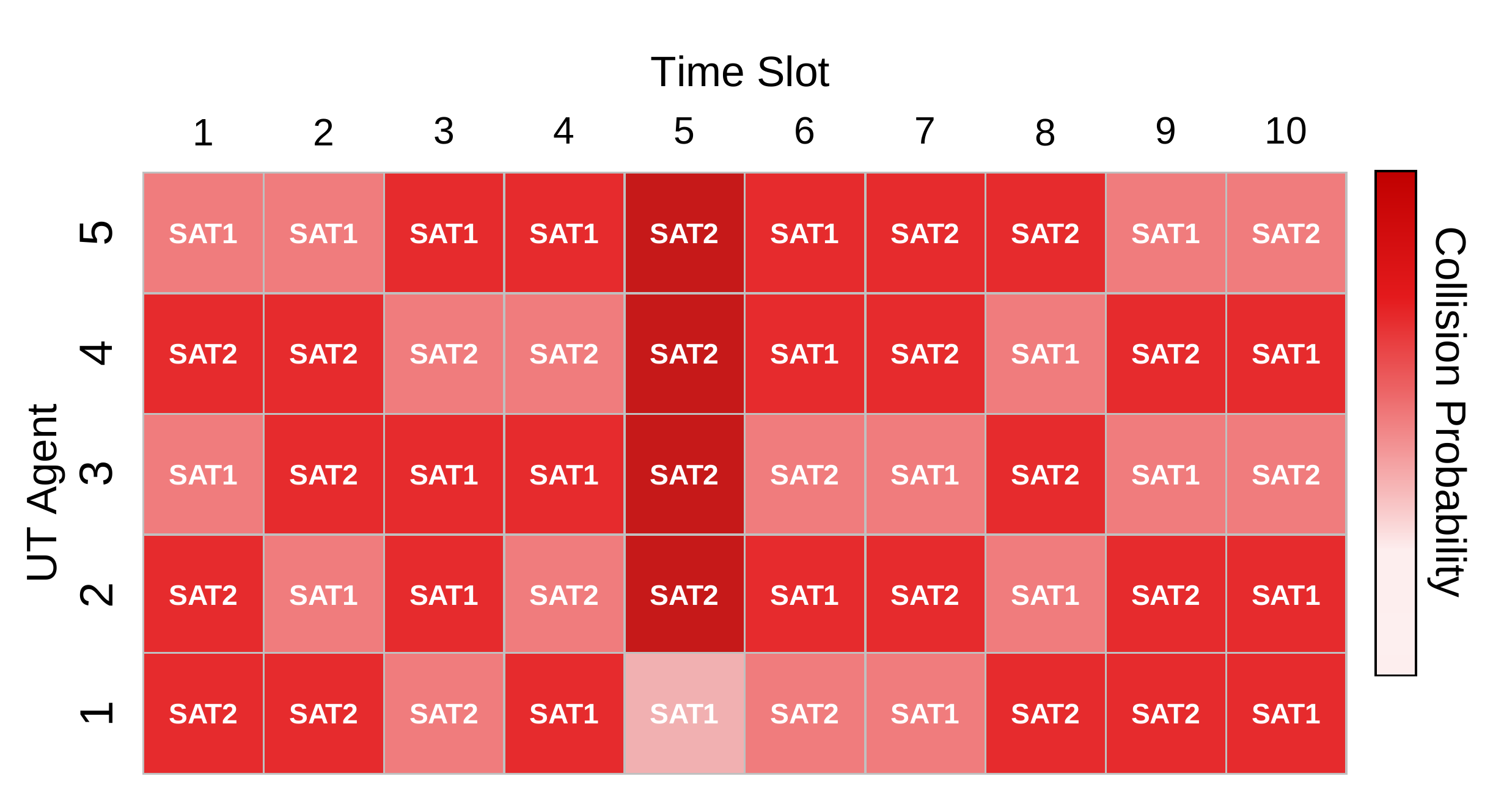}}  \hfill \\
\subfloat[\textit{Collision-Aware} ($\rho=2$). \label{CollisionAware}]{\includegraphics[width=0.75\linewidth]{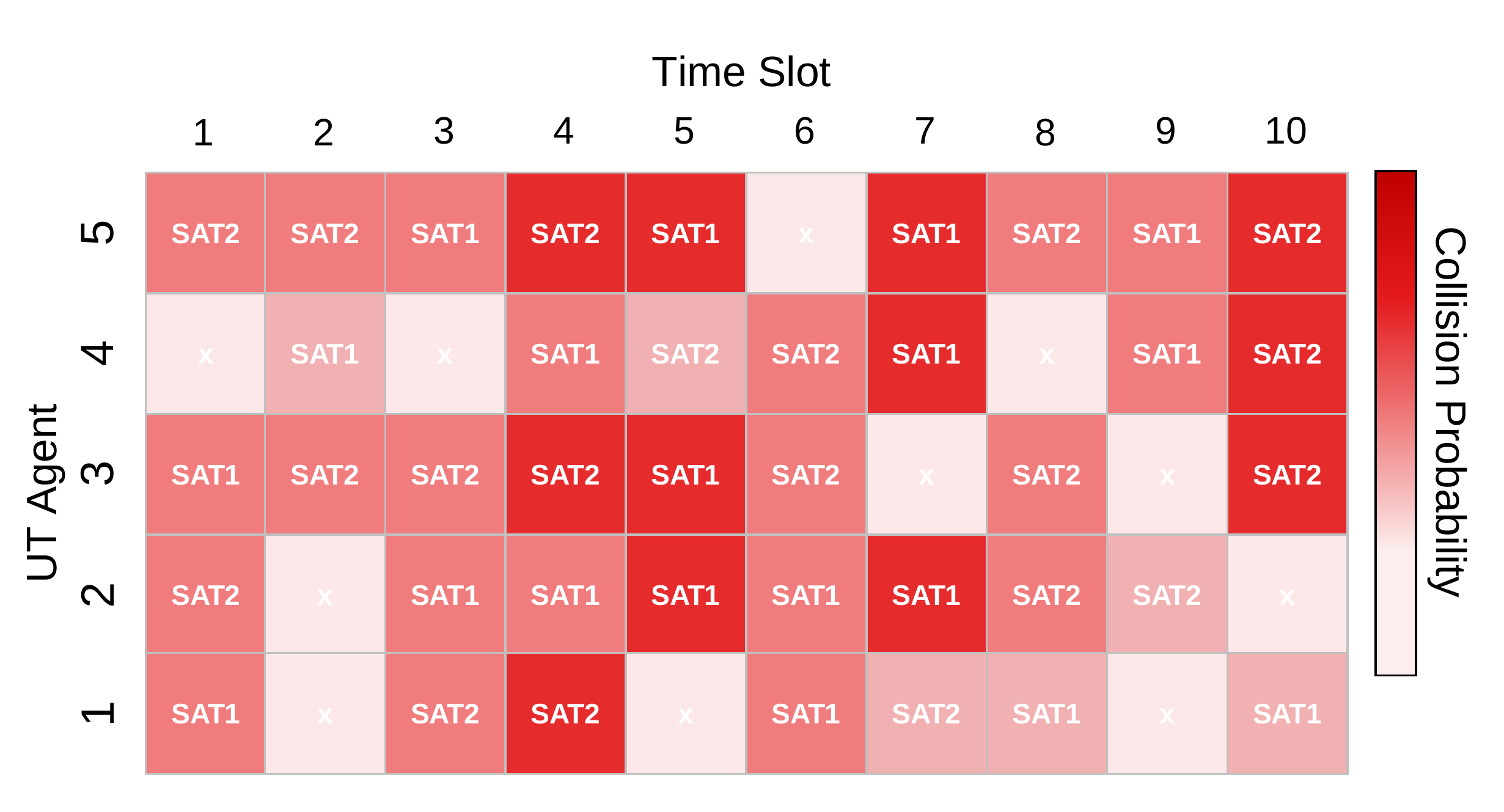}}\hfill
\caption{Impact of the collision aversion factor $\rho$ on access action policy of eRACH.}
% \tred{Snapshot of access action policy of eRACH. The color of each box in the figures indicates the probability of collision, and the darker the color, the higher the collision rate. }
\label{Snapshot_RateMax}
% \vspace{.2em}
\end{figure}

\textbf{Impact of the Collision Aversion Factor $\rho$}.\quad
In the LEO SAT network, which suffers from a long one-way propagation delay, the contention resolution is challenging, as discussed in Sec. \ref{Background}.
Hence, consideration for collision can be another significant issue.
Regarding this, we further present the numerical results of the collision aversion case.

To clearly see the collision aversion case, we additionally present \textit{Rate Max}, which only considers the throughput while training of eRACH by setting $\rho = 0$.
Here, the reward function in \eqref{Reward_D} is rewritten for \textit{Rate Max} case as follows.
\begin{align}
r_{j}[n] = g(R_{j}[n]), \ \forall j, n. \label{Reward_D_Objective}  
\end{align}
% Note thatThen, the framework can optimize collision as well as throughput, by adjusting $$ and $\rho$.

\begin{figure}[]
  \centering
  \includegraphics[width=.85\columnwidth, clip]{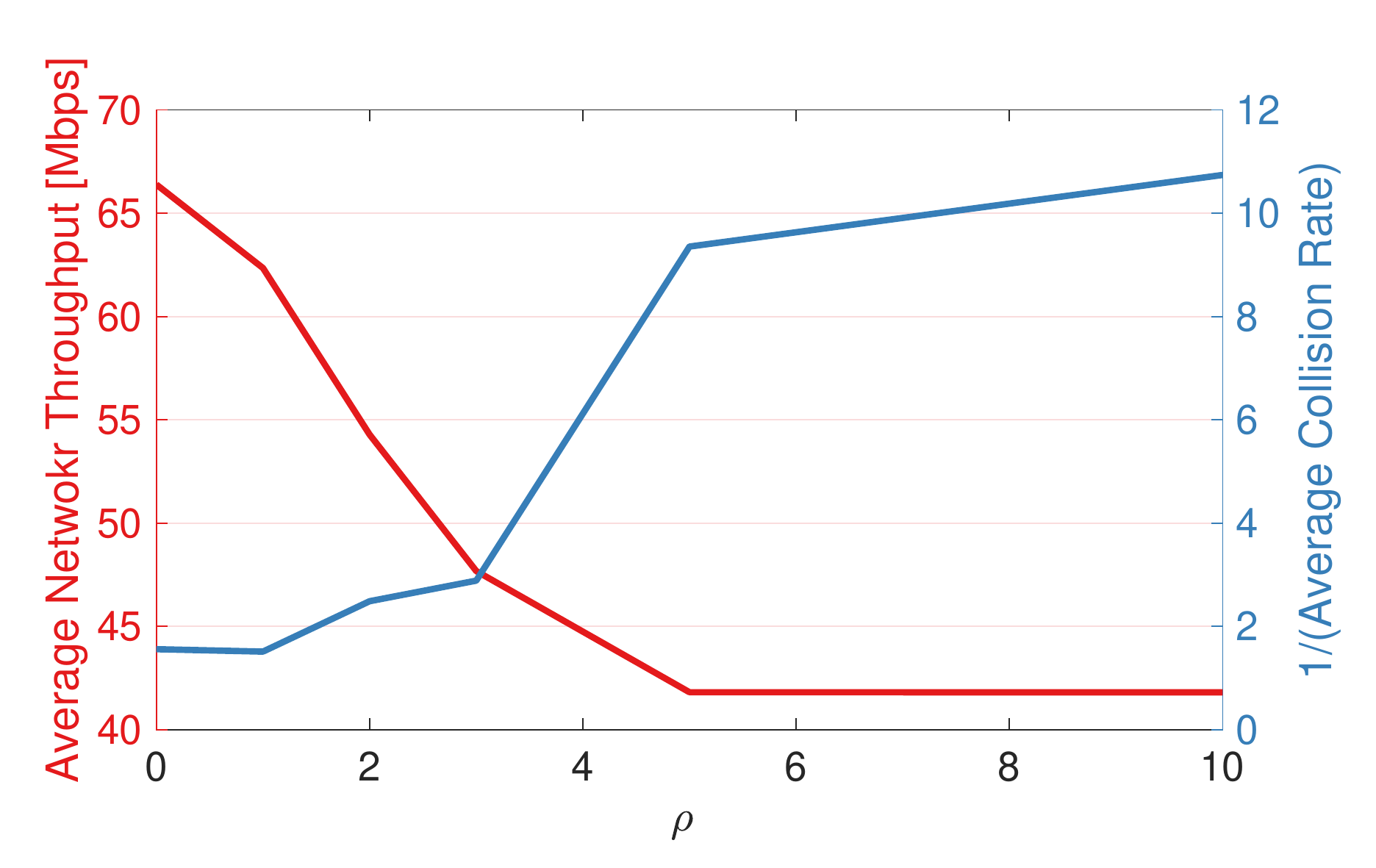}
  \caption{Comparison of eRACH over the collision aversion factor $\rho$.}
  \label{fig_Objective}
% \vskip -5pts
\end{figure}

As shown in Table \ref{Table_Objective} and Figs. \ref{Snapshot_RateMax} and \ref{fig_Objective}, the eRACH agent learns mainly maximizing throughput or mainly minimizing collision, depending on the collision aversion factor $\rho$.
% Here, \textit{Rate Max} is conducted with $\rho=0$ while \textit{Collision Min} with $\rho=2$.
% Those results show the tradeoff over throughput and collision with the coefficient ratio, $\sigma = \frac{\rho}{\rho_{\mathrm{R}}}$. 
Those results verify that eRACH method can be flexibly applied to various applications. 
% \tred{[JH: Explain more details, as we have spent one figure and one table for this.]}

\begin{table}[]
  %   \hspace{5pt}
    \centering          
    \caption{
    Impact of $\rho$ on average network throughput and average collision rate of eRACH ($K=2$, $I=22$, $J=5$, $P=2$).
    % Comparison of proposed eRACH over different objectives.
    }
    % \tred{[JH: Briefly specify some key hyperparameters here as in the above]}
    % \tred{[JH: If possible, remove this table, and draw Fig. 12 with shaded areas as in Fig. 9.]}
    \resizebox{1\columnwidth}{!}{\begin{minipage}[h]{.96\columnwidth}
    \centering
    \label{Table_Objective}
    \input{Table/Table_Objective}
    \end{minipage}}
\end{table}

%% file: Table/Table_Parameter.tex
\begin{tabular} {l l}
    % \centering
	%\begin{tabular} {P{1.2cm} P{0.001cm} P{0.001cm} P{0.001cm} P{0.001cm}P{0.001cm} P{0.001cm}}
	\toprule[1pt]
	\textbf{Parameter} & \textbf{Value} \\
% 	Location of UT & $\mathbf{q}_{j}=[0,0,0]^T$ \\  
% 	Location of Dst & $\mathbf{q}_{\mathcal{D}}=[4000,4000,0]^T$ \\
	\midrule
	Altitude of SAT & $H_{\mathrm{L}}=550$ {[}km{]} (from \cite{Starlink}) \\ 
	Velocity (Speed) of SAT1 ($k=1$) & $\mathbf{v}^{1}_{\mathrm{L}, \mathrm{I}}=[0, 7590, 0]^T$ ($7590$ {[}m/s{]}) \\
	Velocity of SAT2 ($k=2$) & $\mathbf{v}^{2}_{\mathrm{L}, \mathrm{I}}=[0, -7590, 0]^T$\\
	Radius of an orbit & $r_{\mathrm{E}} = 6921$ {[}km{]} \\ 
	Circumference of an orbital lane & $c_{\mathrm{E}} = 43486$ {[}km{]} \\ 
	Number of SATs per orbital lane & $I = 22$ \\ 
	Inter-SAT distance & $1977$ {[}km{]}  \\
	Number of UT & $J = 5$ \\
% 	Length an orbital line segment & $c_{\mathrm{C}} = 6000$ {[}km{]} \\ 
% 	Number of SATs per orbital line segment & $I' = 3$ \\ 
	\midrule
	Bandwidth & $B = 10^{8}$ {[}Hz{]} \\   
% 	Reference SNR ($d=1$ {[}m{]}) & $\gamma_{0} = 10^{11}$ \\ 
    Path loss exponent & $\Tilde{\alpha} = 2.1$    \\
    Parameter for LoS probability & $L_{1}=10$, $L_{2}=0.6$ \\
    Attenuation factor & $\kappa = 0.2$ \\
	\midrule
% 	Time step size & $\delta_t = \delta_{\mathrm{A}} + \delta_{\mathrm{D}}$  \\
	Random access duration & $\tau_{s} = 10$ {[}ms{]} \\
	Data transmission duration & $\tau_{d} = 90$ {[}ms{]}   \\
	Preamble signatures for each SAT & $P = 2 - 54$ \\
	Orbital period & $T = 5728$ {[}s{]} \\
	Number of time slots per episode & $N=2604$ \\
% 	Altitude of UAV & $H_{\mathrm{U}}=50$ {[}km{]} (from \cite{Intro_4}) \\
% 	Maximum acceleration of UAV & $A_{\mathrm{max}} = 5$ [m/s$^2$] \\
% 	Discretization level for $a_{j}^{\mathcal{A}}[n]$ & $D = 5$ \\
% 	\midrule
% 	\midrule
% 	Learning rate, Discount factor & $10^{-4}$, 0.995 \\ 
% 	Batch size, \# of iterations/update & 505, 505 \\ 
% 	\# of iterations/episode & 101 \\ 
% 	Training iterations of episode & 5000 \\
	\bottomrule[1pt]
\end{tabular}

%% file: Table/Table_Proposed.tex
\begin{tabularx}{1\linewidth}{l c c c c c}
\toprule[1pt]
{RA Scheme} & Avg. Thr. [Mbps] & Avg. Collision Rate & Avg. Access Delay [ms] & Jain's Fairness \cite{JainsFairness} & Cheap Talk \\
\cmidrule(lr){1-1} \cmidrule(lr){2-2} \cmidrule(lr){3-3} \cmidrule(lr){4-4} \cmidrule(lr){5-5} \cmidrule(lr){6-6}

Slotted ALOHA & $12.249$\; \tikz{
\draw[gray,line width=.3pt] (0,0) -- (1.1,0);
\draw[white, line width=0.01pt] (0,-2pt) -- (0,2pt);
\draw[black,line width=1pt] (0.05,0) -- (0.15,0);
\draw[black,line width=1pt] (0.05,-2pt) -- (0.05,2pt);
\draw[black,line width=1pt] (0.15,-2pt) -- (0.15,2pt);} 
& $0.9339$\; \tikz{
\draw[gray,line width=.3pt] (0,0) -- (1.1,0);
\draw[white, line width=0.01pt] (0,-2pt) -- (0,2pt);
\draw[black,line width=1pt] (0.9,0) -- (1.0,0);
\draw[black,line width=1pt] (0.9,-2pt) -- (0.9,2pt);
\draw[black,line width=1pt] (1.0,-2pt) -- (1.0,2pt);} 
& $1441.1$\; \tikz{
\draw[gray,line width=.3pt] (0,0) -- (1.3,0);
\draw[white, line width=0.01pt] (0,-2pt) -- (0,2pt);
\draw[black,line width=1pt] (1.2,0) -- (1.3,0);
\draw[black,line width=1pt] (1.2,-2pt) -- (1.2,2pt);
\draw[black,line width=1pt] (1.3,-2pt) -- (1.3,2pt);} 
& $0.8921$\; \tikz{
\draw[gray,line width=.3pt] (0,0) -- (1.1,0);
\draw[white, line width=0.01pt] (0,-2pt) -- (0,2pt);
\draw[black,line width=1pt] (0.68,0) -- (0.92,0);
\draw[black,line width=1pt] (0.68,-2pt) -- (0.68,2pt);
\draw[black,line width=1pt] (0.92,-2pt) -- (0.92,2pt);}
& $\times$
\\ 

RACH & $47.679$\; \tikz{
\draw[gray,line width=.3pt] (0,0) -- (1.1,0);
\draw[white, line width=0.01pt] (0,-2pt) -- (0,2pt);
\draw[black,line width=1pt] (0.36,0) -- (0.45,0);
\draw[black,line width=1pt] (0.36,-2pt) -- (0.36,2pt);
\draw[black,line width=1pt] (0.45,-2pt) -- (0.45,2pt);} 
& $0.1338$\; \tikz{
\draw[gray,line width=.3pt] (0,0) -- (1.1,0);
\draw[white, line width=0.01pt] (0,-2pt) -- (0,2pt);
\draw[black,line width=1pt] (0.05,0) -- (0.2,0);
\draw[black,line width=1pt] (0.05,-2pt) -- (0.05,2pt);
\draw[black,line width=1pt] (0.2,-2pt) -- (0.2,2pt);} 
& $293.1$\; \tikz{
\draw[gray,line width=.3pt] (0,0) -- (1.1,0);
\draw[white, line width=0.01pt] (0,-2pt) -- (0,2pt);
\draw[black,line width=1pt] (0.45,0) -- (0.55,0);
\draw[black,line width=1pt] (0.45,-2pt) -- (0.45,2pt);
\draw[black,line width=1pt] (0.55,-2pt) -- (0.55,2pt);} 
& $0.9917$\; \tikz{
\draw[gray,line width=.3pt] (0,0) -- (1.1,0);
\draw[white, line width=0.01pt] (0,-2pt) -- (0,2pt);
\draw[black,line width=1pt] (0.90,0) -- (1.01,0);
\draw[black,line width=1pt] (0.90,-2pt) -- (0.90,2pt);
\draw[black,line width=1pt] (1.01,-2pt) -- (1.01,2pt);}
& $\times$
\\ 

\textbf{eRACH}  & $62.347$\; \tikz{
\draw[gray,line width=.3pt] (0,0) -- (1.1,0);
\draw[white, line width=0.01pt] (0,-2pt) -- (0,2pt);
\draw[black,line width=1pt] (0.64,0) -- (0.82,0);
\draw[black,line width=1pt] (0.64,-2pt) -- (0.64,2pt);
\draw[black,line width=1pt] (0.82,-2pt) -- (0.82,2pt);}
& $0.6614$\; \tikz{
\draw[gray,line width=.3pt] (0,0) -- (1.1,0);
\draw[white, line width=0.01pt] (0,-2pt) -- (0,2pt);
\draw[black,line width=1pt] (0.55,0) -- (0.70,0);
\draw[black,line width=1pt] (0.55,-2pt) -- (0.55,2pt);
\draw[black,line width=1pt] (0.70,-2pt) -- (0.70,2pt);}
& $197.1$\; \tikz{
\draw[gray,line width=.3pt] (0,0) -- (1.1,0);
\draw[white, line width=0.01pt] (0,-2pt) -- (0,2pt);
\draw[black,line width=1pt] (0.19,0) -- (0.32,0);
\draw[black,line width=1pt] (0.19,-2pt) -- (0.19,2pt);
\draw[black,line width=1pt] (0.32,-2pt) -- (0.32,2pt);} 
& $0.9887$\; \tikz{
\draw[gray,line width=.3pt] (0,0) -- (1.1,0);
\draw[white, line width=0.01pt] (0,-2pt) -- (0,2pt);
\draw[black,line width=1pt] (0.88,0) -- (0.99,0);
\draw[black,line width=1pt] (0.88,-2pt) -- (0.88,2pt);
\draw[black,line width=1pt] (0.99,-2pt) -- (0.99,2pt);}
& $\times$
\\

\textbf{eRACH-Coop} & $73.716$\; \tikz{
\draw[gray,line width=.3pt] (0,0) -- (1.1,0);
\draw[white, line width=0.01pt] (0,-2pt) -- (0,2pt);
\draw[black,line width=1pt] (0.90,0) -- (1.04,0);
\draw[black,line width=1pt] (0.90,-2pt) -- (0.90,2pt);
\draw[black,line width=1pt] (1.04,-2pt) -- (1.04,2pt);}
& $0.4000$\; \tikz{
\draw[gray,line width=.3pt] (0,0) -- (1.1,0);
\draw[white, line width=0.01pt] (0,-2pt) -- (0,2pt);
\draw[black,line width=1pt] (0.31,0) -- (0.43,0);
\draw[black,line width=1pt] (0.31,-2pt) -- (0.31,2pt);
\draw[black,line width=1pt] (0.43,-2pt) -- (0.43,2pt);}
& $151.2$\; \tikz{
\draw[gray,line width=.3pt] (0,0) -- (1.1,0);
\draw[white, line width=0.01pt] (0,-2pt) -- (0,2pt);
\draw[black,line width=1pt] (0.1,0) -- (0.25,0);
\draw[black,line width=1pt] (0.1,-2pt) -- (0.1,2pt);
\draw[black,line width=1pt] (0.25,-2pt) -- (0.25,2pt);} 
& $0.8000$\; \tikz{
\draw[gray,line width=.3pt] (0,0) -- (1.1,0);
\draw[white, line width=0.01pt] (0,-2pt) -- (0,2pt);
\draw[black,line width=1pt] (0.62,0) -- (0.77,0);
\draw[black,line width=1pt] (0.62,-2pt) -- (0.62,2pt);
\draw[black,line width=1pt] (0.77,-2pt) -- (0.77,2pt);}
& $\circ$
\\

\bottomrule[1pt]
\end{tabularx}

%% file: Table/Table_Snapshot.tex
\newcolumntype{R}{>{\raggedleft\arraybackslash}X}

% \tablefontsize
\begin{tabularx}{1\linewidth}{Xllll}
    \toprule[1pt] 
    &\hspace{30pt} Time Slot = 1  &\hspace{30pt} Time Slot = 2 &\hspace{30pt} Time Slot = 3 &\hspace{30pt} Time Slot = 4 \\
    \cmidrule(lr){2-2} \cmidrule(lr){3-3} \cmidrule(lr){4-4} \cmidrule(lr){5-5}
    RACH
        & \raisebox{-.5\height}{\includegraphics[width=.214\linewidth]{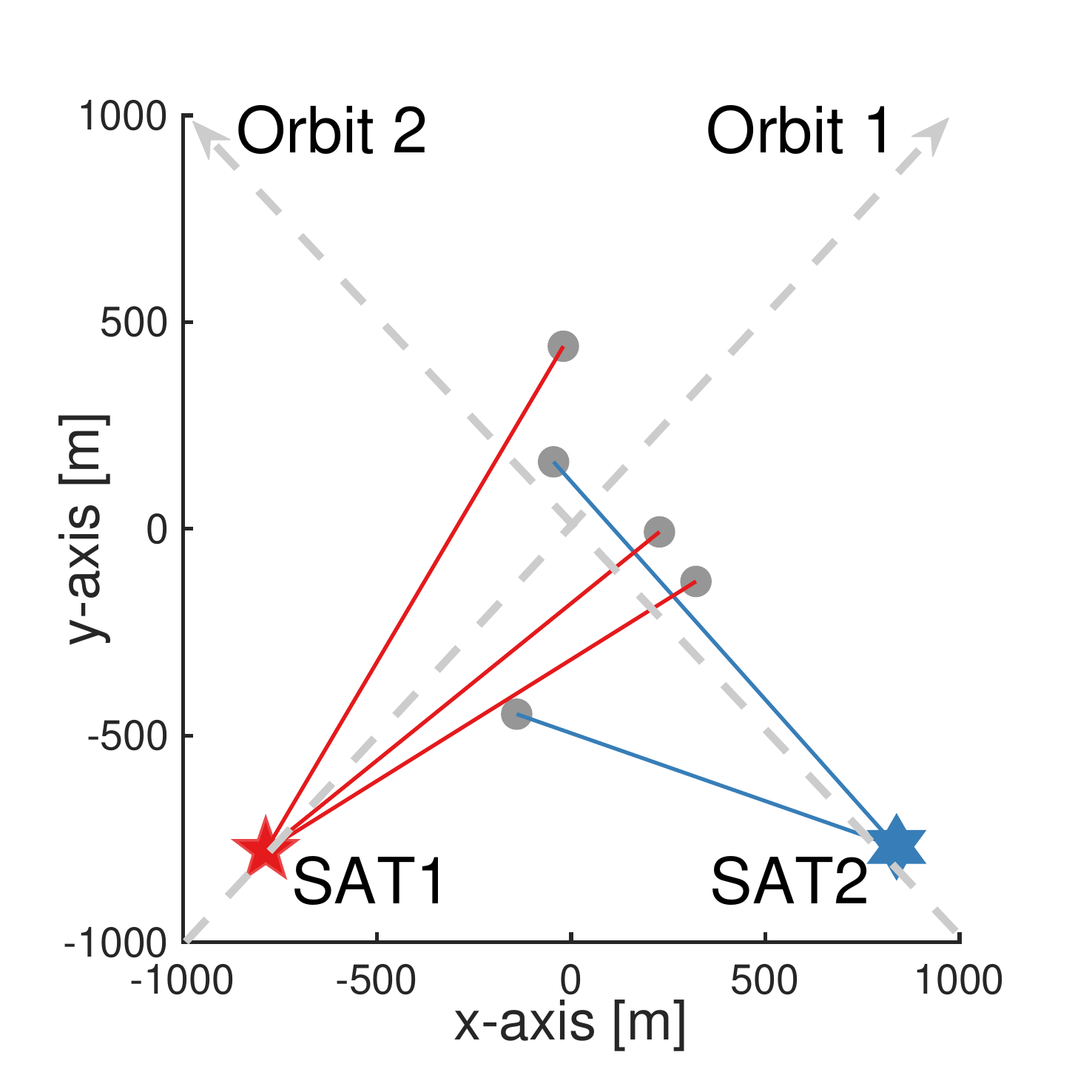}}
        & \raisebox{-.5\height}{\includegraphics[width=.214\linewidth]{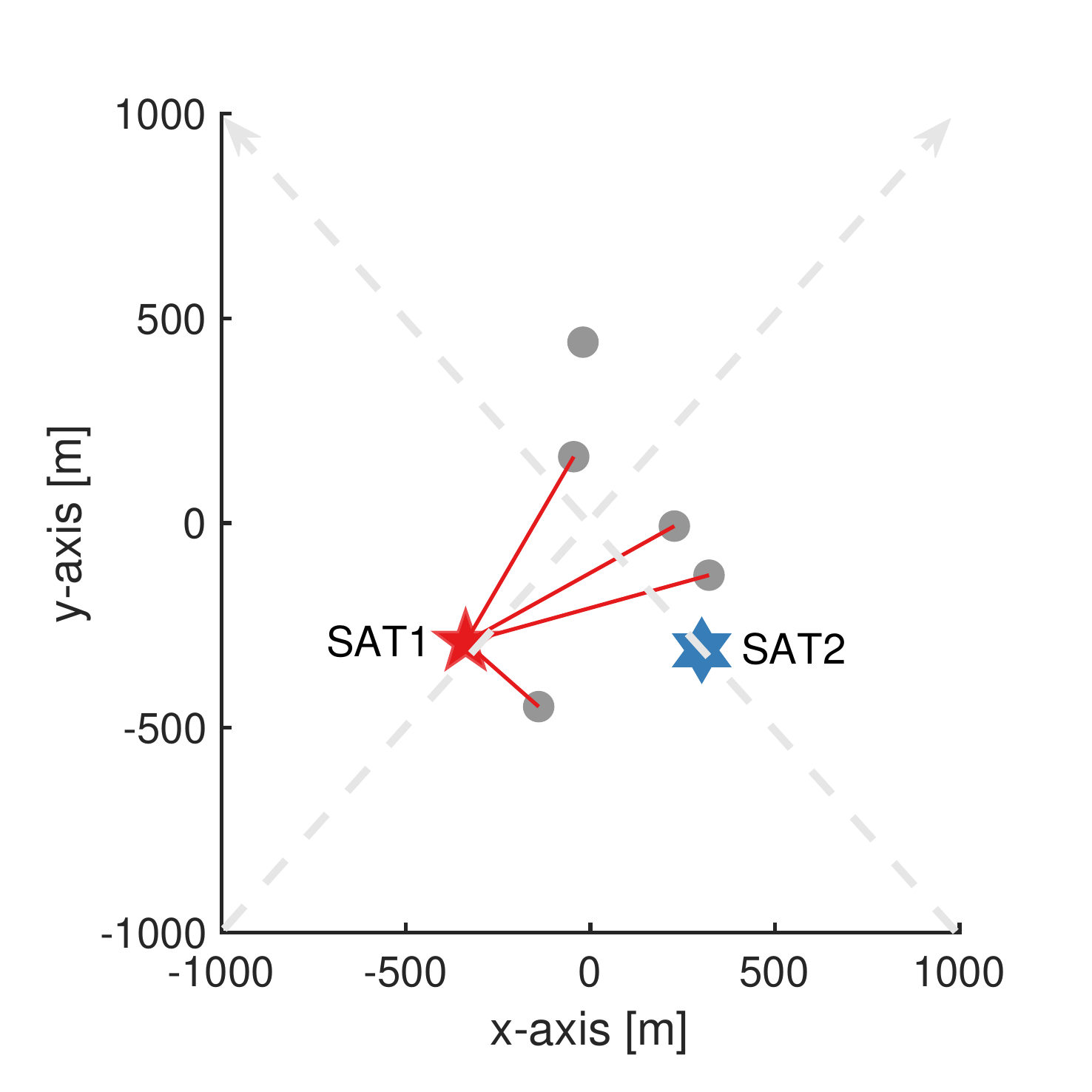}}
        & \raisebox{-.5\height}{\includegraphics[width=.214\linewidth]{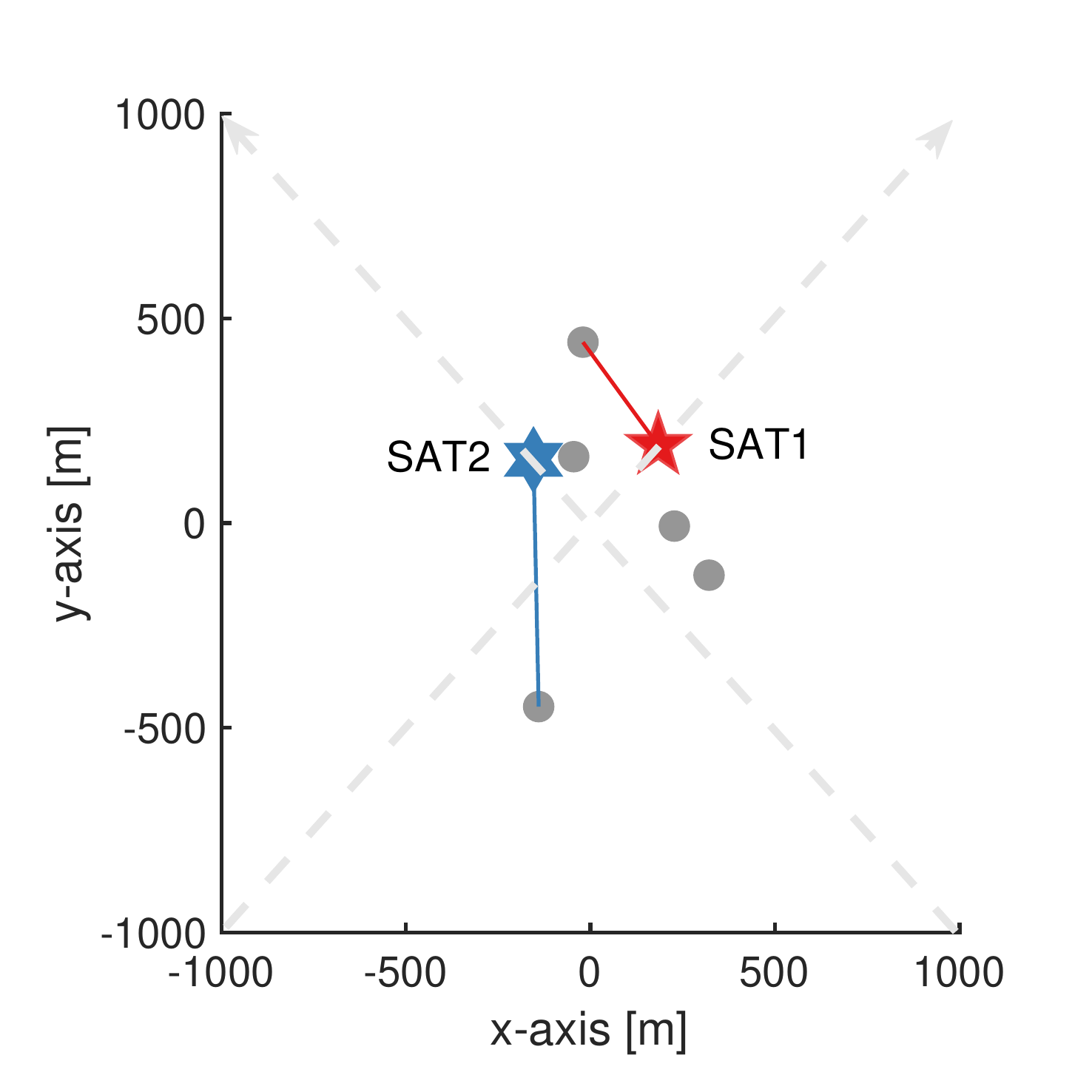}}
        & \raisebox{-.5\height}{\includegraphics[width=.214\linewidth]{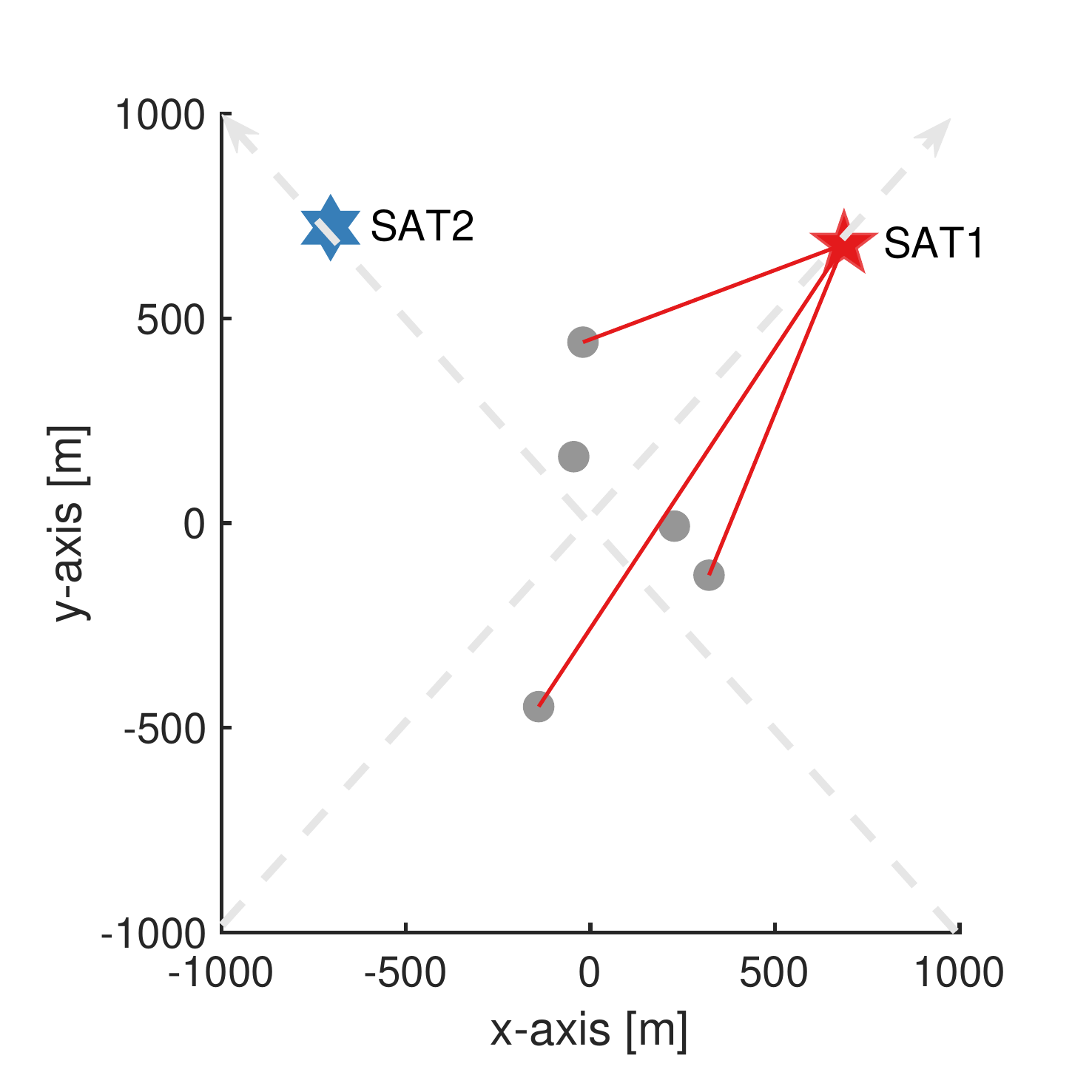}}
        \\
    \cmidrule(lr){2-2} \cmidrule(lr){3-3} \cmidrule(lr){4-4} \cmidrule(lr){5-5}
    % \scriptsize{Collision~Pr.}
    %     &\hspace{20pt} \tikz{
    %     \fill[fill=color1] (0.0,0) rectangle (1.5,0.2);
    %     \fill[pattern=north west lines, pattern color=black!30!color1] (0.0,0) rectangle (1.5,0.2);}
    %     &\hspace{20pt} \tikz{
    %     \fill[fill=color1] (0.0,0) rectangle (2.3,0.2);
    %     \fill[pattern=north west lines, pattern color=black!30!color1] (0.0,0) rectangle (2.3,0.2);}
    %     &\hspace{20pt} \tikz{
    %     \fill[fill=color1] (0.0,0) rectangle (0.6,0.2);
    %     \fill[pattern=north west lines, pattern color=black!30!color1] (0.0,0) rectangle (0.6,0.2);}
    %     &\hspace{20pt} \tikz{
    %     \fill[fill=color1] (0.0,0) rectangle (1.5,0.2);
    %     \fill[pattern=north west lines, pattern color=black!30!color1] (0.0,0) rectangle (1.5,0.2);}
    %     \\
    \scriptsize{Resource~Util.}
        &\hspace{15pt} \scriptsize{67\%} \tikz{
        \fill[fill=color2] (0.0,0) rectangle (1.3,0.2);
        \fill[pattern=north west lines, pattern color=black!30!color2] (0.0,0) rectangle (1.3,0.2);}
        &\hspace{15pt} \scriptsize{27\%} \tikz{
        \fill[fill=color2] (0.0,0) rectangle (0.5,0.2);
        \fill[pattern=north west lines, pattern color=black!30!color2] (0.0,0) rectangle (0.5,0.2);}
        &\hspace{15pt} \scriptsize{42\%} \tikz{
        \fill[fill=color2] (0.0,0) rectangle (0.9,0.2);
        \fill[pattern=north west lines, pattern color=black!30!color2] (0.0,0) rectangle (0.9,0.2);}
        &\hspace{15pt} \scriptsize{24\%} \tikz{
        \fill[fill=color2] (0.0,0) rectangle (0.5,0.2);
        \fill[pattern=north west lines, pattern color=black!30!color2] (0.0,0) rectangle (0.5,0.2);}
        \\
        &\hspace{23pt} \tikz{\draw[black] (1.1,0) -- (3.20,0);
        \draw[black] (1.1,-2pt) -- (1.1,2pt)node[anchor=north] {\tiny{0 \%}};
        \draw[black] (1.8,-2pt) -- (1.8,2pt)node[anchor=north] {};
        \draw[black] (2.5,-2pt) -- (2.5,2pt)node[anchor=north] {};
        \draw[black] (3.20,-2pt) -- (3.20,2pt)node[anchor=north] {\tiny{100 \%}};} 
        &\hspace{23pt} \tikz{\draw[black] (1.1,0) -- (3.20,0);
        \draw[black] (1.1,-2pt) -- (1.1,2pt)node[anchor=north] {\tiny{0 \%}};
        \draw[black] (1.8,-2pt) -- (1.8,2pt)node[anchor=north] {};
        \draw[black] (2.5,-2pt) -- (2.5,2pt)node[anchor=north] {};
        \draw[black] (3.20,-2pt) -- (3.20,2pt)node[anchor=north] {\tiny{100 \%}};} 
        &\hspace{23pt} \tikz{\draw[black] (1.1,0) -- (3.20,0);
        \draw[black] (1.1,-2pt) -- (1.1,2pt)node[anchor=north] {\tiny{0 \%}};
        \draw[black] (1.8,-2pt) -- (1.8,2pt)node[anchor=north] {};
        \draw[black] (2.5,-2pt) -- (2.5,2pt)node[anchor=north] {};
        \draw[black] (3.20,-2pt) -- (3.20,2pt)node[anchor=north] {\tiny{100 \%}};} 
        &\hspace{23pt} \tikz{\draw[black] (1.1,0) -- (3.20,0);
        \draw[black] (1.1,-2pt) -- (1.1,2pt)node[anchor=north] {\tiny{0 \%}};
        \draw[black] (1.8,-2pt) -- (1.8,2pt)node[anchor=north] {};
        \draw[black] (2.5,-2pt) -- (2.5,2pt)node[anchor=north] {};
        \draw[black] (3.20,-2pt) -- (3.20,2pt)node[anchor=north] {\tiny{100 \%}};} 
        \\
    %%%%%%%%%%%%%%%%%%%%%%%%%%%%%%%%%%%%%%%%%%%%%%%%%%%%%%%%%%%%%%%%%%%%%%%%%%%%
    % \cmidrule(lr){2-2} \cmidrule(lr){3-3} \cmidrule(lr){4-4} \cmidrule(lr){5-5}
    \cmidrule(lr){1-5}
    eRACH
        & \raisebox{-.5\height}{\includegraphics[width=.214\linewidth]{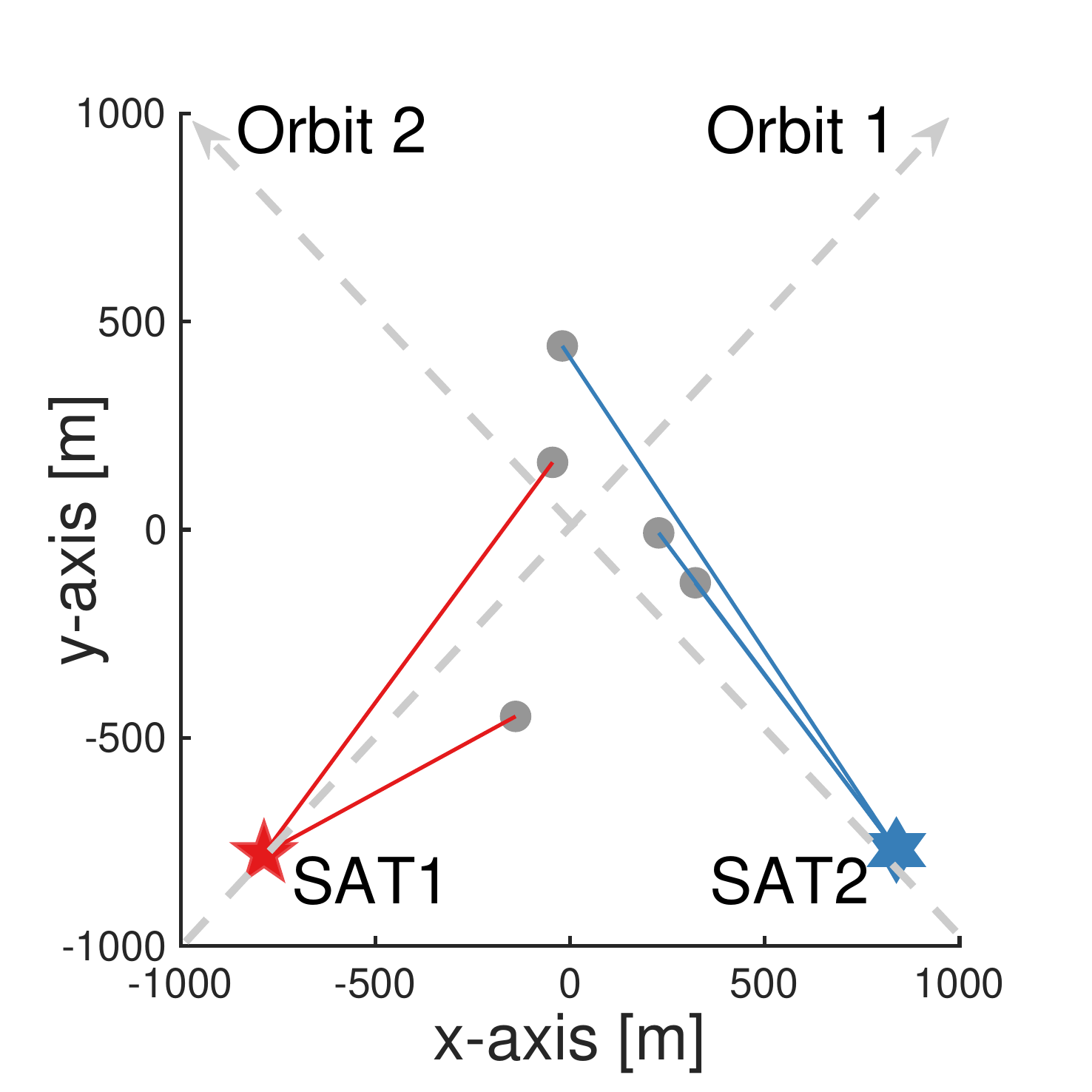}}
        & \raisebox{-.5\height}{\includegraphics[width=.214\linewidth]{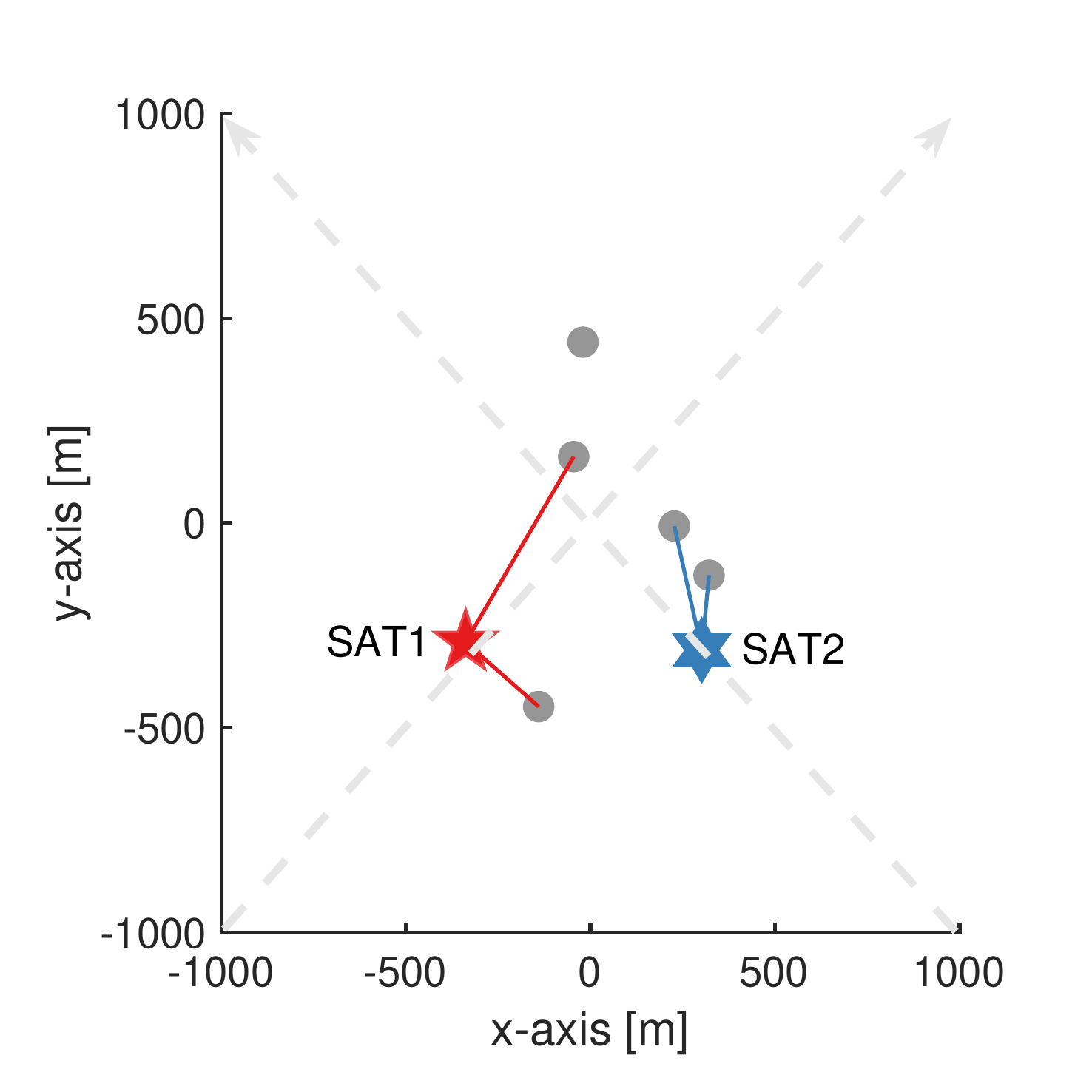}}
        & \raisebox{-.5\height}{\includegraphics[width=.214\linewidth]{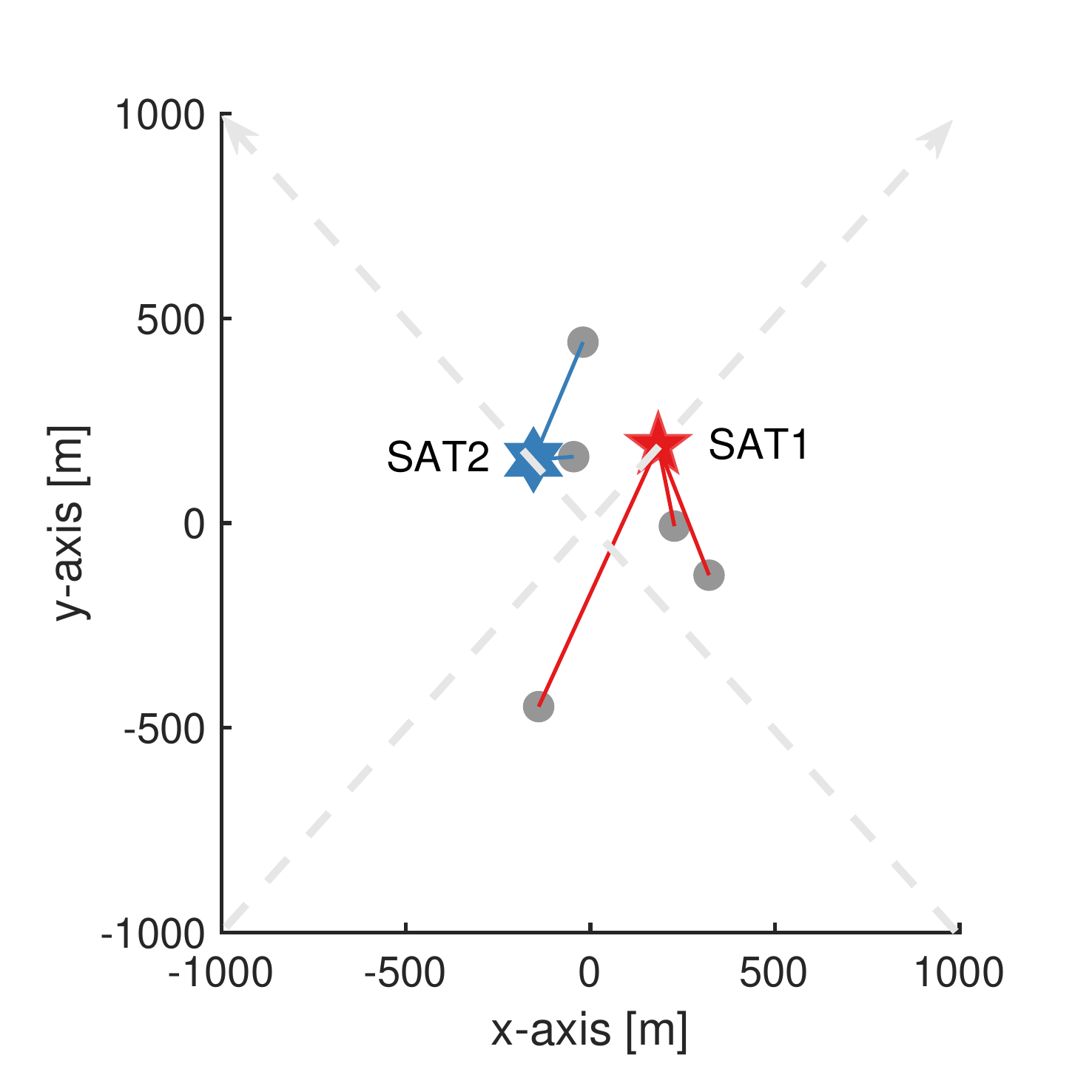}}
        & \raisebox{-.5\height}{\includegraphics[width=.214\linewidth]{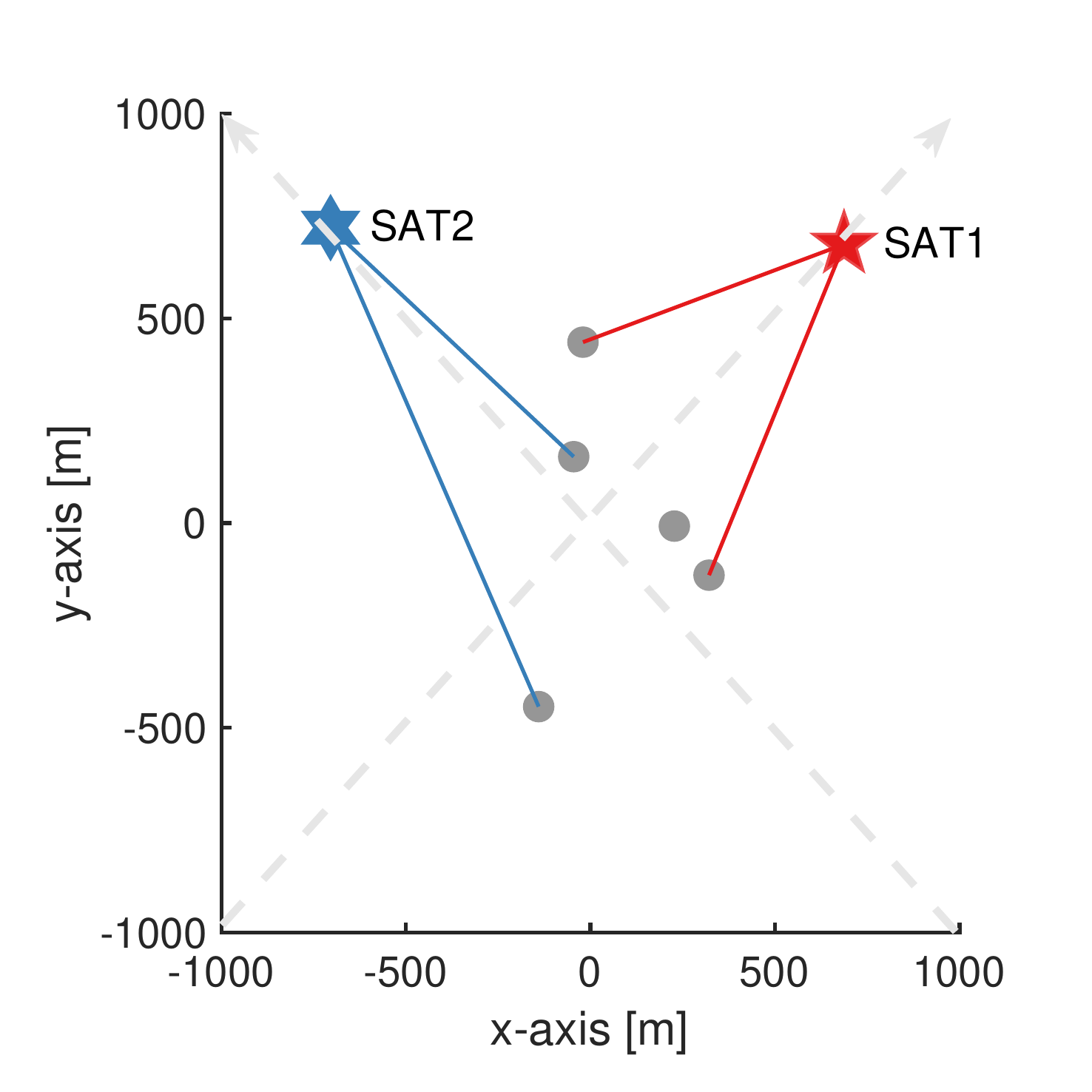}}
        \\
    \cmidrule(lr){2-2} \cmidrule(lr){3-3} \cmidrule(lr){4-4} \cmidrule(lr){5-5}
    % \scriptsize{Collision~Pr.}
    %     &\hspace{20pt} \tikz{
    %     \fill[fill=color1] (0.0,0) rectangle (1.5,0.2);
    %     \fill[pattern=north west lines, pattern color=black!30!color1] (0.0,0) rectangle (1.5,0.2);}
    %     &\hspace{20pt} \tikz{
    %     \fill[fill=color1] (0.0,0) rectangle (0.9,0.2);
    %     \fill[pattern=north west lines, pattern color=black!30!color1] (0.0,0) rectangle (0.9,0.2);}
    %     &\hspace{20pt} \tikz{
    %     \fill[fill=color1] (0.0,0) rectangle (1.5,0.2);
    %     \fill[pattern=north west lines, pattern color=black!30!color1] (0.0,0) rectangle (1.5,0.2);}
    %     &\hspace{20pt} \tikz{
    %     \fill[fill=color1] (0.0,0) rectangle (0.9,0.2);
    %     \fill[pattern=north west lines, pattern color=black!30!color1] (0.0,0) rectangle (0.9,0.2);}
    %     \\
    \scriptsize{Resource~Util.}
        &\hspace{15pt} \scriptsize{77\%} \tikz{
        \fill[fill=color2] (0.0,0) rectangle (1.55,0.2);
        \fill[pattern=north west lines, pattern color=black!30!color2] (0.0,0) rectangle (1.55,0.2);}
        &\hspace{15pt} \scriptsize{99\%} \tikz{
        \fill[fill=color2] (0.0,0) rectangle (2.0,0.2);
        \fill[pattern=north west lines, pattern color=black!30!color2] (0.0,0) rectangle (2.0,0.2);}
        &\hspace{15pt} \scriptsize{79\%} \tikz{
        \fill[fill=color2] (0.0,0) rectangle (1.6,0.2);
        \fill[pattern=north west lines, pattern color=black!30!color2] (0.0,0) rectangle (1.6,0.2);}
        &\hspace{15pt} \scriptsize{97\%} \tikz{
        \fill[fill=color2] (0.0,0) rectangle (1.95,0.2);
        \fill[pattern=north west lines, pattern color=black!30!color2] (0.0,0) rectangle (1.95,0.2);}
        \\
        &\hspace{23pt} \tikz{\draw[black] (1.1,0) -- (3.20,0);
        \draw[black] (1.1,-2pt) -- (1.1,2pt)node[anchor=north] {\tiny{0 \%}};
        \draw[black] (1.8,-2pt) -- (1.8,2pt)node[anchor=north] {};
        \draw[black] (2.5,-2pt) -- (2.5,2pt)node[anchor=north] {};
        \draw[black] (3.20,-2pt) -- (3.20,2pt)node[anchor=north] {\tiny{100 \%}};} 
        &\hspace{23pt} \tikz{\draw[black] (1.1,0) -- (3.20,0);
        \draw[black] (1.1,-2pt) -- (1.1,2pt)node[anchor=north] {\tiny{0 \%}};
        \draw[black] (1.8,-2pt) -- (1.8,2pt)node[anchor=north] {};
        \draw[black] (2.5,-2pt) -- (2.5,2pt)node[anchor=north] {};
        \draw[black] (3.20,-2pt) -- (3.20,2pt)node[anchor=north] {\tiny{100 \%}};} 
        &\hspace{23pt} \tikz{\draw[black] (1.1,0) -- (3.20,0);
        \draw[black] (1.1,-2pt) -- (1.1,2pt)node[anchor=north] {\tiny{0 \%}};
        \draw[black] (1.8,-2pt) -- (1.8,2pt)node[anchor=north] {};
        \draw[black] (2.5,-2pt) -- (2.5,2pt)node[anchor=north] {};
        \draw[black] (3.20,-2pt) -- (3.20,2pt)node[anchor=north] {\tiny{100 \%}};} 
        &\hspace{23pt} \tikz{\draw[black] (1.1,0) -- (3.20,0);
        \draw[black] (1.1,-2pt) -- (1.1,2pt)node[anchor=north] {\tiny{0 \%}};
        \draw[black] (1.8,-2pt) -- (1.8,2pt)node[anchor=north] {};
        \draw[black] (2.5,-2pt) -- (2.5,2pt)node[anchor=north] {};
        \draw[black] (3.20,-2pt) -- (3.20,2pt)node[anchor=north] {\tiny{100 \%}};} 
        \\    
    \bottomrule[1pt] 
\end{tabularx}

%% file: Table/Table_PositionError.tex
\newcolumntype{R}{>{\raggedleft\arraybackslash}X}
\begin{tabularx}{1\linewidth}{l l l}
    \toprule[1pt]
    Position. Err. & Norm. Reward & \# of Ep. for Convergence \\     
    \cmidrule(lr){1-1} \cmidrule(lr){2-2} \cmidrule(lr){3-3}
    \scriptsize{$\sigma^2 \simeq 0$} & \scriptsize{$0$} \hspace{18pt}  \tikz{
        % \fill[fill=lightgray,fill opacity=0.1] (0,0) rectangle (2.2,0.2);
        \fill[fill=color1] (0.0,0) rectangle (1.2,0.2);
        \fill[pattern=north west lines, pattern color=black!30!color1] (0.0,0) rectangle (1.2,0.2);
    } & \scriptsize{$\simeq 200$} \hspace{1pt}  \tikz{
        % \fill[fill=lightgray,fill opacity=0.1] (0,0) rectangle (2.2,0.2);
        \fill[fill=color1] (0.0,0) rectangle (0.1,0.2);
        \fill[pattern=north west lines, pattern color=black!30!color1] (0.0,0) rectangle (0.1,0.2);
    } \\
    \scriptsize{$\sigma^2 < 10^2$} & \scriptsize{$0$} \hspace{18pt}  \tikz{
        % \fill[fill=lightgray,fill opacity=0.1] (0,0) rectangle (2.2,0.2);
        \fill[fill=color2] (0.0,0) rectangle (1.2,0.2);
        \fill[pattern=north west lines, pattern color=black!30!color2] (0.0,0) rectangle (1.2,0.2);
    } & \scriptsize{$< 300$} \hspace{1pt}  \tikz{
        % \fill[fill=lightgray,fill opacity=0.1] (0,0) rectangle (2.2,0.2);
        \fill[fill=color2] (0.0,0) rectangle (0.15,0.2);
        \fill[pattern=north west lines, pattern color=black!30!color2] (0.0,0) rectangle (0.15,0.2);
    } \\
    \scriptsize{$10^2 \!< \sigma^2 \!< 10^3$} & \scriptsize{$<\!-5$} \hspace{5pt}  \tikz{
        \fill[fill=color3] (0.0,0) rectangle (1.1,0.2);
        \fill[pattern=north west lines, pattern color=black!30!color3] (0.0,0) rectangle (1.1,0.2);
    } & \scriptsize{$<700$} \hspace{1pt}  \tikz{
        % \fill[fill=lightgray,fill opacity=0.1] (0,0) rectangle (2.2,0.2);
        \fill[fill=color3] (0.0,0) rectangle (0.5,0.2);
        \fill[pattern=north west lines, pattern color=black!30!color3] (0.0,0) rectangle (0.5,0.2);
    } \\
    \scriptsize{$10^3 \!< \sigma^2 \!< 10^4$} & \scriptsize{$<\!-45$} \hspace{1pt}  \tikz{
        \fill[fill=color4] (0.0,0) rectangle (0.1,0.2);
        \fill[pattern=north west lines, pattern color=black!30!color4] (0.0,0) rectangle (0.1,0.2);
    } & \scriptsize{$<2000$} \hspace{-3pt}  \tikz{
        % \fill[fill=lightgray,fill opacity=0.1] (0,0) rectangle (2.2,0.2);
        \fill[fill=color4] (0.0,0) rectangle (1.4,0.2);
        \fill[pattern=north west lines, pattern color=black!30!color4] (0.0,0) rectangle (1.4,0.2);
    } \\
    \scriptsize{$\sigma^2 > 10^4$} & \scriptsize{Diverge} \hspace{1pt}  \tikz{
        \fill[fill=color5] (0.0,0) rectangle (0,0.2);
        \fill[pattern=north west lines, pattern color=black!30!color5] (0.0,0) rectangle (0.0,0.2);
    } & \scriptsize{Diverge} \hspace{0pt}  \tikz{
        % \fill[fill=lightgray,fill opacity=0.1] (0,0) rectangle (2.2,0.2);
        \fill[fill=color6] (0.0,0) rectangle (1.9,0.15);
        \fill[pattern=north west lines, pattern color=black!30!color6] (0.0,0) rectangle (1.9,0.15);
    } \vspace{-.3em} \\
    & \ \ \ \ \ \ \tikz{
        \draw[black] (0.0,0) -- (1.2,0);
        \draw[black] (0.0,-2pt) -- (0.0,2pt)node[anchor=north] {};
        \draw[black] (-0.1,-0.1pt) -- (-0.1,0.1pt)node[anchor=north] {\tiny$-50$};
        \draw[black] (0.6,-2pt) -- (0.6,2pt)node[anchor=north] {};
        \draw[black] (0.5,-0.1pt) -- (0.5,0.1pt)node[anchor=north] {\tiny$-25$};
        \draw[black] (1.2,-2pt) -- (1.2,2pt)node[anchor=north] {};
        \draw[black] (1.15,-0.1pt) -- (1.15,0.1pt)node[anchor=north] {\tiny$0$};
    } 
    & \ \ \ \ \ \ \tikz{
        \draw[black] (0.0,0) -- (1.3,0);
        \draw[color6] (1.3,0) -- (1.9,0);
        \draw[black] (0.0,-2pt) -- (0.0,2pt)node[anchor=north] {};
        \draw[black] (-0.1,-0.1pt) -- (-0.1,0.1pt)node[anchor=north] {\tiny$200$};
        \draw[black] (0.6,-2pt) -- (0.6,2pt)node[anchor=north] {};
        \draw[black] (0.5,-0.1pt) -- (0.5,0.1pt)node[anchor=north] {\tiny$1000$};
        \draw[black] (1.3,-2pt) -- (1.3,2pt)node[anchor=north] {};
        \draw[black] (1.2,-0.1pt) -- (1.2,0.1pt)node[anchor=north] {\tiny$2000$};
        \draw[black] (1.9,-2pt) -- (1.9,2pt)node[anchor=north] {};
        \draw[black] (1.85,-0.1pt) -- (1.85,0.1pt)node[anchor=north] {\tiny$\infty$};
    } \vspace{-.6em} \\
    \bottomrule[1pt]
\end{tabularx}

%% file: Table/Table_Objective.tex
% \begin{tabularx}{1\linewidth}{l c c c}
% 			\toprule[1pt]
% 			{Objective} & Avg. Thr. [Mbps] & Avg. Collision Rate & Avg. Access Delay [ms] \\
% 			\cmidrule(lr){1-1} \cmidrule(lr){2-2} \cmidrule(lr){3-3} \cmidrule(lr){4-4} 
\begin{tabularx}{1\linewidth}{l c c }
			\toprule[1pt]
			{Objective} & Avg. Thr. [Mbps] & Avg. Collision Rate  \\
			\cmidrule(lr){1-1} \cmidrule(lr){2-2} \cmidrule(lr){3-3}

\textit{Rate-Max} ($\rho = 0$)  & $66.391$\; \tikz{
\draw[gray,line width=.3pt] (0,0) -- (1.1,0);
\draw[white, line width=0.01pt] (0,-2pt) -- (0,2pt);
\draw[black,line width=1pt] (0.71,0) -- (0.86,0);
\draw[black,line width=1pt] (0.71,-2pt) -- (0.71,2pt);
\draw[black,line width=1pt] (0.86,-2pt) -- (0.86,2pt);}
& $0.6713$\; \tikz{
\draw[gray,line width=.3pt] (0,0) -- (1.1,0);
\draw[white, line width=0.01pt] (0,-2pt) -- (0,2pt);
\draw[black,line width=1pt] (0.60,0) -- (0.75,0);
\draw[black,line width=1pt] (0.60,-2pt) -- (0.60,2pt);
\draw[black,line width=1pt] (0.75,-2pt) -- (0.75,2pt);}
% & $204.2$\; \tikz{
% \draw[gray,line width=.3pt] (0,0) -- (1.1,0);
% \draw[white, line width=0.01pt] (0,-2pt) -- (0,2pt);
% \draw[black,line width=1pt] (0.21,0) -- (0.37,0);
% \draw[black,line width=1pt] (0.21,-2pt) -- (0.21,2pt);
% \draw[black,line width=1pt] (0.37,-2pt) -- (0.37,2pt);} 
\\

\textit{Collision-Aware} ($\rho = 2$) & $54.285$\; \tikz{
\draw[gray,line width=.3pt] (0,0) -- (1.1,0);
\draw[white, line width=0.01pt] (0,-2pt) -- (0,2pt);
\draw[black,line width=1pt] (0.35,0) -- (0.51,0);
\draw[black,line width=1pt] (0.35,-2pt) -- (0.35,2pt);
\draw[black,line width=1pt] (0.51,-2pt) -- (0.51,2pt);}
& $0.4020$\; \tikz{
\draw[gray,line width=.3pt] (0,0) -- (1.1,0);
\draw[white, line width=0.01pt] (0,-2pt) -- (0,2pt);
\draw[black,line width=1pt] (0.211,0) -- (0.33,0);
\draw[black,line width=1pt] (0.211,-2pt) -- (0.211,2pt);
\draw[black,line width=1pt] (0.33,-2pt) -- (0.33,2pt);}
% & $241.2$\; \tikz{
% \draw[gray,line width=.3pt] (0,0) -- (1.1,0);
% \draw[white, line width=0.01pt] (0,-2pt) -- (0,2pt);
% \draw[black,line width=1pt] (0.47,0) -- (0.59,0);
% \draw[black,line width=1pt] (0.47,-2pt) -- (0.47,2pt);
% \draw[black,line width=1pt] (0.59,-2pt) -- (0.59,2pt);} 
\\

\bottomrule[1pt]
\end{tabularx}